%% file: main.tex
\documentclass[twocolumn]{aastex631} % Use for publication
\usepackage[figure,figure*]{hypcap}
\usepackage{longtable, hyperref, graphicx, subfigure, comment}
\usepackage[fleqn]{amsmath}
\usepackage{amstext, float}
\usepackage{newtxmath} %use times font for math
\usepackage{lmodern}
\usepackage[T1]{fontenc}
\usepackage{url, footmisc}
\usepackage{soul}
\usepackage{booktabs}
\usepackage{todonotes}

\newcounter{column_number}
\newcommand{\Msun}{\ifmmode {M_{\odot}}\else${M_{\odot}}$\fi}
\newcommand{\Rsun}{\ifmmode {R_{\odot}}\else${R_{\odot}}$\fi}
\newcommand{\Lsun}{\ifmmode {L_{\odot}}\else${L_{\odot}}$\fi}
\newcommand{\lapprox }{{\lower0.8ex\hbox{$\buildrel <\over\sim$}}}
\newcommand{\gapprox }{{\lower0.8ex\hbox{$\buildrel >\over\sim$}}}
\newcommand{\degree}{\ifmmode {^\circ}\else$^\circ$\fi}
\newcommand{\Ro}{\ifmmode {R_\mathrm{o}}\else$R_\mathrm{o}$\fi}

\def\amin{\ifmmode^{\prime}\else$^{\prime}$\fi}
\def\asec{\ifmmode^{\prime\prime}\else$^{\prime\prime}$\fi}
\def\ergcms{erg~cm$^{-2}$~s$^{-1}$}

\def\halpha{$\mathrm{H}\alpha$}
\def\fx{$f_\mathrm{X}$}
\def\ffx{$f_{\mathrm{X}}$/$f_{\mathrm{bol}}$}
\def\LX{$L_{\mathrm{X}}$}
\def\Lbol{$L_{\mathrm{bol}}$}
\def\LLX{$L_{\mathrm{X}}$/$L_{\mathrm{bol}}$}

\def\prot{$P_{\mathrm{rot}}$}
\def\Pmem{$P_{\mathrm{mem}}$}
\def\gminusk{($G - K$)}
\def\bpminusrp{($G_\mathrm{BP} - G_\mathrm{RP}$)}

\defcitealias{Douglas2016}{D16}
\defcitealias{Douglas2017}{D17}

\shorttitle{X-ray Activity and Rotation in Praesepe and Hyades}
\shortauthors{N\'u\~nez et al.}

\begin{document}

\title{\sc The Factory and the Beehive.~IV.~A Comprehensive Study of the Rotation X-ray Activity Relation in Praesepe and the Hyades}

\correspondingauthor{Alejandro N\'u{\~n}ez}
\email{alejo.nunez@gmail.com}

\newcommand{\columbia}{Department of Astronomy, Columbia University, 550 West 120th Street, New York, NY 10027, USA}

\newcommand{\lafayette}{Department of Physics, Lafayette College, 730 High St, Easton, PA 18042, USA}

\newcommand{\wwu}{Department of Physics \& Astronomy, Western Washington University, Bellingham, WA 98225, USA}

\newcommand{\cfa}{Center for Astrophysics $\vert$ Harvard $\&$ Smithsonian, 60 Garden St, Cambridge, MA 02138, USA}

\newcommand{\unc}{Department of Physics and Astronomy, University of North Carolina, Chapel Hill, NC 27599, USA}

\newcommand{\ut}{Department of Astronomy, The University of Texas at Austin, Austin, TX 78712, USA}

\newcommand{\exeter}{Department of Physics and Astronomy, Stocker Road, University of Exeter, EX4 4QL, UK}

\author[0000-0002-8047-1982]{Alejandro N\'u{\~n}ez}
\altaffiliation{NSF MPS-Ascend Postdoctoral Research Fellow}
\affil{\columbia}

\author[0000-0001-7077-3664]{Marcel A.~Ag\"{u}eros}
\affil{\columbia}

\author[0000-0001-6914-7797]{Kevin R.~Covey}
\affiliation{\wwu}

\author[0000-0001-7371-2832]{Stephanie T.\ Douglas}
\affiliation{\lafayette}

\author[0000-0002-0210-2276]{Jeremy J.\ Drake}
\affiliation{\cfa}

\author[0000-0001-7337-5936]{Rayna Rampalli}
\affiliation{Department of Physics and Astronomy, Dartmouth College, Hanover, NH 03755, USA}

\author{Emily~C.~Bowsher}
\affiliation{\columbia}

\author[0000-0002-1617-8917]{Phillip A.~Cargile}
\affiliation{\cfa}

\author[0000-0001-9811-568X]{Adam L.~Kraus}
\affiliation{\ut}

\author[0000-0001-9380-6457]{Nicholas M.~Law}
\affiliation{\unc}

\begin{abstract}
% max 250 words
X-ray observations of low-mass stars in open clusters are critical to understanding the dependence of magnetic activity on stellar properties and their evolution. Praesepe and the Hyades, two of the nearest, most-studied open clusters, are among the best available laboratories for examining the dependence of magnetic activity on rotation for stars with masses $\lapprox$1 \Msun. We present an updated study of the rotation--X-ray activity relation in the two clusters. We updated membership catalogs that combine pre-Gaia catalogs with new catalogs based on Gaia Data Release 2. The resulting catalogs are the most inclusive ones for both clusters: 1739 Praesepe and 1315 Hyades stars. We collected X-ray detections for cluster members, for which we analyzed, re-analyzed, or collated data from ROSAT, the Chandra X-ray Observatory, the Neil Gehrels Swift Observatory, and XMM-Newton. We have detections for 326 Praesepe and 462 Hyades members, of which 273 and 164, respectively, have rotation periods---an increase of 6$\times$ relative to what was previously available. We find that at $\approx$700 Myr, only M~dwarfs remain saturated in X-rays, with only tentative evidence for supersaturation. We also find a tight relation between the Rossby number and fractional X-ray luminosity \LLX\ in unsaturated single members, suggesting a power-law index between $-3.2$ and $-3.9$. Lastly, we find no difference in the coronal parameters between binary and single members. These results provide essential insight into the relative efficiency of magnetic heating of the stars' atmospheres, thereby informing the development of robust age-rotation-activity relations.
\end{abstract}


%stars:~coronae -- stars:~evolution --  stars:~rotation}
%\maketitle

\section{Introduction}

%As stars age, their rotation period and tracers of magnetic activity decrease with time. This was first presented in \citet{skumanich72}, and subsequent studies have continued to support it. The existence of an age-rotation-activity relation (ARAR) has generated hope that measurements of rotation and/or activity can be used to obtain the ages of field stars.  However, an accurate, quantitative description of the ARAR, and its dependence on stellar mass, still eludes us. Angular momentum lost through stellar winds is thought to be responsible for the spin down described by Skumanich, but the exact dependence of rotation on age is not known, and in any case relies on the assumed magnetic field geometry and degree of core-envelope coupling \citep[e.g.,][]{kawaler1988}. Furthermore, later-type, fully convective stars have longer active lifetimes than their early-type brethren \citep[e.g.,][]{andy08}, indicating that they are capable of generating magnetic fields even in the absence of solar-type dynamos \citep{Browning2008}.

In the interiors of low-mass stars (late-F dwarfs and later types), rotational shear at the tachocline, the boundary between the radiative and convective zones, is thought to power a dynamo \citep{Parker1993, Charbonneau2014}. This dynamo generates magnetic field that rises to the stellar surface where a fraction of  the magnetic energy is dissipated and heats the corona to the $\gapprox$10$^6$ K temperatures required to produce thermal X-rays. % \citep{vaiana1981}.
Observations of GKM dwarfs with the Einstein Observatory found that their X-ray luminosity (\LX) is proportional to their rotational velocity, confirming this connection between \LX\ and rotation \citep{Pallavicini1981, Pizzolato2003}. Subsequent measurements of \LX, generally expressed as a fraction of the bolometric luminosity, \LLX, to remove the mass dependence, have shown that \LLX\ increases as the stellar rotation period (\prot) decreases. This relation was refined by \citet{Noyes1984}, who examined chromospheric fluxes in terms of the Rossby number, \Ro\ = \prot/$\tau$, where $\tau$ is the convective turnover time. \LLX\ then has a power-law dependence on \Ro\ such that \LLX\ $\propto R_\mathrm{o}^{\beta}$, with $\beta \approx -2$ \citep[e.g.,][]{Randich2000b, Wright2011, Nunez2015, Thiemann2020}. However, this holds only up to some threshold \Ro, below which \LLX\ is $\approx$constant. For faster rotators, magnetic activity is saturated, i.e., it no longer depends on rotation speed \citep[e.g.,][]{Stauffer1994a,Pizzolato2003, Wright2011, Nunez2015}. 

The cause(s) of the saturation of X-ray activity are not well established, though several competing hypotheses have been formulated. A fundamental challenge is determining whether saturation is caused by changes in the dynamo efficiency, or is instead a consequence of fast rotation. The former could result in saturation of the dynamo \citep[e.g.,][]{Gilman1983, Blackman2015}, although this is contradicted by observations that not all activity indicators appear to saturate, and that different indicators can saturate at different \Ro\ \citep[e.g,][]{Cardini2007,mamajek2008,marsden2009}. The latter could suggest that saturation instead occurs in the coronal filling factor \citep{Vilhu1984}, although here again, the observational evidence is not strong, as studies have found that the coronal filling factor can be small in saturated X-ray emitters \citep[e.g.,][]{Testa2004}. Very rapid rotation could also cause coronal loops to become unstable due to the centrifugal force, a situation known as coronal stripping, as a result of the Keplerian corotation radius %($R_{\rm Kepler}$)
getting close to the stellar surface \citep{Jardine1999}. Other work has explored additional potential explanations, including changes in the underlying dynamo mechanism \citep[from a convective dynamo to an interface dynamo; see][]{Barnes2003a, Barnes2003b}. %; for a review, see \citet{Testa2015}. We are still far from a consensus view of physics of saturation.

Meanwhile, some studies have also found that, in the fastest rotators, X-ray activity decreases relative to the saturation level, a phenomenon known as supersaturation \citep{Randich1996}. These authors found a decrease in \LLX\ for the fastest rotators ($v \sin i >$ 100 km s$^{-1}$) in the $\approx$60 Myr-old open cluster $\alpha$ Persei. Subsequent studies have claimed to observe supersaturation in FGK dwarfs \citep{Stauffer1997b,Jeffries2011,Argiroffi2016}, K dwarfs \citep{Thiemann2020}, M dwarfs, and ultracool dwarfs \citep{Alexander2012,Cook2014}.
% I changed the sentence to streamline it - Stephanie
% For example, \citet{Stauffer1997b}, \citet{Jeffries2011}, and \citet{Argiroffi2016} found evidence of supersaturation in young F, G, and K dwarfs, while \citet{Alexander2012} and \citet{Cook2014} did so in M dwarfs and ultracool dwarfs. And recently, \citet{Thiemann2020} found evidence for supersaturation in the X-ray emission from 60 K dwarfs that are among the fastest rotators in their sample. 
% 60 very fast rotating dwarfs within the supersaturated regime, characterized by a power law with $\beta$ = 1.42$\pm$0.22.

Supersaturation is even less well understood than saturation. \citet{james2000} and \citet{jardine2004} proposed that coronal stripping would explain both saturation and supersaturation, while \citet{stepien2001} argued that supersaturation is caused by a decrease in the filling factor due to the poleward migration of active regions in very fast rotators. \citet{Wright2011} assembled a sample of 824 stars to study the dependence of \LX\ on \Ro\ and found that their data favored the \citet{stepien2001} hypothesis, even though the Keplerian corotation radius and the excess polar updraft were both better predictors of \LLX\ than either \prot\ or \Ro\ for their supersaturated stars. However, supersaturation does not appear to occur for chromospheric activity \citep[e.g.,][]{marsden2009, jackson2010}, which argues against it being due to e.g., the migration of active regions, and favors the coronal-stripping explanation.

%evidence of supersaturation by comparing \LLX\ $R_\mathrm{Kepler}$/$R_\mathrm{star}$ and the excess polar updraft acceleration. %The latter two quantities were calculated to explore the two main hypotheses explaining supersaturation: centrifugal stripping of the corona and the poleward migration of active regions.

Further complicating our understanding of the coronal rotation-activity relation is the posited transition from a solar-type $\alpha\Omega$ dynamo in stars with a radiative core and a convective outer layer to a turbulent or $\alpha^2$ dynamo in fully convective M dwarfs. \citet{Wright2016} and \citet{Wright2018} found that fully convective M dwarfs follow the same \LLX--\Ro\ relation as their partly convective counterparts, suggesting that fully and partly convective stars have very similar dynamos, driven mostly by the interaction of rotation and turbulent convection. This conclusion impugns the relevance of the shear at the tachocline in driving the magnetic dynamo in partly convective stars.

Studies of the dependence of \LLX\ on rotation have mostly focused on mixed-aged samples of open cluster and/or of field stars, where age effects are difficult to quantify \citep[e.g.,][]{Randich2000a,Wright2011,Wright2018,Thiemann2020}. Other studies have instead focused on the single-aged populations in the handful of open clusters currently observable with X-ray telescopes,\footnote{The small fields-of-view of the current flagship X-ray observatories are poorly matched to the size of nearby open clusters, while the low-mass members in more distant, smaller-on-the-sky clusters require excessive exposure times.} where the number of detected stars is generally small \citep[e.g.,][]{Douglas2014, Nunez2015, Nunez2017}. There is a clear need for more data before we can reach a consensus view on the age-dependence of the rotation--coronal activity relation or on the reason(s) for saturation and supersaturation. 

The approximately coeval Praesepe and Hyades clusters, each about 700 Myr old, are the oldest open clusters within 250 pc, and thus the oldest accessible ensembles of low-mass stars with a well-constrained age.\footnote{\citet{Douglas2019} combined literature estimates for the age of Praesepe to assign the cluster an age of 670$\pm$67 Myr, and used gyrochronology to determine that the Hyades is 57 Myr older, or 727$\pm$75 Myr old.}
Fully characterizing rotation and activity for the low-mass members of these two clusters is therefore critical to constraining the age-rotation-activity relation, particularly as Praesepe and the Hyades serve as a bridge of knowledge between younger cluster stars and their older field-age cousins.

This paper is the fourth in a series focused on Praesepe and the Hyades. In \citet[][Paper~I]{Agueros2011}, we presented \prot\ for 40 Praesepe members measured using Palomar Transient Factory \citep[PTF;][]{Law2009, Rau2009} light curves. In \citet[][Paper~II]{Douglas2014}, we combined literature \prot\ and X-ray data with new and archival optical spectra for members of both clusters to show that \LX\ and \halpha\ emission, a proxy for chromospheric activity, depend differently on \Ro. In \citet[][Paper~III]{Kraus2017}, we characterized a newly discovered eclipsing binary in Praesepe. 
Parallel to this paper series, in \citet[][hereafter D16 and D17]{Douglas2016, Douglas2017} we added 48 new \prot\ for Hyades members and 677 for Praesepe members from an analysis of K2 \citep{howell2014} photometry. %We also examined discrepancies between our observed color-period distributions for the clusters and the predictions of gyrochronology models. 
And in \citet{Rampalli2022} we used two additional K2 campaigns to complete the census of rotation in Praesepe, measuring new \prot\ for 220 stars and bringing the total number of \prot\ for the cluster to 1013.

We begin below by revisiting the membership catalogs for the two clusters in light of the information provided by the Gaia data, and particularly by its Data Release 2 \citep[GDR2;][]{GaiaDR2}, in Section~\ref{sec:cat}. We also discuss the \prot\ available for both clusters, using these to calculate \Ro\ for their members, as well as binary information for stars in the two clusters. In Section~\ref{sec:xray} we present our analysis of the extensive set of X-ray observations we have assembled for Praesepe and the Hyades from the R\"ontgen Satellite (ROSAT), the Chandra X-ray Observatory (Chandra), the Neil Gehrels Swift Observatory (Swift), and the X-ray Multi-Mirror MissionNewton (XMM). We derive several stellar parameters for cluster stars in Section~\ref{sec:properties}. We present our results in Section~\ref{sec:res} and conclude in Section~\ref{sec:concl}.

\input{tbl_mems}

\section{Constructing Our Membership Catalogs}\label{sec:cat}
%We continue to use the HyPra membership catalogs assembled in Paper II and used in \citetalias{Douglas2016}, \citetalias{Douglas2017}, and \citetalias{Douglas2019}. We now build upon them using membership studies that use Gaia-DR2 data.

\subsection{Legacy Membership Catalogs}\label{sec:legacy}
\citet{Douglas2014} and \citetalias{Douglas2017} based their Praesepe catalog on the cluster membership probabilities \Pmem\ calculated by \citet{Kraus2007}. These \Pmem\ were found for $>$10$^6$ sources in the field of view of Praesepe using photometric and astrometric data. \citetalias{Douglas2017} considered the \citet{Kraus2007} 1130 stars with \Pmem\ $\geq$ 50\% as bona fide cluster members. To these stars, \citetalias{Douglas2017} added 39 previously identified Praesepe members too bright to be included in the \citet{Kraus2007} catalog, assigning these stars \Pmem~=~100\%. As a starting point, we took all stars from \citet{Kraus2007} with \Pmem~$\geq$~10\% to be potential Praesepe members; we also added the 39 additional stars from \citetalias{Douglas2017} as bona fide members.

As in \citetalias{Douglas2016}, we adopted the \citet{Goldman2013} catalog for the Hyades and supplement it with members identified through analysis of Hipparcos data \citep{hip}. %by Cargile et al.\ (in prep). 
%These authors follow the method outlined by \citet{Vanleeuwen2009} and consider stars within 26\degree\ and 20 pc of the cluster center. Cargile et al. identify 170 cluster members with reduced proper motions  ($\mu$) satisfying $-170<\mu_{\parallel}<-60$ and $-20<\mu_{\perp}<20$~mas~yr$^{-1}$ and distances obtained by {\it Hipparcos}. 
All but 13 Hyads identified through this analysis were also identified by \citet{Goldman2013}. %; we add these 13 stars to our catalog.
% See table 6 in \citetalias{douglas2014} for a list of members identified through our {\it Hipparcos} analysis, and section 2.1 of \citetalias{Douglas2016} for our calculation of \Pmem~values for \citet{Goldman2013}. %blergh
% In \citet{Goldman2013, stars have discrete \Pmem\ values (0, 70, 92.5, or 99\%); for our current work, we considered all Hyads with \Pmem~$\geq$ 70\% to be cluster members.
In \citet{Goldman2013}, stars have discrete values for field contamination (100, 30, 17.5, or 1\%). \citetalias{Douglas2016} adopted the invert of these values as \Pmem\ and considered stars with \Pmem\ $\geq$ 70\% (corresponding to contamination fraction $\leq$~30\%) as bona fide members. As a starting point, we took all stars in \citetalias{Douglas2016} with \Pmem\ $>$ 0\% to be potential Hyades members.

\subsection{Gaia-based Membership Catalogs}\label{sec:gaiacatalogs}
Successive Gaia data releases have transformed our view of the membership of stellar clusters. Table~\ref{tbl_mems} lists the Gaia-based studies of Praesepe and the Hyades that we considered, in addition to our legacy catalogs, to produce our definitive membership catalogs, and the number of members included in each. Our starting assumption was that the Gaia-based catalogs are correct, in that they are likely to contain fewer contaminants than earlier catalogs---thanks in part to Gaia's higher spatial resolution---but are incomplete, in that Gaia data either lack five-parameter solutions or have poor fits for unresolved binaries \citep[e.g.,][]{Belokurov2020}. Gaia-based studies, therefore, exclude most potential binaries from published cluster catalogs.
%data releases to date deliberately do not include all of the data collected by the mission, notably for binaries. 

\begin{figure}[!t]
\centerline{\includegraphics{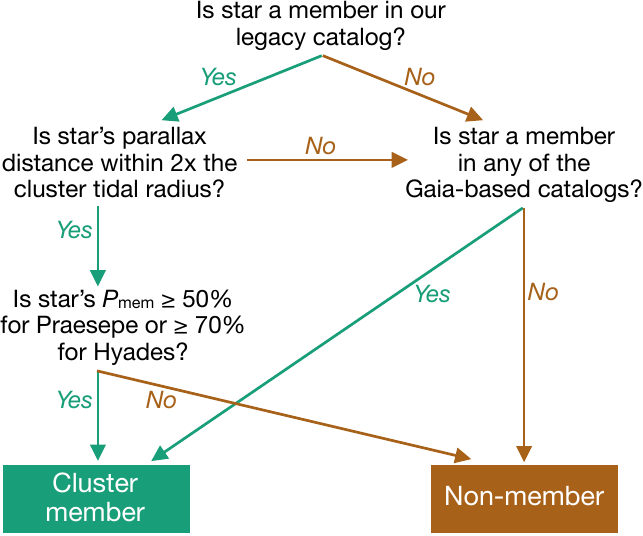}}
%\centerline{\includegraphics[width=\columnwidth]{decision_tree.pdf}}
%, trim=0.05cm 0cm 0cm 0cm, clip=true]\
\caption{Decision process to consolidate our Praesepe and Hyades legacy catalogs (described in Section~\ref{sec:legacy}) with the GDR2-based catalogs listed in Table~\ref{tbl_mems}.}
\label{fig_tree}
\end{figure}

To produce our updated catalogs, we first cross-matched our legacy catalogs against the GDR2-based catalogs using a 5\arcsec\ matching radius. To account for the large proper motions of Hyads, we also searched for GDR2 counterparts using a matching radius between 5\arcsec\ and 7\arcsec\ and found 17 additional matches; we inspected these visually using Aladin Sky Atlas.\footnote{We also tracked these 17 stars in our subsequent analysis. Ten of them have X-ray detections; none of which stand apart from their spectral type cohort in their X-ray measurements.} We found two or more GDR2 stars within the matching radius of 161 Hyads and 80 Praesepe stars. We assumed the closest match to be the best match, except when the radial distances of the potential matches were too similar. In such instances, we guided our best match selection using GDR2 parallax, proper motion, and photometry for each potential match. We resolved ambiguous matches using this approach for 23 Hyads and 5 Praesepe stars. All in all, the median closest matching radius was 1\farcs6 for Hyads and 0\farcs6 for Praesepe stars. 

Stars that appeared as members in our legacy catalogs and are included in any of the GDR2 catalogs in Table~\ref{tbl_mems} form the bulk of our final catalogs. There are 746 Hyads and 1123 Praesepe stars in this category. If, however, a star was not in our legacy catalogs but is included as a member in any of the GDR2 catalogs, we added that star to our list of bona fide members for that cluster. In Praesepe, there are 292 such stars in the core, and 312 in the tidal tails. In the Hyades, the numbers are 143 and 405, respectively.

We then matched our catalog stars to Gaia Early Data Release 3 \citep[GEDR3;][see Section~\ref{photometry}]{GaiaEDR3} to obtain GEDR3 photometry and GEDR3-based distances from \citet{BailerJones2021}. For stars without such distances, we found either parallax or distance calculations in the literature. For Hyades, we found 2 Hipparcos-based parallaxes in \citet{vanLeeuwen2007}, 51 PPMXL-based parallaxes in \citet{Roser2011}, and one trigonometric parallax in \citet{Dittmann2014}. We also found eight spectroscopic distances in \citet{Lodieu2014}, one photometric distance in \citet{Robert2016}, and two photometric distances in \citet{Schneider2017}. For all other stars missing distances (68 Praesepe stars and five Hyads), we assigned them a distance equal to that of the cluster as a whole. To estimate the distance of each cluster as a whole, we calculated the median, 16$^\mathrm{th}$, and 84$^\mathrm{th}$ percentiles of the \cite{BailerJones2021} distances.\footnote{We excluded the tidal tail stars cataloged for both clusters \citep{Meingast2019, Roser2019a,Roser2019b} to calculate these cluster distances.} The resulting cluster distances are 183.2$^{+11.1}_{-13.7}$~pc for Praesepe and 49.1$^{+25.4}_{-7.7}$~pc for the Hyades; the 1$\sigma$ distance errors were approximated as the mean of the 16$^\mathrm{th}$ and 84$^\mathrm{th}$ percentiles.

Figure~\ref{fig_tree} illustrates as a decision tree the steps described next to build our cluster catalogs. To start, we compared the one-dimensional parallax distance ($\pm$2$\sigma$) of each star in our legacy catalogs to its cluster's center $\pm$twice the tidal radius, which we took to be 11.5 pc for Praesepe \citep{Kraus2007} and 10 pc for Hyades \citep[][]{Roser2011}. If the star is closer to or farther from us than these distances, indicating that it is beyond twice the tidal radius for its cluster, we categorized that star as a non-member. 

We then checked whether a star that was rejected as a  non-member in the step above is considered a member in any of the GDR2 catalogs. We assumed that such stars are bona fide members and therefore added them back in to our final catalogs.

Lastly, we rejected stars from our legacy catalogs that passed the above radial distance check but that have \Pmem\ lower than the thresholds used in \citetalias{Douglas2016} and \citetalias{Douglas2017}: 70\% for Hyades and 50\% for Praesepe, respectively (see Section~\ref{sec:legacy}). As these stars were not present in any of the GDR2 catalogs and they have low \Pmem\ in the legacy catalogs, we consider them to have a high probability of being field contaminants. 

Table~\ref{tbl_catalogcols} includes our consolidated catalog of Praesepe and Hyades stars. Column 7 of Table~\ref{tbl_catalogcols} specifies if a star is a member in our legacy catalogs, and Columns 8--14 specify if a star is considered a cluster member in each of the GDR2 catalogs. Our resulting catalogs include 1727 Praesepe members and 1294 Hyads. To these, we add ultracool dwarfs from the literature, as we explain next.

\input{tbl_catalogcols}

\subsection{Ultracool Dwarfs}\label{sec:ultracool}
To our consolidated cluster catalogs, we added ultracool dwarfs that were not already included in our legacy catalogs or in the GDR2-based catalogs. These are primarily late M and early-to-mid L dwarfs. We found twelve such Praesepe stars in the literature: one M9 and four L0 type dwarfs from \citet{Boudreault2010}, one L0 found by \cite{Zhang2010}, and four late M and two early-L dwarfs from \cite{Manjavacas2020}. To the Hyades catalog we added 21 objects: two late M and four early-L dwarfs cataloged by \citet{Lodieu2014}, one L3 found by \citet{Robert2016}, two mid L found by \citet{Schneider2017}, and three late M and nine L dwarfs found by \citet{PerezGarrido2018}. Column 15 of Table~\ref{tbl_catalogcols} indicates whether a cluster star is an ultracool dwarf from the literature.

Only eight ultracool dwarfs (one Praesepe and seven Hyads) have \citet{BailerJones2021} distances. For the others, we adopted either the distances from the literature---which was the case for one Praesepe and 14 Hyads---or the cluster distance as a whole (see Section~\ref{sec:gaiacatalogs}).

The final tallies in our cluster catalogs are 1739 Praesepe stars and 1315 Hyads.

\subsection{Photometry}\label{photometry}
As mentioned in Section~\ref{sec:gaiacatalogs}, we matched our final catalog from Sections~\ref{sec:gaiacatalogs} and \ref{sec:ultracool} against GEDR3 using a 3\arcsec\ matching radius for stars in both clusters. When we found more than one GEDR3 star within the matching radius, we compared the $G$ magnitudes, when available, to determine the better match. In most cases, the closest match was the best match. 

\begin{figure*}[t]
\centerline{\includegraphics{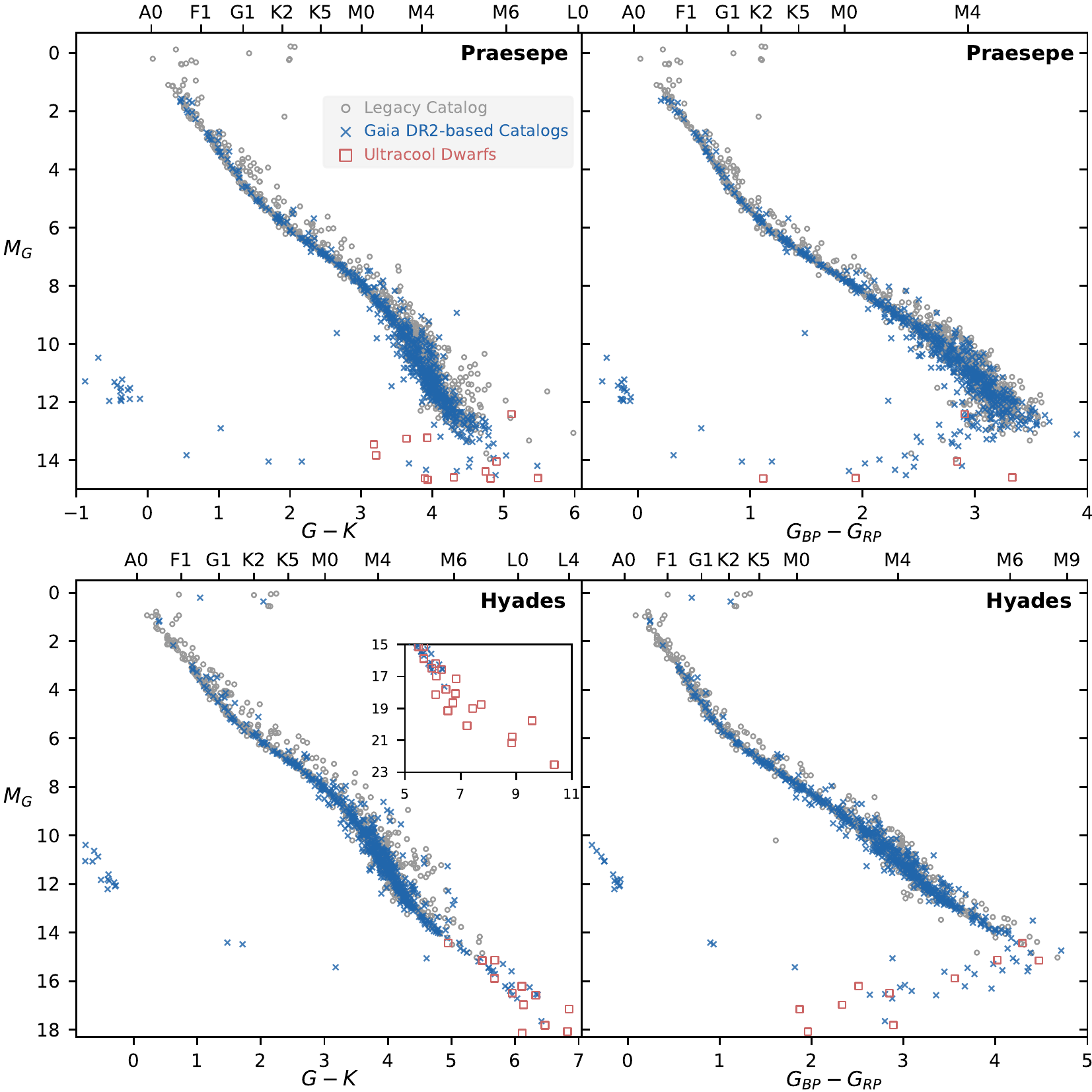}}
\caption{CMDs for Praesepe (top panels) and Hyades (bottom). The left column combines GEDR3 and either 2MASS, UKIDSS, or GEDR3-derived $K$ photometry to produce a \gminusk\ color. The right column uses the \bpminusrp\ color from GEDR3. Our full catalog consists of stars from our legacy catalog (gray circles, see Section~\ref{sec:legacy}), Gaia DR2-based studies of cluster membership (blue $\times$ symbols, see Section~\ref{sec:gaiacatalogs}), and ultracool dwarfs from the literature (red squares, see Section~\ref{sec:ultracool}). The bottom left panel includes an inset to show the full red end of the main sequence for Hyades in \gminusk\ color. We have a total of 1739 Praesepe members and 1315 Hyades members in our full catalog.}
\label{fig_cmd}
\end{figure*}

\begin{figure*}[t]
\centerline{\includegraphics{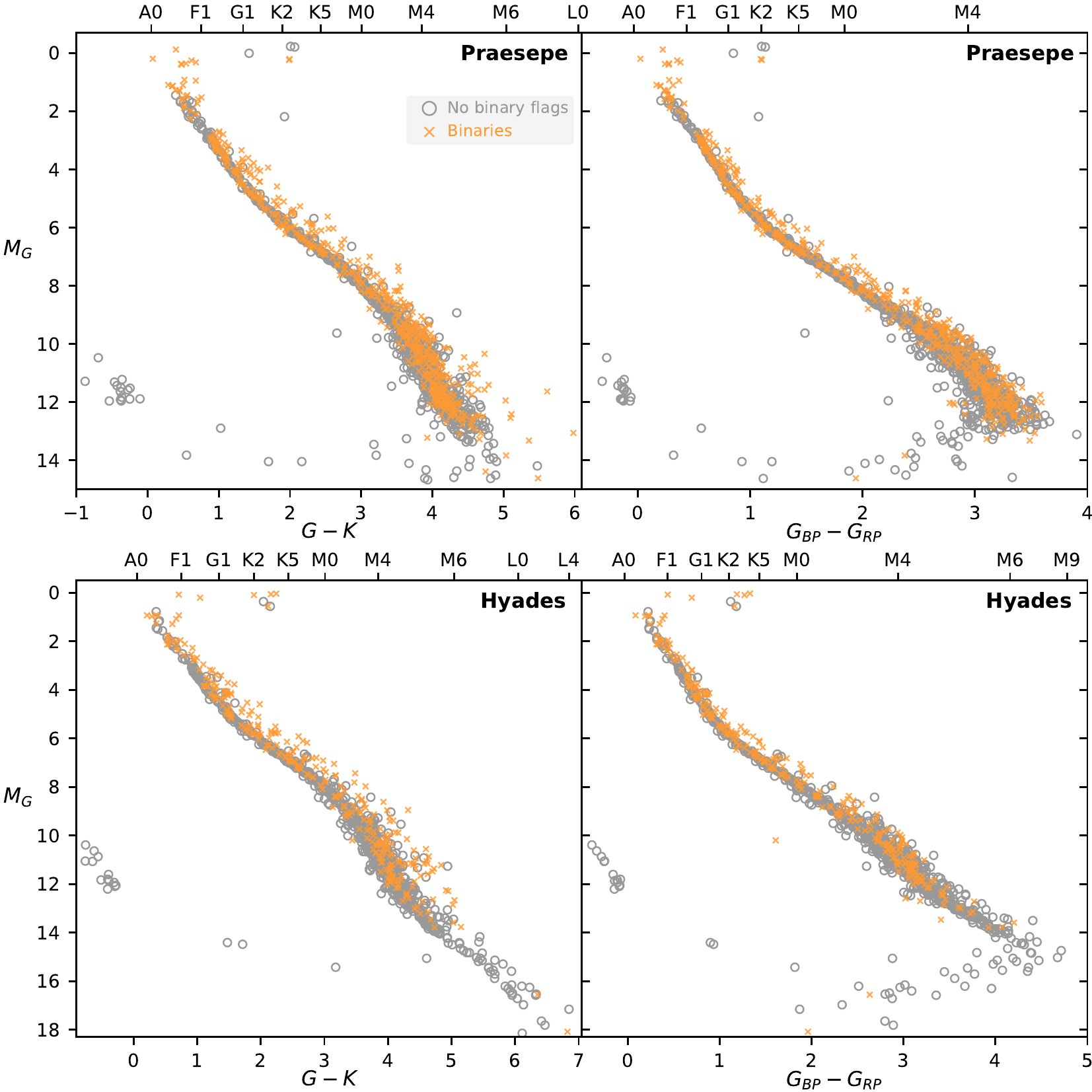}}
\caption{Same as in Figure~\ref{fig_cmd}, but this time indicating stars with no binary flags with gray circles, and candidate/confirmed binaries with orange crosses. Candidate/confirmed binaries constitute 31\% of our Praesepe catalog and 23\% of our Hyades catalog.}
\label{fig_cmdbin}
\end{figure*}

We obtained GEDR3 magnitudes for 2684 members of the two clusters (1732 for Praesepe, 1304 for the Hyades), ranging from $G$ = 2.7 to 21.1 mag. For seven ultracool dwarf Praesepe members and eleven ultracool dwarf Hyads lacking a GEDR3 counterpart, we derived $G$ magnitudes from Two Micron All-Sky Survey \citep[2MASS;][]{2mass} $J$ and $K$ magnitudes using the transformations given by \citet{Riello2021}. These ultracool dwarfs cover the magnitude range $14.5 < K < 17.1$, and the derived $G$ magnitudes are in the range $19.6 < G < 25.2$.

We also collected good quality ({\sc ph$_-$qual} A, B, or C) 2MASS $K$ magnitudes for 1669 Praesepe and 1274 Hyades members. The resulting coverage is $1.5 < K < 15.7$ mag. For the 111 Praesepe and Hyades stars lacking a good quality 2MASS $K$, we searched for $K$ photometry from the United Kingdom InfraRed Telescope (UKIRT) Infrared Deep Sky Survey \citep[UKIDSS;][]{Lawrence2007} data release 9, both from the Galaxy Clusters Survey (GCS) and Large Area Survey (LAS). We found UKIDSS $K$ photometry for 41 additional Praesepe members and nine Hyads, covering the magnitude range $9.8 < K < 18.8$.

There are nine Praesepe stars lacking good quality 2MASS/UKIDSS $K$ photometry. For these stars, we derived $K$ magnitudes from GEDR3 $G$ and \bpminusrp\ color using the transformations given in \citet{Riello2021}. There are also 20 Praesepe stars unmatched in 2MASS/UKIDSS. Ten of these are newly identified white dwarfs in the GDR2 catalogs, and five others are secondary stars in binaries unresolved by 2MASS/UKIDSS. Except for one of the white dwarfs and three of the secondary stars---all lacking a \bpminusrp\ color, we derived their $K$ magnitudes using the same transformations as above. The remaining five unmatched stars are ultracool dwarfs, for which we adopt the $K$ magnitudes measured in \citet{Boudreault2010}.

In the Hyades, there are 27 stars lacking good quality 2MASS/UKIDSS $K$ photometry. We derived their $K$ magnitudes using the transformations in \citet{Riello2021} mentioned above. There are also five Hyads unmatched in 2MASS/UKIDSS. Three of these are newly identified white dwarfs in the GDR2 catalogs, and two are secondary stars in binaries unresolved by 2MASS/UKIDSS. We derived their $K$ magnitudes using the same transformations as above. The magnitude range of all derived $K$ magnitudes is $1.0 < K < 19.7$. Column 21 of Table~\ref{tbl_catalogcols} identifies the source of the $K$ magnitude for each star.
%Lastly, we used the $K$ magnitudes measured in \citet{Boudreault2010} for five ultracool Praesepe dwarfs that lacked 2MASS or UKIDSS match.

All in all, we have $G$ photometry for all stars in our catalogs, and $K$ photometry for all but four Praesepe stars. The combined photometry covers the color range $-0.9 <$ \gminusk\ $<$ 10.4 mag. We calculated absolute $G$ magnitudes ($M_G$) for our cluster stars using the distances we adopted for each, and an extinction correction for the Gaia $G$ band. We obtained the latter by first calculating the total absorption in $V$ ($A_V$) using the extinction tables by \citet{Schlafly2011} assuming $R_V$ = 3.1 and adopting a reddening of $E$($B-V$) = 0.035 for Praesepe \citep{Douglas2019} and 0.001 for the Hyades \citep{Taylor2006}. We then obtained the total absorption in $G$ ($A_G$) using $A_G$/$A_V$ = 0.789 \citep{Wang2019}. The left panels of Figure~\ref{fig_cmd} show the $M_G$ -- \gminusk\ color-magnitude diagram (CMD), and the right panels, the $M_G$ -- \bpminusrp\ CMD\footnote{We collected \bpminusrp\ for 1724 Praesepe and 1301 Hyades stars, although we do not use this Gaia color in our analysis.} for both clusters.

\subsection{Rotation Periods}\label{sec:rot}

The most up-to-date catalogs of rotation periods for the two clusters are those of \citet{Rampalli2022} for Praesepe and \citet{Douglas2019} for the Hyades. Both of these catalogs supplement ground-based \prot\ measurements with large numbers of \prot\ obtained from the five K2 campaigns dedicated to the two clusters (C5, C16, and C18 to Praesepe, and C4 and C13 to the Hyades). 
%For the most up-to-date catalog of HyPra rotators, we used the assembled list of rotators in \citetalias{Douglas2019} for both clusters. The \citetalias{Douglas2019} catalogs include \prot\ values from \citet{Radick1987, Radick1995}, \citet{Prosser1995}, \citet{Delorme2011}, \citet{Hartman2011}, \citet{Kovacs2014}, \citetalias{douglas2014}, \citetalias{Douglas2016}, and \citetalias{Douglas2019} for Hyades, and from \citet{Scholz2007}, \citet{Delorme2011}, \citetalias{Agueros2011}, \citet{Scholz2011}, \citet{Kovacs2014}, \citetalias{Douglas2017}, and \citetalias{Douglas2019} for Praesepe. 
In total, we have \prot\ measurements for 1052 Praesepe members and 233 Hyads. Column 28 in Table~\ref{tbl_catalogcols} includes these \prot\ values.

\subsection{Binary Flags}\label{sec:bin}

To identify binaries and multiple systems in the two clusters, we relied on the binary flags from several previous studies. \citetalias{Douglas2016} and \citetalias{Douglas2017} obtained and tabulated binary flags for members of Hyades and Praesepe, respectively. \citet{Douglas2019} updated and complemented those flags and reported the following individual flags: visual identification, UVW kinematic deviations (for Hyades only), vertical distance from the main sequence on a Gaia CMD, multiple periodicity from K2 periodograms, GDR2 radial velocity and proper motion deviations, high GDR2 astrometric excess noise ($\epsilon_i > 1$), and confirmed binarity from the literature. \citet{Rampalli2022} further updated these flags for Praesepe and replaced the $\epsilon_i > 1$ flag with GEDR3's re-normalized unit weight error (RUWE) $>$ 1.2 to identify likely unresolved binaries (see binary indicators in their Table 4).

In our catalog, we consider as \emph{candidate} binaries (Binary Flag = 1) stars having at least one of the kinematics, CMD distance, radial velocity, proper motion, or multiple periodicity flags. We consider as \emph{confirmed} binaries (Binary Flag = 2) those confirmed in the literature. However, we ignore the $\epsilon_i$ and the RUWE flags in \citet{Douglas2019} and \citet{Rampalli2022}, respectively, as we implement our own approach to using the most recent GEDR3 spurious astrometry indicators (see our RUWE treatment in the next paragraph). Also, we found ten Praesepe stars and seven Hyads for which GEDR3 resolved two point sources and 2MASS only one. We assigned Binary Flag = 1 to these 17 stars. All in all, we have 456 candidate and 82 confirmed binaries in Praesepe and 60 candidate and 238 confirmed binaries in Hyades. Figure~\ref{fig_cmdbin} shows the same CMDs as in Figure~\ref{fig_cmd}, this time highlighting with orange crosses candidate/confirmed binaries. Column 26 of Table~\ref{tbl_catalogcols} includes our binary flag for each star.

Lastly, we collected RUWE values for our stars\footnote{Both $\epsilon_i$ and RUWE have their pros and cons in identifying spurious Gaia astrometry; we opted to use only RUWE by virtue of its more sound and guaranteed distribution (with peak at 1.0) across the full color-magnitude range of Gaia  \citep[see discussions in][]{Belokurov2020, Penoyre2020}}. The RUWE is a goodness-of-fit measure of the single-star model fit to the source's astrometry. If a star is an unresolved binary and its center of light deviates from the assumed single-star model, its RUWE will deviate significantly from 1.0 \citep[e.g.,][]{Jorissen2019, Belokurov2020}. Gaia cannot resolve separations $\lapprox$0\farcs7 \citep{Ziegler2018}, which corresponds to a semimajor axis $a~\approx~35$ au at the typical Hyad distance and $\approx$130 au at the typical Praesepe star distance. Therefore, most stars in our catalogs with high RUWE are likely to be ``intermediate'' binaries, i.e., with small enough separation ($0.1~\lapprox~a~\lapprox~80$ au) for the binary components to have affected each other's protoplanetary disks in the first 10 Myr \citep[][S. T. Douglas et al. 2022, in preparation]{Rebull2006, Meibom2007, Kraus2016, Messina2017}. They are unlikely to be tight, tidally interacting binaries ($a~\lapprox~0.1$ au) because their center of light will deviate minimally from the assumed single-star model ($a = 0.1$ au corresponds to $\delta\theta~\approx~2$ mas at the typical Hyad distance and $\approx$0.5 mas at the typical Praesepe star distance), thus only negligibly deviating RUWE from 1.0.

Typically, stars with RUWE $>$ 1.4 are considered to have high probability of being unresolved binaries \citep[e.g.,][]{Deacon2020, Ziegler2020, Kervella2022}, although some studies have used a more conservative $>$1.2 threshold to identify potential binaries \citep[e.g.,][]{Pearce2020}. We use RUWE $>$ 1.4 to determine which stars with Binary Flag = 0 (i.e., assumed to be single) are potentially unresolved binaries. However, we do not necessarily exclude them from the sample of single stars in our analysis, because we do not have any additional information on their potential binarity. Column 27 of Table~\ref{tbl_catalogcols} includes the RUWE for each star. In Praesepe, 52 (4\%) stars with Binary Flag = 0 and 129 (24\%) with Binary Flag > 0 have RUWE $>$ 1.4. For Hyades, the numbers are 152 (15\%) and 131 (44\%), respectively. 

\section{X-Ray data}\label{sec:xray}
Over the past few decades, Praesepe and the Hyades have been regularly targeted by X-ray missions, and members of both clusters have also been detected serendipitously. We collected data from ROSAT, Chandra, XMM-Newton, and Swift. In some cases this required extracting sources from the X-ray observations ourselves. In others, we relied on previously compiled lists of X-ray detections, typically serendipitous source catalogs, which are constructed from automated processing of archival pointings. We homogenized these data to build a complete and uniform set of X-ray detections. To achieve this, our approach was to use the reported count rates and our own energy conversion factors (ECF; see Section~\ref{lx}) to obtain unabsorbed X-ray energy fluxes, \fx, in the 0.1--2.4 keV energy band.\footnote{We chose the most restrictive energy bandpass out of the four X-ray observatories to homogenize our data, which happens to be the ROSAT bandpass.}. We then converted \fx\ to \LX\ by using our adopted individual distances. For Chandra, XMM, or Swift sources with enough X-ray counts, we performed spectral analysis to extract unabsorbed \fx\ as well as coronal temperature, coronal metal abundance, and, in some cases, neutral hydrogen column density along the line of sight.

\begin{figure}
\centerline{\includegraphics[scale=1.5]{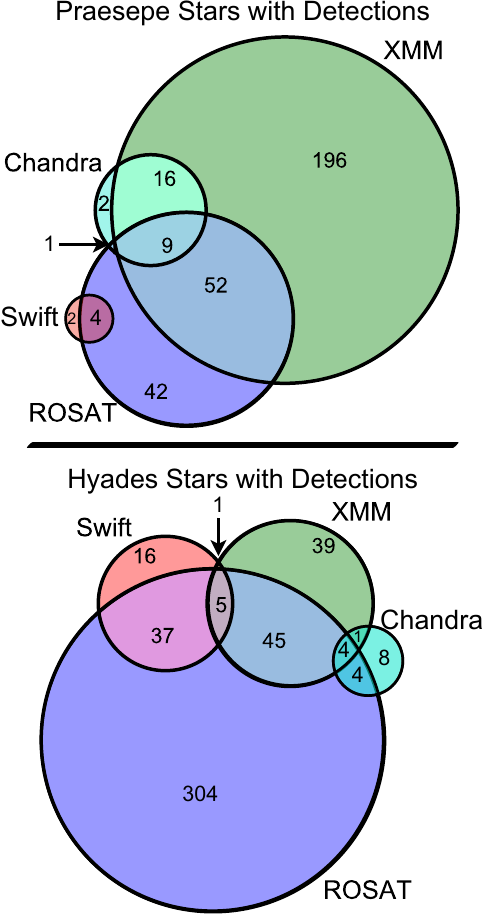}}
\caption{Venn diagrams illustrating the number of Praesepe (top panel) and Hyades (bottom panel) stars detected in X-rays with the four X-ray observatories. In our Praesepe catalog, 19\% of stars are detected in X-rays. In our Hyades catalog, the number is 35\%.}
\label{fig_venn}
\end{figure}

In the few instances in which the count rates were not reported, namely, some ROSAT-based studies, we worked backwards from the reported \LX, distance, and ECF to estimate the count rates and then applied the procedure described above to obtain our own \fx\ and \LX\ values. Figure~\ref{fig_venn} illustrates the number of Praesepe and Hyades stars detected with at least one of the X-ray observatories.

Our full list of X-ray detections for Praesepe and Hyades stars is in Table~\ref{tbl_Xcols}. If a star had more than one X-ray detection, we calculated the error-weighted average of the \fx\ values and adopted it as the bona fide \fx\ for that star. Columns 29 and 30 of Table~\ref{tbl_catalogcols} include the \fx\ and 1$\sigma$ uncertainty for each cluster star with at least one X-ray detection. In total, we have 504 individual X-ray detections for 326 Praesepe stars (two of which are ultracool dwarfs), and 603 for 464 Hyads (three ultracool dwarfs). Only 3\% of X-ray detections have questionable quality (see Column 28 in Table~\ref{tbl_Xcols}).

In the next subsections, we explain in detail the X-ray source detection and photometric extraction we performed with data from each X-ray observatory. 

\input{tbl_Xcols}

\subsection{ROSAT Observations}\label{archxrays}

\subsubsection{ROSAT Sources from the Publicly Released Catalogs}\label{sec:Rosat_surveys}
We searched for X-ray counterparts to our cluster stars in the latest versions of the three publicly released ROSAT catalogs: the Second ROSAT Source Catalog of Pointed Observations with the Position Sensitive Proportional Counters (PSPC) catalog \citep[2RXP,][]{2RXP}, the Second ROSAT All-Sky survey (RASS) catalog \citep[2RXS,][]{2RXS}, and the ROSAT High-Resolution Imager Pointed Observations catalog \citep[1RXH,][]{1RXH}. We cross-matched these ROSAT catalogs and our cluster catalogs using a 50\arcsec\ radius; this value was chosen to be large enough to account for the large ROSAT positional errors. To estimate the number of likely false matches, we shifted the X-ray source positions in steps of 50\arcsec\ out to 10\arcmin\ in all directions and re-matched them to the cluster catalogs using the same 50\arcsec\ radius used before. All resulting matches are assumed to be false. We found that the median number of false matches is 30 ($\approx$8\% of our matched ROSAT sources) with a median offset of 34\arcsec. In our final ROSAT-cluster catalogs matches, we consider those with radii $>$30\asec\ in Praesepe and $>$40\asec\ in Hyades to be likely false matches, and we exclude those from our X-ray analysis, although we still include them in our consolidated catalog of X-ray sources, with a ``likely false match'' flag (Quality Flag = ``x''; Column 28 of Table~\ref{tbl_Xcols}).

From the ROSAT catalogs, we obtained count rates and some other information about the X-ray point sources, such as hardness ratios, detection likelihoods, and variability flags, when available (see Table~\ref{tbl_Xcols}). We used the published count rates, our own ECFs (see Section~\ref{lx}), and our adopted individual distances to calculate \fx\ and \LX\ in the 0.1--2.4 keV energy band.

In the 2RXP catalog, we found X-ray counterparts for 78 Praesepe stars and 56 Hyads. Five Praesepe stars have a matching radius $>$30\asec, and we flagged these matches as likely false matches. All 56 Hyads, on the other hand, have a matching radius $<$20\asec. We also found six additional Praesepe and 90 Hyades stars with X-ray counterparts at large ($>$20\amin) offset angles from the instrument aimpoint. For these X-ray sources, 2RXP did not present count rate uncertainties, and so we opted to ignore these X-ray sources. Also, one of the sources matched to a Praesepe star has a source extent maximum likelihood higher than its source detection maximum likelihood; we therefore flag this source as a likely extended source in Table~\ref{tbl_Xcols}. Lastly, we found six additional 2RXP X-ray sources with ambiguous matches to stars in our catalogs. In these cases, each X-ray source had either two Praesepe stars within a 15\asec\ radius or two Hyads within a 40\asec\ radius roughly equidistant, and we opted to ignore these X-ray sources, as it was not possible to determine the true optical source of the X-ray detection.

In the 2RXS catalog, we found X-ray counterparts for nine Praesepe stars and 312 Hyads. One  Praesepe star has a matching radius $>$30\asec, and 19 Hyads have a matching radius $>40$\asec. We flagged these 20 matches as likely false matches. Furthermore, five of the Praesepe matches and 35 of the Hyades matches were also found in the 2RXP catalog, and we opted to adopt the X-ray information for these 40 sources from the 2RXP catalog instead\footnote{The 2RXP catalog was curated for point sources by both automatic algorithms and detailed visual inspection. As such, we assume 2RXP to be a more reliable source of X-ray information than 2RXS.}. Lastly, we found nine additional 2RXS X-ray sources with ambiguous matches to Hyads. In each case, the X-ray source had two roughly equidistant Hyads within a 30\asec\ radius, and we opted to ignore these X-ray sources, as it was not possible to determine the true optical source of the X-ray detection. All in all, we include 2RXS sources for four Praesepe stars and 277 Hyads, and their X-ray information is included in Table~\ref{tbl_Xcols}. 

Lastly, in the 1RXH catalog, we found X-ray counterparts for one Praesepe star and 22 Hyads. All matches to 1RXH have a matching radius $<$15\asec, and none of them were found in the 2RXP or the 2RXS catalogs.

\subsubsection{ROSAT Sources from Independent Studies}\label{sec:Rosat_lit}

To complement the ROSAT X-ray sources we found in the publicly released ROSAT catalogs, we also searched for X-ray counterparts from ROSAT studies of Praesepe and Hyades in the literature. Such studies sometimes included sources that were otherwise rejected by the automatic pipelines of the ROSAT catalogs, but that were otherwise acceptable knowing the nature of the X-ray emission (i.e., hot plasma from the atmospheres of low-mass stars). We describe below the individual ROSAT studies and the data we used from each one, which is included in Table~\ref{tbl_Xcols}.

\citet{Stelzer2001} used PSPC data publicly available as of 1998 Oct and analyzed by \citet{Stelzer2000} to report 0.1--2.0 keV quiescent\footnote{\citet{Stelzer2000} found X-ray flares in the detections of 12 Hyads. The \LX\ values \citet{Stelzer2001} published for these 12 stars exclude X-ray counts during the flares.} \LX\ and hardness ratio (HR) values for 181 Hyads in our catalog. %Of these, 72 were not in \citet{Reid1995} or \citet{Stern1995}. 
We used \LX\ and HR values published by \citet{Stelzer2001} and the distances adopted by these authors (from Hipparcos when available; otherwise, 46~pc) to calculate the corresponding X-ray count rates, using the formula ECF = ($8.31+5.30~\times$ HR) $\times$~$10^{-12}$ erg cm$^{-2}$ count$^{-1}$, as adopted in their work.

Of the 181 X-ray sources in \citet{Stelzer2001}, 126 are also in the ROSAT catalogs, and we opted to adopt the X-ray information for these 126 sources from the ROSAT catalogs instead. Therefore, we used X-ray information for only 55 Hyads from the \citet{Stelzer2001} catalog.

\citet{Stern1995} used data from the RASS \citep{Voges1999} to extract X-ray point sources and match them to an optical catalog of 440 Hyads put together by these authors. They found 187 matches to cluster stars and 24 matches to stars they considered nonmembers. For the former, they only published \LX\ values in the 0.1--1.8 keV band, which were calculated assuming a distance of 45 pc and an ECF of $6.0 \times 10^{-12}$ erg cm$^{-2}$ count$^{-1}$. For the latter, they instead published count rates in the 0.1--1.8 keV band. We converted the published \LX\ for cluster stars back into count rates.

\input{tbl_Xlog}

Of the 211 X-ray sources in \citet{Stern1995}, 174 are also in the ROSAT catalogs, and we opted to adopt the X-ray information for these 174 sources from the ROSAT catalogs instead. Seven additional sources are also in the \citet{Stelzer2001} catalog, and we opted to adopt the X-ray information for these seven sources from the latter catalog instead. Furthermore, we found 18 of their sources to be nonmembers. Finally, we found five sources with ambiguous optical counterparts, and we opted to ignore those, as it was not possible to determine the true optical counterparts to these five X-ray sources. In summary, we used X-ray information for only seven Hyads from the \citet{Stern1995} catalog.

\citet{Randich1995} surveyed Praesepe using the PSPC instrument aboard ROSAT. These authors obtained 42 0.4--2.0~ks fields in a raster scan pattern over two years of observations, detecting 68 stars in the \citet{KleinWassink1927}, \cite{Jones1983}, and \citet{Jones1991} Praesepe catalogs. For these 68 stars, \citet{Randich1995} published X-ray count rates in the 0.4--2.0 keV band. 

Of the 68 X-ray sources in \citet{Randich1995}, 39 are in the ROSAT catalogs, and we opted to adopt the X-ray information for these 39 sources from the ROSAT catalogs instead. We also found two of their sources to be nonmembers. Finally, we found one source with an ambiguous optical counterpart, and we opted to ignore this source, as it was not possible to determine its true optical counterpart. Therefore, we used X-ray information for only 26 Praesepe stars from the \citet{Randich1995} catalog.

\citet{Reid1995} published the results of three 30--40~ks PSPC observations offset by 3--4\degree\ from the Hyades cluster center. These authors detected 20 Hyads, 19 of which are also considered members in our updated membership catalog. However, we did not use any X-ray information from this X-ray catalog. Nine of their sources are in the ROSAT catalogs, and we opted to adopt the X-ray information for these sources from the ROSAT catalogs instead. The remaining ten X-ray sources are all present in the \citet{Stelzer2001} catalog, and we opted to adopt the X-ray information for these sources from the latter catalog.

\subsection{Chandra Observations}\label{chandraobs}

\subsubsection{Data from the Chandra Source Catalog}\label{sec:CSC}

We searched for X-ray counterparts to our cluster stars in the latest version of the Chandra Source Catalog \citep[CSC 2.0,][]{CSC2}. %To create this catalog, observations are processed with a series of automated data analysis pipelines, which extract uniformly calibrated properties for both point and extended X-ray sources. The sensitivity limit for point sources in the CSC 2.0 is $\approx$5 net counts.
We cross-matched the CSC and our cluster catalogs using a 15\arcsec\ radius. We found CSC counterparts to ten Praesepe stars and nine Hyads, all have a matching radius $<$ 8\arcsec. Eleven of these CSC sources were observed with an ACIS-I chip, six with an ACIS-S chip, and two with the High Resolution Camera instrument. In Figure~\ref{fig_footprints} we draw in blue the ACIS footprints of archival observations that include sources matched to Praesepe stars.\footnote{We do not show a figure with X-ray observations of Hyads because they are are too scattered in the sky.}

Like with ROSAT (see Section~\ref{sec:Rosat_surveys}), we estimated the number of likely false matches by shifting the X-ray source positions in steps of 25\arcsec\ out to 3\arcmin\ in all directions and re-matched them to the cluster catalogs using the same 15\arcsec\ radius used before. All resulting matches are assumed to be false. We found that the median number of false matches is 2 ($\approx$10\% of our matched CSC sources) with a median offset of 10\arcsec. As such, we consider all our final CSC-cluster catalogs matches to be true matches.

From the CSC, we obtained count rates in the 0.5--2.0 (soft), 2.0--7.0 (hard), and 0.5--7.0 (broad) keV bands, plus some other information about the X-ray point sources, such as detection likelihoods, and variability flags (see Table~\ref{tbl_Xcols}). We used the published count rates, our own ECFs (see Section~\ref{lx}), and our adopted individual distances to calculate \fx\ and \LX\ in the 0.1--2.4 keV band. Information about the original Chandra observations that collected the data for these 19 CSC sources is shown in Table~\ref{tbl_Xlog}.

Three of the CSC sources matched to Hyads have $>$300 net counts in the broad band. For these sources we performed spectral analysis to extract unabsorbed \fx\ as well as plasma parameters (see Section~\ref{sec:spec}). The extracted spectroscopic information is included in Table~\ref{tbl_spec}.

\subsubsection{Chandra Source Extraction with CIAO}\label{sec:ciao}

The CSC 2.0 only includes observations through the end of 2014. For Chandra pointings obtained after this cutoff date, we performed our own data reduction and source extraction using the Chandra Interactive Analysis of Observations \citep[CIAO,][we used CIAO v.4.13 and CALDB v.4.9.5]{Fruscione2006} tools. Table~\ref{tbl_Xlog} provides the basic information about all the Chandra pointings described next.

\paragraph{New Chandra Observations}

\begin{figure*}[t]
\centerline{\includegraphics[scale=1.03]{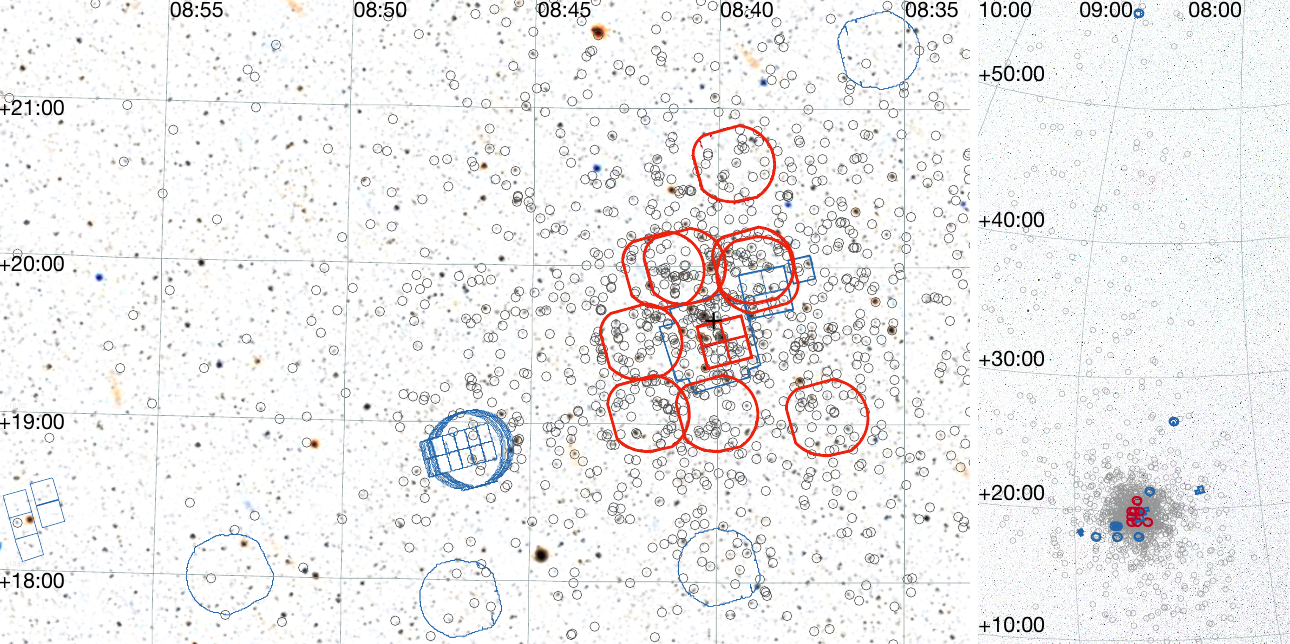}}
\caption{{\it Left}: Footprints of X-ray observations in the field of view of Praesepe: three with the Chandra ACIS camera and 24 with the XMM EPIC camera. Red footprints indicate observations for which one of the co-authors is the P.I., and blue footprints, other publicly available observations. The gray circles indicate cluster members. A black plus symbol indicates the cluster center. Each XMM footprint represents the combined contours of the usable parts of the EPIC pn and MOS cameras. {\it Right:} Zoomed out view of Praesepe, to display eight additional footprints (one Chandra and seven XMM) along a tidal tail of Praesepe, including one at $\delta >$ +50$^{\circ}$, and six that completely overlap. A PanSTARRS color image serves as the background on both panels.}
\label{fig_footprints}
\end{figure*}

The central field of Praesepe was observed four separate times with the same pointing and similar roll angle between 2015 May and 2015 Jun with the Advanced CCD Imaging Spectrometer \citep[ACIS;][]{Garmire2003} for a total of 191.6 ks (Proposal 16200863, PI: J. Drake). The four ACIS-I chips were used in Very Faint telemetry mode to improve the screening of background events and thus increase the sensitivity of ACIS to faint sources \citep{Vikhlinin2001}. The exposure-weighted average aimpoint of the 16$\farcm$9$\times$16$\farcm$9 ACIS-I field of view is $\alpha=08^{\mathrm{h}}39^{\mathrm{m}}50.^{\mathrm{s}}2$, $\delta=+19^{\circ}31'05\farcs0$0 (J2000). Figure~\ref{fig_footprints} shows the combined footprint of this observation, plus an archival Chandra observation and new and archival XMM observations (see Section~\ref{sec:xmm}).

Two M dwarf Hyads, LP 415-19 (2MASS J04214435+2024105) and 2MASS J04214586+2023446, half an arcminute apart, were observed 12 separate times with the same pointing and varying roll angles between 2020 Nov and 2020 Dec, for a total of 218.7 ks (Proposal 22200351, PI: Ag\"ueros). We used the ACIS-S3 chip, plus optional S2, I0, and I1 chips, in Very Faint telemetry mode. The exposure-weighted average aimpoint of the combined field of view is $\alpha=04^{\mathrm{h}}21^{\mathrm{m}}45.^{\mathrm{s}}2$, $\delta=+20^{\circ}23'55\farcs8$ (J2000). In addition to these two M dwarfs, we detected another M dwarf, 2MASS J04213829+2018102, that serendipitously fell within the field of view of the combined pointings. 

The M dwarf Hyad 2MASS J04351354+2008014 was observed six separate times with the same pointing and differing roll angles during between 2021 Jan and 2021 Oct, for a total of 175.1 ks (Proposal 22200351, PI: Ag\"ueros). We used the ACIS-S3 chip, plus optional S2, I0, and I1 chips, in Very Faint telemetry mode. The exposure-weighted average aimpoint of the combined field of view is $\alpha=04^{\mathrm{h}}35^{\mathrm{m}}13.^{\mathrm{s}}6$, $\delta=+20^{\circ}07'59\farcs9$ (J2000).

Lastly, three K dwarf Hyads, StKM 1-393 (2MASS J03390791+2822560), BD+05 526 (2MASS J03400754+0552286), and HD 286363 (2MASS J03550142+1229081), and one G0 dwarf Hyad, HD 265537 (2MASS J06531311+2119128), were observed individually in 2020 and 2021 as part of the Chandra Cool Targets program\footnote{\url{https://cxc.harvard.edu/proposer/CCTs.html}} (Proposal 20201075, PI: Ag\"ueros). In each observation, we used the ACIS-S3 chip in Very Faint telemetry mode. Each star was observed for 10.0 ks.

\paragraph{Using CIAO on the New Chandra Observations}
We performed the following steps on each one of the recent Chandra observations. We started by running \texttt{chandra\_repro}. We did not use very faint correction, as we wanted to avoid excluding potentially good events in modestly bright point sources. Next, we corrected the absolute astrometry of the observations. To do this, we ran \texttt{wavdetect} with a false-probability threshold of 10$^{-6}$ on the observations to produce a conservative list of point sources. We then used a list of high quality ({\sc ph\_qual} = AAA) 2MASS sources and the tool \texttt{reproject\_aspect} to register the astrometry of the observations to the astrometric frame of 2MASS. We used a 3\arcsec\ matching radius and residual rejection limit of 0\farcs6. 

In the cases where observations were broken down into several pointings, we ran the steps above on the longest pointing, and then we ran \texttt{wavdetect} and \texttt{reproject\_aspect} on the rest of the pointings to register their astrometry to that of the longest pointing. Next, we merged the individual pointings using \texttt{merge\_obs} and produced merged events in the soft, hard, and broad bands. 

We created point-spread function (PSF) maps for the observations. In the cases with more than one pointing, we combined the individual PSF maps using \texttt{dmimgfilt} and selecting the minimum PSF map size out of the individual pointings at each pixel. This last step allows us to detect point sources that may be smaller than the mean size, but still larger than the local PSF in the individual pointing maps.

Subsequently, we ran \texttt{wavdetect} for the three energy bands using the PSF map and using a less stringent false-probability threshold of 10$^{-5}$, to allow for a non-negligible number of spurious sources to be included in the initial source candidate list. We consolidated the resulting three lists of sources by matching them against each other using a 2\arcsec\ matching radius for sources with off-axis angle $\theta <$5 \arcmin\ and 4\arcsec\ for sources with $\theta >$ 5\arcmin. For matched sources, we adopted the source region description, in order of priority, from the broad, soft, or hard band list. We then visually inspected the consolidated source list to discard sources that overlapped significantly with others or sources characterized by ellipse regions that deviated significantly from the local PSF (e.g., minor axis $\approx$0\arcsec). Next, we used a 15\arcsec\ matching radius to find optical counterparts in our cluster catalogs for the X-ray sources. The largest offset radius was 6\farcs5 (for a source at an off-axis angle of $\approx$9\arcmin); all other offset radii were below 3\farcs5.

We ran \texttt{srcflux} to extract X-ray source count rates for the three energy bands. We then used our own ECFs (see Section~\ref{lx}) and our adopted individual distances to calculate \fx\ and \LX\ in the 0.1--2.4 keV band. Three of our sources extracted with CIAO have $>$300 net counts in the broad band. As for the bright CSC sources described above, we performed spectral analysis on these sources to extract unabsorbed \fx\ values as well as plasma parameters (see Section~\ref{sec:spec}). The extracted spectroscopic information is included in Table~\ref{tbl_spec}.

Lastly, we ran \texttt{glvary} on each source, which searches for X-ray variability using the Gregory-Loredo algorithm. We inspected the light curves of the six sources we have with a probability of variability $\geq$0.9 (considered to be definitely variable). We found X-ray flares in five of the six light curves. In such cases, we filtered them out (see Section~\ref{sec:flares}) and then re-ran \texttt{srcflux} to re-extract the X-ray photometry and properties of the sources.

In total, we matched X-ray sources from these recent Chandra observations to 19 Praesepe stars and eight Hyads in our catalogs. Table~\ref{tbl_Xcols} includes the X-ray information for these 27 X-ray sources.

\subsection{XMM-Newton Observations}\label{sec:xmm}

\subsubsection{Data from the XMM Serendipitous Source Catalog}\label{4XMM}

We searched for X-ray counterparts to our cluster stars in the latest version of the XMM Serendipitous Source Catalog \citep[4XMM-DR11,][]{4XMM} from observations with the EPIC instrument. We cross-matched the list of individual detections (which use the {\sc detid} naming convention) in the 4XMM catalog and our cluster catalogs using a 15\arcsec\ radius. We found 4XMM counterparts to 315 Praesepe stars and 110 Hyads. Like with CSC (see Section~\ref{sec:CSC}), we estimated the number of likely false matches by shifting the X-ray source positions in steps of 25\arcsec\ out to 5\arcmin\ in all directions and re-matched them to the cluster catalogs using the same 15\arcsec\ radius used before. We found the median number of false matches to be 21 ($\approx$5\% of our matched 4XMM sources) with a median offset of 10\arcsec. In our final 4XMM-cluster catalogs matches, we flag those with radii $>$10\arcsec---there are eight Praesepe matches (all for the same star) in this category---as likely false matches. All 110 Hyad matches, on the other hand, have matching radii $<$6\arcsec.

From 4XMM, we obtained count rates in the basic energy bands 1 through 5, which together span the energy range 0.2--12.0 keV, and then combined bands 1 through 3 to create a soft band, 4 and 5 to create a hard band, and 1 through 5 to create the broad band. Additionally, we obtained some other information about the X-ray sources, such as detection likelihoods and variability flags (see Table~\ref{tbl_Xcols}). We used the published count rates, our own ECF (see Section~\ref{lx}), and our adopted individual distances to calculate \fx\ and \LX\ in the 0.1--2.4 keV band. Information about the original XMM observations that collected the data for these 425 4XMM sources is shown in Table~\ref{tbl_Xlog}.

Of the 4XMM X-ray sources matched to cluster stars, 129 had $>$300 net counts in the broad band for the pn camera or $>$500 net counts in the broad band for the combined pn and MOS cameras. For these sources we performed spectral analysis to extract unabsorbed \fx\ as well as plasma parameters (see Section~\ref{sec:spec}). The extracted spectroscopic information is included in Table~\ref{tbl_spec}.

We inspected the light curves for all 4XMM sources with flags indicating variability. We found 28 sources with X-ray flares in their light curves. In such cases, we filtered the flare events out (see Section~\ref{sec:flares}) and then re-extracted the X-ray photometry and properties of the sources. Table~\ref{tbl_Xcols} includes the X-ray information for all the 4XMM sources.

\subsubsection{XMM Source Extraction with SAS}\label{SAS}

The 4XMM-DR11 catalog only includes observations through the end of 2020. For the XMM pointing we obtained after this cutoff date, we performed our own data reduction and source extraction using the XMM Science Analysis System \citep[SAS v19.0,][]{SAS} tools. Table~\ref{tbl_Xlog} provides the basic information about this XMM recent pointing.

Furthermore, due to the automated nature of the 4XMM-DR11 catalog pipeline, some faint point-like sources may have been discarded as spurious detections. As some of our cluster stars are expected to be very X-ray faint, we performed our own data reduction of ten archival XMM observations, all targeting the cluster core, to verify that no X-ray faint source be left behind.

\paragraph{New XMM Observation}

The slow-rotating Praesepe K3 dwarf JS 297 (2MASS J08393203+2039203) was observed in 2021 Apr for 32.0 ks (Proposal 86371, PI: N\'u\~nez). We used the EPIC cameras with the medium filter. The aimpoint of the field of view is $\alpha=08^{\mathrm{h}}39^{\mathrm{m}}32.^{\mathrm{s}}0$, $\delta=+20^{\circ}39'20\farcs3$ (J2000) and the roll angle is 283\fdg6. In addition to this K dwarf, we detected ten other Praesepe dwarfs (one F, one G, one K, and seven M types) that serendipitously fell within the field of view of this observation, as we describe next.

\paragraph{Using SAS on New and Archival XMM Observations}

We used SAS to reduce the recent observation---plus ten archival observations targeting the core of Praesepe---with the most up-to-date calibration. We started by applying standard filters to event files. Next, we created filtered event files for the three EPIC cameras by excluding times when the global count rate in the MOS1 camera increased beyond $\approx$2.5 counts sec$^{-1}$. Finally, we used the \texttt{edetect\_chain} task to perform source detection on both MOS and pn images simultaneously. The algorithm in \texttt{edetect\_chain} runs a sliding box source detection (\texttt{eboxdetect}), then computes maximum likelihood analysis to prune the initial source list (\texttt{emldetect}). 

The 4XMM SAS pipeline uses the five basic energy bands 0.2--0.5, 0.5--0.1, 1.0--2.0, 2.0--4.5, and 4.5--12.0 keV (bands 1 through 5) for source detection and extraction. We used instead the broader energy bands 0.2--2.0 and 2.0--12.0, plus the total band 0.2--12.0 keV. We adopted these three bands, which correspond to 4XMM bands 6, 7, and 8, as the soft, hard, and broad bands, respectively. An image with a broader energy band will include a higher number of source counts than one with a narrower band; for a very faint source, this increase in source counts may be significant. Therefore, the source detection routines may be able to pick up fainter sources in broader bands. At the same time, a broadband image will also include a higher number of background counts. As a compromise, we increased the detection likelihood threshold in the detection algorithm from 6---the value used in the automatic SAS pipeline---to 8. Furthermore, although \texttt{edetect\_chain} performs simultaneous source detection across all bands and detectors, we did not require sources to be detected in all three cameras. As, for example, there are several dead chips on the MOS1 camera, this would severely limit our available area for detection.

We cross-matched the resulting X-ray source list with our cluster catalogs using a 15\arcsec\ tolerance radius. From the new XMM observation (not included in the 4XMM catalog), we matched eleven X-ray sources to Praesepe counterparts. One of these matched sources had enough source counts to perform spectral analysis (see Section~\ref{sec:spec}). Table~\ref{tbl_Xcols} includes information for these eleven sources.

From our SAS reductions of the archival observations of the core of Praesepe, we found 34 X-ray counterparts to Praesepe stars that are not present in the 4XMM catalog. The maximum likelihood values for these 34 sources are in the range 8.2--140.0; the source with the highest likelihood was detected only with the pn camera. Table~\ref{tbl_Xcols} includes information for these 34 sources. One of the 34 sources has a likelihood of being an extended source of 9.1, and we therefore flag this source as a likely extended source in this Table.

\subsection{Data from the Swift XRT Point Source Catalog}

We searched for X-ray counterparts to our cluster stars in the latest version of the Swift XRT Point Source Catalog \citep[2SXPS,][]{2SXPS}. This catalog includes observations through 2018 Aug, which covers eleven observations from our proposal to observe low-mass Hyads (Proposal 1215128, PI: Ag\"ueros). %The 2SXPS catalog includes 2MASS counterparts to its X-ray sources; this was done by cross-correlating the X-ray sources with 2MASS using 99.7\% confidence levels and Rayleigh statistics. Nonetheless, 
We cross-matched 2SXPS and our cluster catalogs using a 15\arcsec\ radius. We found 2SXPS counterparts to six Praesepe stars and 59 Hyads. Like with ROSAT (see Section~\ref{sec:Rosat_surveys}), we estimated the number of likely false matches by shifting the X-ray source positions by 25\arcsec\ out to 3\arcmin\ in all directions and re-matched them to the cluster catalogs using the same 15\arcsec\ radius used before. We found that the median number of false matches is one ($\approx$2\% of our matched 2SXPS sources) with a median offset of 10\arcsec. None of our 2SXPS matches merited a ``likely mismatch'' flag (Column 28 of Table~\ref{tbl_Xcols}).%In the latter cohort, there are two, with matching radii $\approx$7\arcsec, with no 2MASS counterpart in the 2SXPS catalog. %We flagged these two matches as likely false matches.
%Both of them, however, are matched to Hyads with X-ray detections from other observatories: one has a 2RXP counterpart, and the other has a 2RXP and a 4XMM counterpart. Our derived \fx\ values for these two 2SXPS sources are well within 20\% of the \fx\ values of the X-ray sources from the other observatories. As such, we opted to include these two 2SXPS matches to our Hyades catalog.

From the 2SXPS catalog we obtained count rates in the 0.3--1.0, 1.0--2.0, and 2.0-10.0 keV bands. We then combined the first two bands to create the soft band, adopted the third band as the hard band, and combined all three bands to create the broad band. Additionally, we obtained some other information about the X-ray sources, such as hardness ratios, detection likelihoods, and variability flags (see Table~\ref{tbl_Xcols}). We used the published count rates, our own ECF (see Section~\ref{lx}), and our adopted individual distances to calculate \fx\ and \LX\ in the 0.1--2.4 keV band.

Three of the 2SXPS sources had $>$300 source counts in the broad band. For these sources, we performed spectral analysis to extract unabsorbed \fx\ as well as plasma parameters (see Section~\ref{sec:spec}). The extracted spectroscopic information is included in Table~\ref{tbl_spec}.

\subsection{X-ray Spectral Fitting and Properties}\label{sec:spec}
We performed spectral analysis on six Chandra, three Swift, and 130 XMM sources, for a total of 139. 
To be included in spectral fitting, we required an X-ray source to have at least 300 counts\footnote{We performed spectral fitting on sources between 100 and 300 counts, but obtained poor statistical results. We therefore excluded sources with $<$300 counts from our spectral analysis.} in either Chandra ACIS, Swift XRT, or XMM EPIC pn cameras; we also included XMM sources with at least 500 combined counts in the three EPIC cameras.
% We required an X-ray source to have at least 300 counts in either Chandra ACIS, Swift XRT, or XMM EPIC pn cameras (we also included XMM sources with at least 500 counts in the combined three EPIC cameras) for us to perform spectral fitting. 

\begin{figure*}[t]
\centerline{\includegraphics{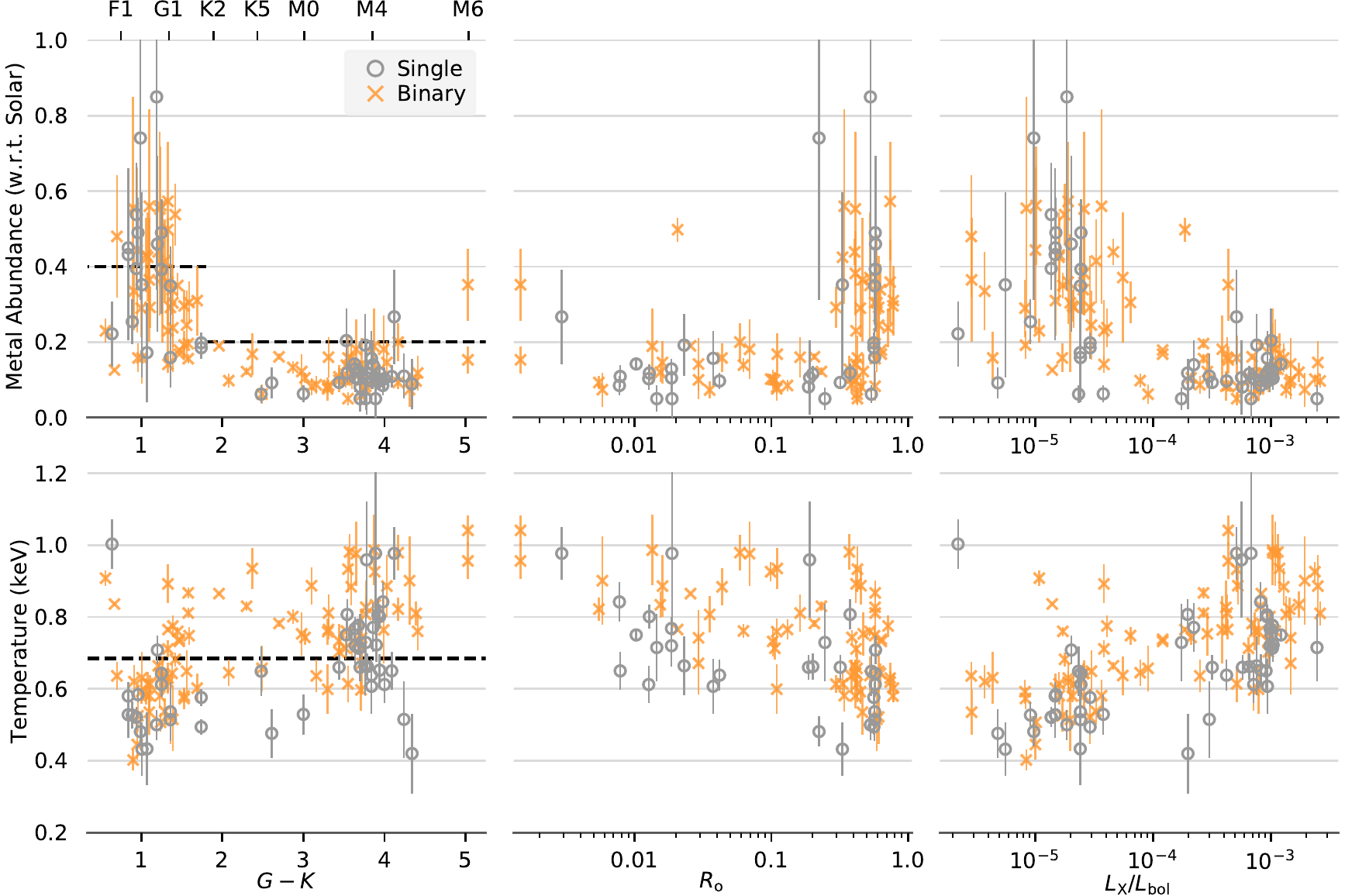}}
\caption{X-ray spectral fitting parameters for X-ray sources with $>$300 counts, using a one-temperature APEC model and setting the neutral hydrogen column density to the typical value for each cluster (see Section~\ref{sec:spec}). The panels show the coronal metal abundance with respect to Solar (top row) and coronal temperature (bottom row) as a function of \gminusk\ color (left column), \Ro\ (middle column), and \LLX\ (right column). Gray circles indicate single members, and orange $\times$ symbols, candidate/confirmed binaries. The dash horizontal lines in the left column indicate the parameter values we assumed for the APEC model we used to derive our own ECFs to apply to low-count X-ray sources (see Section~\ref{lx}). Most K and M dwarfs have metal abundance less than $\approx$0.2, whereas F and G dwarfs have a large spread in abundance, from $\approx$0.1 up to $\approx$0.9. Plasma temperature is constrained between 0.4 and 1.0 keV for all dwarfs.}
\label{fig_Xfits}
\end{figure*}

\input{tbl_spec}

Of the 139 X-ray spectra, 75 are of Praesepe stars and 64 are of Hyads, including 89 total spectra for candidate/confirmed binaries. 
Also, seven of the 139 are cluster giants, and one is a candidate white dwarf--M6 dwarf binary \citep[2MASS J03040207+0045512,][]{Becker2011}. We excluded these eight non-main-sequence stars from our analysis.

We used the CIAO tool Sherpa \citep[v. 4.13.0,][]{Sherpa, Sherpa_software} to perform the spectral analysis. We fitted the spectra with an XSPEC one-temperature (1T) APEC optically thin plasma emission spectrum model. We combined the APEC model with ISM absorption model TBabs to account for extinction by neutral atomic hydrogen ($N_\mathrm{H}$). We binned the spectra by 20 counts per bin for sources with more than 1000 counts, and by 15, for those with fewer than 1000 counts. We then fitted the spectra using the $\chi^2$ statistic with the Gehrels variance function and Sherpa's {\it levmar} optimization method. We obtained 1$\sigma$ confidence intervals of all free parameters by computing the co-variance matrices.

Most XMM sources had up to three spectra from the same observation, one EPIC pn and two MOS spectra. In such cases, we performed the spectral fit for each source using all available spectra simultaneously.

For each source, we ran the fit twice. First, we ran it by setting the $N_\mathrm{H}$ parameter to that of the star's cluster: 1.50$\times10^{20}$ cm$^{-2}$ for Praesepe and 5.50$\times10^{18}$ cm$^{-2}$ for Hyades, derived using our adopted $E$($B$-$V$) values for the two clusters (see Section~\ref{photometry}) and $N_\mathrm{H}$[cm$^{-2}/A_v$] = 1.79 $\times 10^{21}$ \citep{Predehl1995}. Then, we freed the $N_\mathrm{H}$ parameter and ran the fit again. We considered the fit with the reduced $\chi^2$ statistic ($\chi_\nu^2$) closest to unity as the best fit. Only eleven of the 139 X-ray sources had a better spectral fit with a free $N_\mathrm{H}$ parameter. The resulting $N_\mathrm{H}$ in these eleven spectra are between 0.2 and fifteen times the value for the cluster, except one case, for which $N_\mathrm{H}$ is two orders of magnitude larger than that of the Hyades cluster. This X-ray source with unexpectedly large $N_\mathrm{H}$ is matched to the white dwarf--M6 binary mentioned above.

From each spectral fit, we also obtained values for the plasma temperature {\it kT}, metal abundance with respect to the Sun, and the unabsorbed energy flux in the 0.1--2.4 keV band. Column 26 of Table~\ref{tbl_Xcols} identifies the X-ray sources for which we used spectral fitting to calculate their \fx. Table~\ref{tbl_spec} shows the fit results for each of the 139 X-ray sources, including the $\chi_\nu^2$ statistic and the degrees of freedom. Figures of our spectral fits are publicly available from the Columbia University Academic Commons, an online research repository.\footnote{Available at \url{https://doi.org/10.7916/dtws-0x90}. \label{fn:commons}}

Figure~\ref{fig_Xfits} shows the metal abundance (top row) and plasma temperature (bottom row) parameters resulting from our spectral fits, drawn against \gminusk\ color (left column), \Ro\ number (middle column, see Section~\ref{rossby}), and \LLX\ (right column, see Section~\ref{lx}) for main-sequence stars in the two clusters. There is a clear trend for metal abundances to remain less than $\approx$0.2 with respect to Solar for most K and M dwarfs ($G-K > 1.8$). F and G dwarfs ($G-K$ \lapprox\ 1.8), on the other hand, show a large spread in abundance, from $\approx$0.1 up to $\approx$0.9, with an average value of $\approx$0.4. This trend is true for both single (gray circles) and binary stars (orange $\times$ symbols). Plasma temperature, on the other hand, appears to be constrained between 0.4 keV (4.6 MK) and $\approx$1.0 keV (11.6 MK) for all low-mass stars, with an average value of $\approx$0.7 keV (8.1 MK).

\subsection{Removing X-ray Flares}\label{sec:flares}

Some of the X-ray observations used in our work have exposure times long enough ($\gapprox$30 ks) to increase the probability of catching a transient stellar flare. During an X-ray flare, the X-ray emission of a star increases significantly from the quiescent level, in some extreme cases up to 7000 times, for a couple to a few hours \citep[e.g.,][]{Osten2010}. Therefore, including the emission from the flare in the calculation of the energy flux of a source would inflate the flux value from the expected quiescent level---the latter being the more representative measurement of the magnetic activity level of a low-mass star. To obtain the most representative energy flux values for stars in our two clusters, we removed X-ray flares whenever possible from our X-ray source calculations. 

First, we examined the X-ray light curves of sources with variability flags indicating a high probability of a variable source, as these are the ones most likely to include an X-ray flare. Next, we identified the flare time interval of each X-ray flare, which we defined as the time with a rapid increase in the count rate to at least 3$\sigma$ from the quiescent level and lasting for at least $\approx$2 ks (half an hour). We then generated new event files by removing counts that occurred during the flaring time intervals. Finally, we re-extracted X-ray source parameters and photometry from the new event files. 

All in all, we removed flares from five Chandra and 27 XMM sources. The average change in source count rate from pre- to post-flare removing was a decrease of $\approx$12\%. Seven of these 32 sources had an unexpected higher count rate after we removed flares. We found that in all seven cases---all from the 4XMM catalog---we removed, in addition to an X-ray flare, noisy light curve edges caused possibly by a faulty X-ray background extraction in the 4XMM automated reduction pipeline, which resulted in episodes of negative count rates in the original light curve. 

Column 27 of Table~\ref{tbl_Xcols} indicates whether we removed an X-ray flare from each source to calculate its X-ray parameters and photometry. Figures of the X-ray light curves with identified flares are publicly available from the Columbia University Academic Commons.\footref{fn:commons}

\section{Deriving Stellar Properties}\label{sec:properties}

\subsection{Stellar Masses and Bolometric Luminosities}\label{sec:lbol}

We calculated stellar masses $m$ and bolometric luminosities (\Lbol) for main-sequence members of both clusters using the empirical $m$-- and log(\Lbol)--$M_G$ relations of E.~Mamajek.\footnote{Version 2021.03.02. Available at \url{http://www.pas.rochester.edu/~emamajek/EEM_dwarf_UBVIJHK_colors_Teff.txt}. Much of this table comes from \citet{Pecaut2013}.} We linearly interpolated between the $M_G$ values in the empirical relation to obtain $m$ and log(\Lbol) values. The calculated $m$ values are in the range 0.09--2.28 \Msun\ for Praesepe and 0.08--2.47 \Msun\ for Hyades. Column 31 in Table~\ref{tbl_catalogcols} includes $m$, and columns 33 and 34, \Lbol\ and its 1$\sigma$ uncertainty.

For nine of the ultracool dwarfs that we found in the literature (seven in Praesepe and two in the Hyades), we adopted $m$ published by those authors (see Section~\ref{sec:ultracool}), instead of relying on the $m$--$M_G$ relation above. These nine values are in the range 0.05--0.11 \Msun.

Lastly, although we calculated $m$ and \Lbol\ for candidate/confirmed binaries, we recognize that these two quantities are likely over-estimated for these stars, as there is more than one light source contributing to the brightness of the spatially unresolved stellar source.

\subsection{Rossby Numbers} \label{rossby}

Studies of the rotation-activity relation typically express rotation in a mass-independent manner by substituting the Rossby number \Ro\ for \prot, as was first demonstrated by \citep{Noyes1984}. $R_o$ is given by \prot/$\tau$, where $\tau$ is the convective overturn time. We used the empirical mass--log($\tau$) relation of \citet{Wright2018}, which is based on \prot\ and X-ray luminosity measurements for almost 850 stars in the mass range 0.08--1.36~\Msun. Column 32 in Table~\ref{tbl_catalogcols} includes our $\tau$ estimates. With those $\tau$ values, we calculated $R_o$ for cluster stars with a measured \prot.

\subsection{X-ray Luminosities}\label{lx}
Most of the X-ray sources matched to Praesepe and Hyades stars do not have enough source counts to derive \fx\ from spectral fitting (see Section~\ref{sec:spec}). For these low-count X-ray sources, we calculated ECFs using the tool WebPIMMS\footnote{\url{https://heasarc.gsfc.nasa.gov/cgi-bin/Tools/w3pimms/w3pimms.pl}} to convert the instrumental count rates into unabsorbed \fx\ in the 0.1--2.4 keV energy band. We calculated the ECFs using a one-temperature thermal APEC model \citep{Smith2001} with a plasma temperature\footnote{The PIMMS models only have solutions for discrete values of the input parameters. The closest of these discrete values to our average plasma temperature is 0.6845 keV, and to our average metal abundances are 0.2 and 0.4 Solar.\label{fn:pimms}} of 0.6845 keV and setting the following inputs specific for each X-ray source: 
\begin{itemize}
    \item the X-ray instrument used, namely, Chandra's ACIS-I, ACIS-S, and HRC detectors, ROSAT's PSPC and HRI cameras, Swift's XRT camera, and XMM's pn and MOS EPIC cameras;
    \item the Chandra cycle in which the source was detected\footnote{We used \url{https://cxc.harvard.edu/toolkit/pimms.jsp} for Chandra's ECFs, because it accounts for the evolution of the effective areas of the Chandra detectors.};
    \item the filter used with the XMM EPIC cameras (thin, medium, or thick filters);
    \item the energy bands of the instrumental count rates, which are generally instrument-specific and thus different for the different X-ray missions; plus, for the case of the ROSAT catalogs we used the energy bands reported therein;
    \item the $N_\mathrm{H}$ value assumed for each open cluster (see Section~\ref{sec:spec}); and
    \item a plasma metal abundance\footref{fn:pimms} of 0.2 for stars with \gminusk\ $>$ 1.8 and 0.4 for stars with \gminusk\ $\leq$ 1.8 (see Section~\ref{sec:spec}).
\end{itemize}

\input{tbl_ECFs}

Table~\ref{tbl_ECFs} lists all the ECFs we used for the different X-ray point sources in our X-ray catalog. Columns 24 and 25 in Table~\ref{tbl_Xcols} include our \fx\ values for each X-ray source and their 1$\sigma$ uncertainties, and Column 26 identifies the X-ray sources for which we derived \fx\ using our ECFs. We do not calculate \fx\ for X-ray sources with Quality Flag = ``m'' or ``x'' (Column 28 in Table~\ref{tbl_Xcols}).

There are 106 Hyads and 79 Praesepe stars with more than one \fx\ (and up to four). For these stars, we calculated the \fx\ range as a fraction of the mean: (\fx$_\mathrm{,max} - $\fx$_\mathrm{,min}$) / \fx$_\mathrm{,mean}$. The median value is 0.45, with 16$^\mathrm{th}$ and 84$^\mathrm{th}$ percentiles of 0.15 and 1.12. Eleven stars have values beyond 1.6 (and up to 2.8). In all eleven cases, the time span of the detections is between 15 and 25 years. We suspect, therefore, that we may be capturing Solar-type activity cycles in these stars. However, a study of such activity cycles is beyond the scope of this project.

For stars with more than one \fx, we used the error-weighted mean \fx\ as the adopted \fx\ for the star. Lastly, we converted \fx\ into \LX\ using our adopted distances. For each cluster star detected in X-rays, columns 29 and 30 in Table~\ref{tbl_catalogcols} include the adopted \fx\ value and its 1$\sigma$ uncertainty.

\begin{figure}
\centerline{\includegraphics{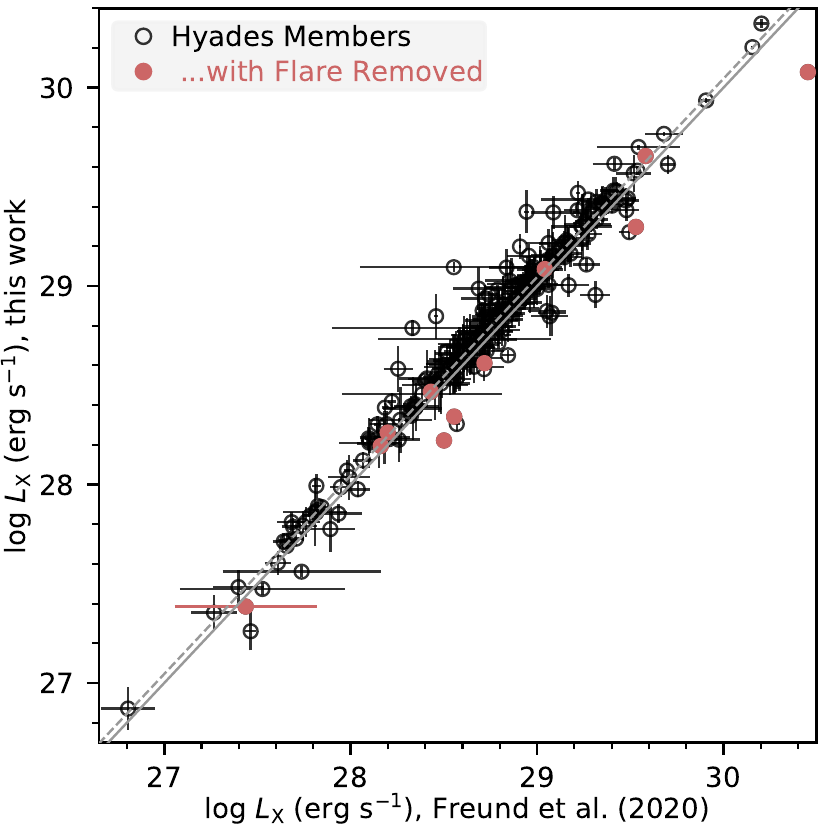}}
\caption{Comparison of \LX\ values derived for Hyads in \citet{Freund2020} (abscissa) vs. our \LX\ values for the same stars (ordinate). We account for the difference in energy range (their 0.2--12.0 keV vs. our 0.1--2.4 keV) by increasing their \LX\ values by 16.3\%. Hyads for which we removed X-ray flares before computing their \LX\ are highlighted with red circles. The gray solid line is the 1:1 relation, and the gray dashed line is offset from the 1:1 relation by 11\%, which corresponds to the median offset in luminosity between our \LX\ values and those in \citet{Freund2020}.}
\label{fig_Freund}
\end{figure}

\citet{Freund2020} calculated \LX\ values for Hyads From Chandra, ROSAT, and XMM detections. If a star had more than one X-ray detection, these authors assigned it a ``best'' X-ray detection to derive its \LX, instead of combining them like we did. This results in our sample having \LX\ uncertainties smaller than theirs for several stars. Also, they did not account for potential flares in their X-ray detections like we did, which results in several of these flaring Hyads having lower \LX\ in our catalog. Furthermore, in their ECFs, these authors assumed an APEC temperature of 0.2725 keV and a metal abundance of 1.0 for all stars, compared to our temperature of 0.6845 keV and abundance of 0.2 or 0.4.

To compare our \LX\ values against those in \citet{Freund2020} for the same stars, we had to account for the difference in energy ranges used to calculate \LX\ (their 0.2--12.0 keV vs. our 0.1--2.4 keV). Using PIMMS, we found that one must increase \fx\ by 16.3\% to go from 0.2--12.0 keV to 0.1--2.4
keV, everything else being equal. After applying this correction, we found a median offset in luminosity of $\approx$11\% between our \LX\ values and theirs, with our values being larger on average. Figure~\ref{fig_Freund} compares our \LX\ to those in \citet{Freund2020}, highlighting stars for which we removed flares \LX\ (filled red circles). The systematic offset between our \LX\ and those of \citet{Freund2020} can be explained by the difference in assumed APEC parameter values: increasing the plasma temperature from 0.2725 keV to 0.6845 keV increases the resulting \fx\ values by up to 20\%, and decreasing the plasma abundance from 1.0 to 0.2 decreases the resulting \fx\ values by up to 5\%.
%We suspect that the difference in assumed APEC parameters drives the systematic offset between our \LX\ values and those of \citet{Freund2020}.

\section{Results and Discussion}\label{sec:res}

\subsection{X-ray Spectral Parameters}\label{sec:spectralparams}

Beginning with the Advanced Satellite for Cosmology and Astrophysics (ASCA) and Extreme Ultraviolet Explorer (EUVE) missions, studies of both low- and high-resolution EUV and X-ray spectra of low-mass stars have revealed that the chemical composition of the corona differs from that of the underlying photosphere \citep[e.g.,][]{Drake1996, Brinkman2001, Guedel2007, Testa2010}. Long documented in the case of the Sun, this chemical abundance anomaly is referred to as the ``First Ionization Potential (FIP) Effect'': elements with low FIP (FIP~$\leq 10$~eV, e.g., Si, Mg, Fe) are enhanced by factors of 2--4  relative to
elements with high FIP (FIP~$\geq 10$~eV, e.g., N, Ne, Ar). In the stellar case, stars with fairly low, solar-like  magnetic activity levels exhibit a solar-like FIP effect \citep[e.g.,][]{Drake1997}. Instead, more active stars exhibit the reverse, and the low FIP metals are generally depleted relative to high-FIP elements \citep[e.g.,][]{Brinkman2001,Drake2001}. 

More recent studies with high-resolution X-ray spectra have also uncovered a spectral type dependence of the FIP effect, and 
it is not yet clear whether the depletion or enhancement of metals is mostly controlled by magnetic activity level  \citep[e.g.,][]{Telleschi2005, Garcia-Alvarez2009}, or by the change in outer convection zone properties with spectral type \citep[e.g.,][]{Wood2010, Wood2012, Wood2013}. 

The FIP and inverse FIP effects are not yet fully understood, but are likely related to one or more mechanisms of coronal heating \citep[e.g.][]{Drake2002,Testa2015}. The most promising model to date cites the ponderomotive force on ions in the chromosphere associated with magnetohydrodynamic waves \citep{Laming2015}. 

Our results show a clear correlation between spectral type and coronal metal abundance. In our sample of Praesepe and Hyades stars, K and M dwarfs ($G-K\ \gapprox\ 1.8$) display very low abundances, whereas F and G dwarfs ($G-K\ \lapprox\ 1.8$) display higher abundances (see upper left panel of Figure~\ref{fig_Xfits}). We find similar correlations with \Ro\ and \LLX: fast rotators (\Ro\ $\lapprox~0.2$) have very low abundances, and vice versa (upper middle panel), and the more active stars (\LLX\ $\gapprox~10^{-4}$) have very low abundances, and vice versa (upper left panel).

As will be shown in Section~\ref{sec:rossbyresults}, spectral type, \Ro, and \LLX\ are tightly correlated in our sample of $\approx$700 Myr old stars, as most fast rotating stars with high activity levels are K and M types, whereas most slowly spinning stars with moderate to low activity are F and G types. Therefore, we cannot completely disentangle activity level, rotation, and spectral type from their relation to coronal abundance in our sample. Nonetheless, we note that abundance has a tighter correlation with spectral type than with activity level, as evidenced by a clear shift from $\lapprox$0.2 to an average $\approx$0.4 abundance in both \gminusk\ color space and \Ro\ space, the latter being a function of the stellar mass dependent $\tau$ parameter. Furthermore, a significant fraction of K dwarfs have similar slow rotation and low activity levels as G dwarfs, and yet in Figure~\ref{fig_Xfits} it is clear that all K dwarfs have abundance levels comparable to those of the fast rotating, highly active M dwarfs instead.

Studies have also found mean coronal temperature to decrease with decreasing magnetic activity levels \citep[see review by][]{Telleschi2005}. In our sample, we observe a correlation between temperature and \LLX\ (see lower right panel of Figure~\ref{fig_Xfits}). Temperature increases by a factor of $\approx$2.5 as \LLX\ increases by $\approx$3 orders of magnitude. In a similar study, \citet{Singh1999} found on a sample of ten low-mass field stars that temperature increased by a factor of $\approx$4 while \LLX\ increased by three orders of magnitude. Thus, in our sample, coronal temperature appears to be slightly less sensitive to activity levels.

A similar correlation can be observed between temperature and \Ro\ (bottom middle panel), where stars with higher \Ro, i.e. slower rotation rates and, consequently, lower activity levels, display lower coronal temperatures, more so for single stars than for binaries. The bottom left panel of Figure~\ref{fig_Xfits}, on the other hand, shows a very weak correlation between coronal temperature and \gminusk\ color. Rising temperature with increasing color (i.e. later spectral types) can be discerned for binaries. For single stars, the mean temperature of later types is higher, but this is accompanied by a larger spread in values.

For the most part, however, single and binary stars display very similar behaviors in Figure~\ref{fig_Xfits}. This result appears to contradict that of \citet{Pye1994}, who showed binary K dwarf Hyads being at least twice as X-ray bright as their single brethren. We note that there is a significant gap in our coverage of K and early M dwarfs (1.8 $\lapprox~G - K~\lapprox~3.5$) with X-ray spectral fitting parameters: in this spectral range we have only four single stars, but we have 17 binaries. And indeed, in that range binaries appear to have higher coronal temperatures than the single cohort, but with only four single stars to compare against, we cannot affirm that with a high level of confidence. %If indeed, coronal parameters between single and binary stars do not differ significantly---as our result shows, then we interpret this as evidence that binaries have no more active coronae than single stars.

\begin{figure}
\centerline{\includegraphics{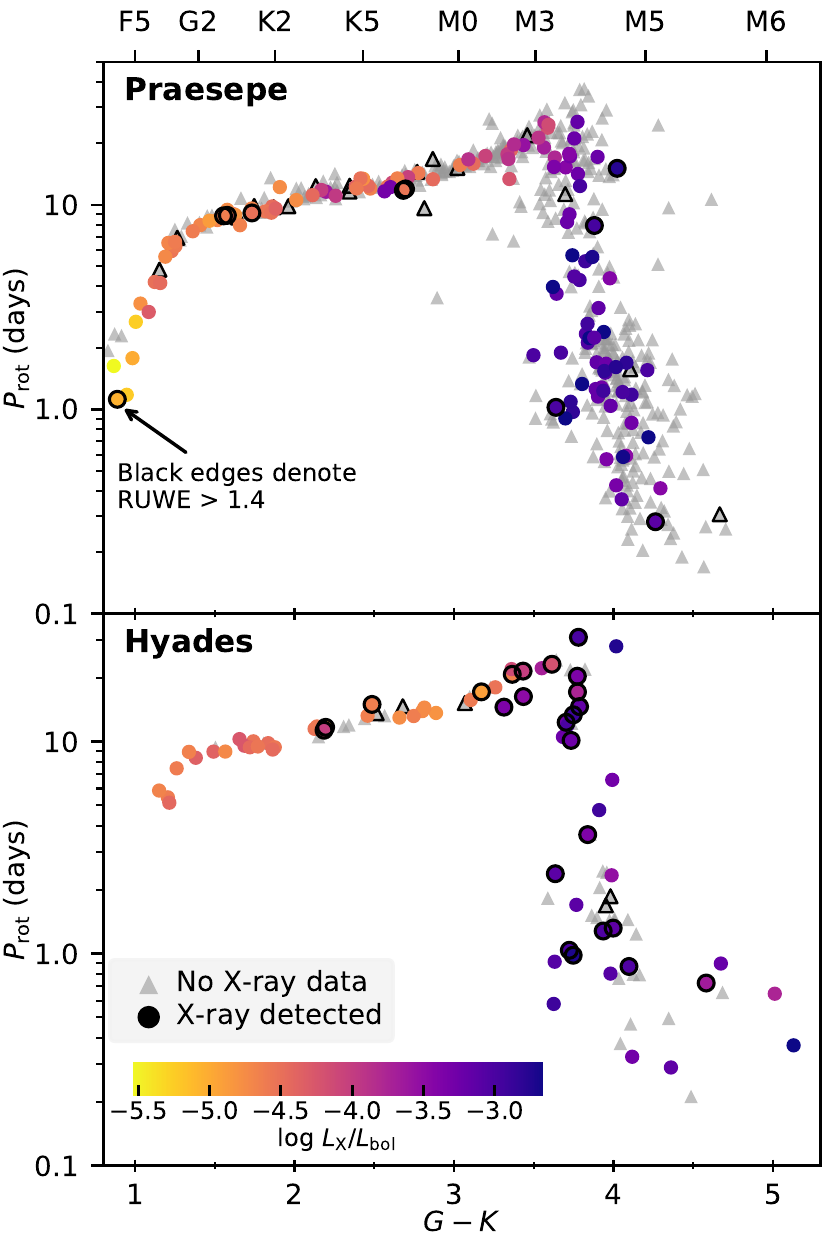}}
\caption{\prot\ vs.~\gminusk\ color for single Praesepe (top panel) and Hyades (bottom panel) stars. Circles indicate stars with an \LLX\ measurement. Gray triangles indicate stars without X-ray data. The circles are color-coded according to their $\log$ \LLX\ following the colorbar at the bottom left. Symbols with black edges are single stars with RUWE > 1.4.}
\label{fig_prots}
\end{figure}

\subsubsection{The Color--Rotation Plane for Single Stars}
In Figure~\ref{fig_prots}, we plot single Praesepe and Hyades members in \gminusk--\prot\ space, with color indicating \LLX. Stars that remain undetected in X-rays are indicated with gray triangles. We note in both clusters (although more so in Praesepe due to its broader \prot\ coverage) that mid F to early G dwarfs can have similar \prot\ values to some early-to-mid M dwarfs ($\approx$M3 to M5), and yet the latter have much higher levels of magnetic activity. 
Furthermore, early-to-mid M dwarfs have \LLX\ spanning up to $\approx$1 order of magnitude, and yet their \prot\ values span over 2 orders of magnitude ($\approx$5 hr up to $\approx$25 d). Evidently, some saturation mechanism is preventing the fastest rotators from increasing their X-ray emission beyond a certain \LLX\ level, as we will see more clearly in Section~\ref{sec:rossbyresults}.

Lastly, stars with RUWE $>$ 1.4 do not stand out in the \gminusk--\prot\ plane (highlighted with black edges in Figure~\ref{fig_prots}). If these high-RUWE stars are indeed unresolved intermediate binaries, Figure~\ref{fig_prots} tells us that their spin down evolution has not been greatly affected, if at all, by the companion.

\subsection{The Distribution of X-Ray Emission}\label{sec:lxcolor}
\begin{figure*}[t]
\centerline{\includegraphics{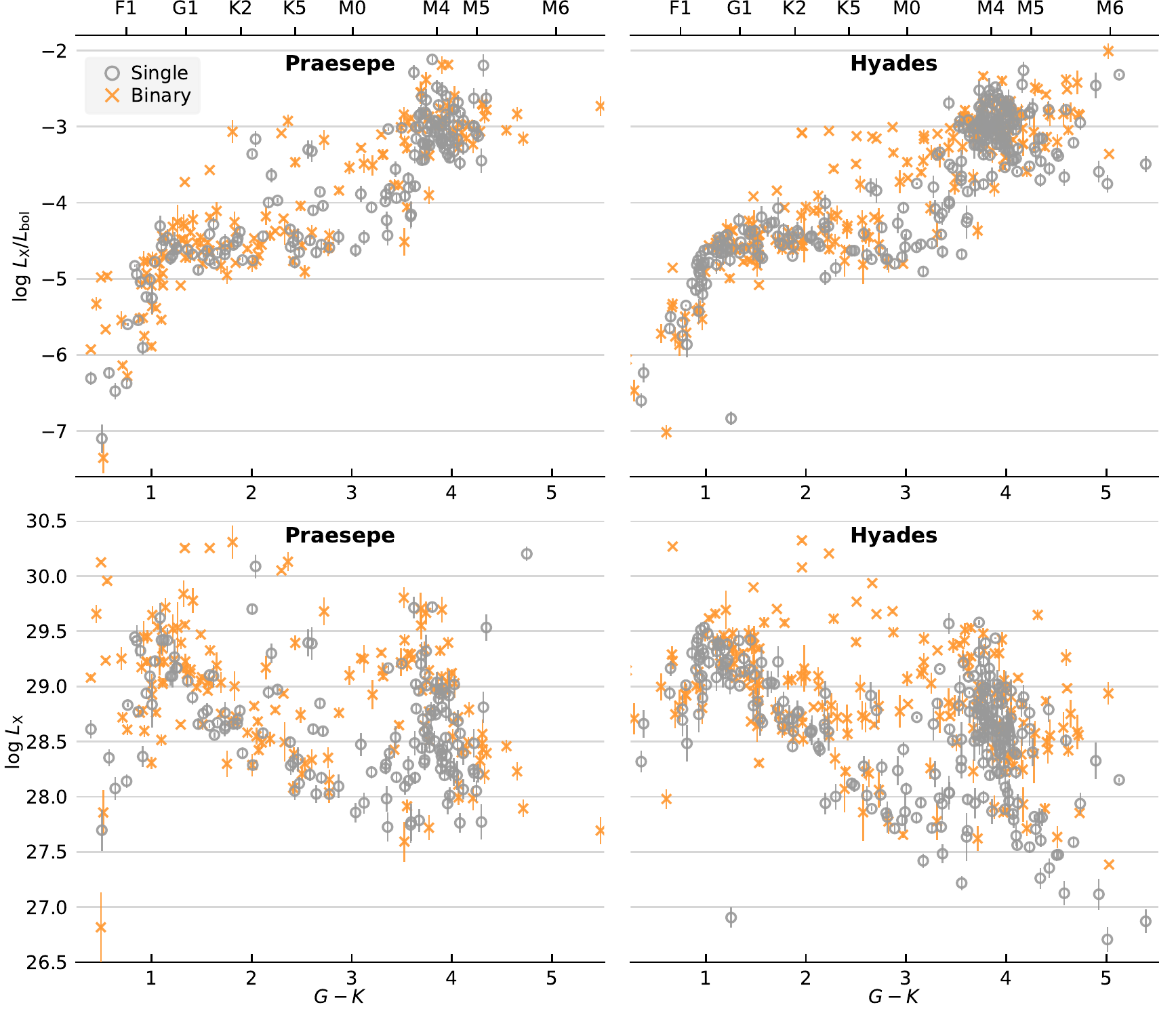}}
\caption{\LLX\ (top panels) and \LX\ (bottom) vs. \gminusk\ color for Praesepe (left panels) and Hyades (right) stars. Gray circles indicate single members, and orange $\times$ symbols, candidate/confirmed binaries. Stars, both single and binary, follow a narrow increasing \LLX\ sequence from F through early G that then spreads between G and early M types, and narrows again beyond early M types.}
\label{fig_LLXb}
\end{figure*}

Figure~\ref{fig_LLXb} shows the relation between X-ray emission, in the form of \LLX\ (upper panels) and \LX\ (lower panels), and \gminusk\ color for Praesepe (left panels) and Hyades (right panels) stars. We calculate in Table~\ref{tbl_llxstats} the \LLX\ and \LX\ median, 16$^\mathrm{th}$, and 84$^\mathrm{th}$ percentiles of several spectral type ranges for both single and binary stars. \LLX\ follows a narrow increasing \LLX-color relation from F through early G types, then spreads---if not remaining flat---between early G and early M types, and then narrows again beyond early M types. This is true for both Praesepe and Hyades stars, as well as single (gray circles) and candidate/confirmed binary (orange $\times$ symbols) stars. For the latter, there is a larger spread in \LLX\ compared to singletons at the relatively flat \LLX\ region between early G and early M types.

In \LX\ space (bottom panels), we find a global trend for later spectral types to have lower \LX, which is not surprising as the latter depends on the stellar luminosity and the size of the stellar corona and, thus, on $m$ and radius. We note that mid M dwarfs can have equivalent \LX\ levels as some F and early G dwarfs, even though F types can have stellar radii $\approx$5 times larger than mid M types, and hence, up to 25 times larger surface areas. 
The small sizes of the highly active M dwarfs in our sample is compensated by their brighter X-ray emission reflected by their smaller \Ro\ and largely saturated activity (see Section~\ref{sec:rossbyresults}). 
As we found that coronal temperatures varied only weakly across spectral types (see bottom left panel of Figure~\ref{fig_Xfits}), this brighter emission cannot simply be due to hotter coronae in the M dwarfs.
The brighter X-ray emission from M dwarfs is therefore likely driven by higher number of active regions and/or a significant difference in magnetic topology \citep[e.g.,][]{Lang2012}.
% Evidently, the small sizes of the highly active M dwarfs in our sample is compensated by their brighter X-ray emission, the latter probably driven by higher number of active regions and/or a significant difference in magnetic topology \citep[e.g.,][]{Lang2012}, and {\emph not} by higher coronal temperatures, as we found a very small difference in this parameter between F, G, K, and M types (see bottom left panel of Figure~\ref{fig_Xfits}).

\input{tbl_LLXstats}

\subsubsection{X-Ray Emission of Singles vs. Binaries}\label{sec:lxcolorbins}
\begin{figure}
\centerline{\includegraphics{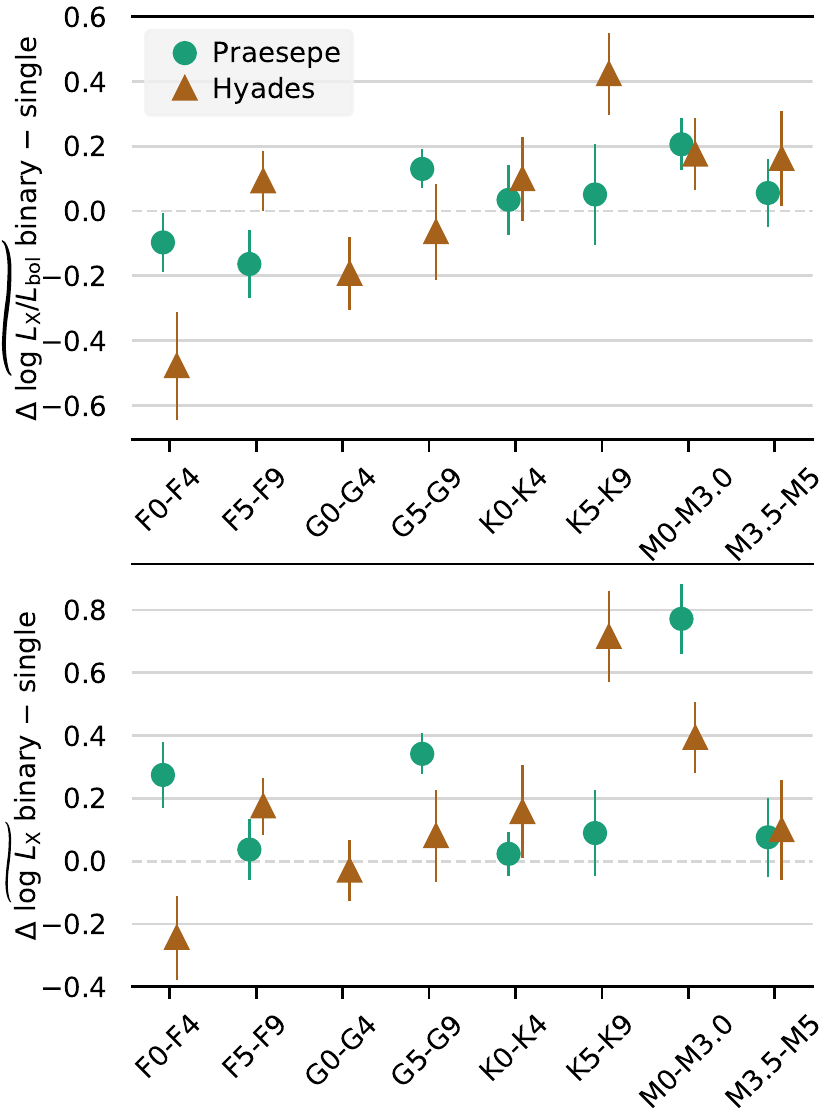}}
\caption{{\it Top}: Mean value, plus 95\% confidence interval, of the median log \LLX\ for binaries minus the median log \LLX\ for single members in Praesepe (green circles) and Hyades (brown triangles) for each spectral type bin of Table~\ref{tbl_llxstats}, using a Monte Carlo simulation. {\it Bottom}: Same as above, but in log \LX\ space. We find only marginal differences in X-ray activity---more evident in \LX\ space and for a couple of spectral type bins---between the single and binary populations.}
\label{fig_deltas}
\end{figure}

In Section~\ref{sec:spectralparams}, we noted that there were only small differences in the X-ray emission characteristics, namely, coronal abundance and temperature, between single and binary stars. To better visualize any systematic difference between singles vs. binaries in \LLX\ and \LX, we performed 1,000 Monte Carlo iterations of $\Delta$ median log \LLX\ = (median log \LLX)$_\mathrm{binaries}~-$ (median log \LLX)$_\mathrm{singles}$ for each spectral type range in Table~\ref{tbl_llxstats}. At each iteration, we added noise to all \LLX\ measurements, drawn from a Gaussian with width equal to the uncertainty of each measurement, and re-measured the median value. We repeated the same process in log \LX\ space. We show in Figure~\ref{fig_deltas} the mean and standard deviation of $\Delta$ from our Monte Carlo results for each cluster and spectral type range.

If we consider the case of unresolved binary systems including two X-ray emitters, their unresolved \LX\ would be higher compared to those for single stars. In the scenario of their total X-ray emission coming from two stars each with activity levels commensurate to those of single stars, their total \LX\ would be twice that of an equivalent single star. On the other hand, in the scenario of the binary components having experienced past and/or present interactions, resulting in inflated levels of activity, their total \LX\ would be even greater than that of the previous scenario. In both scenarios, our measured \Lbol\ would also be inflated for unresolved binaries, as our \Lbol\ are derived from $M_G$ (see Section~\ref{sec:lbol}). In the extreme case of equal mass binary systems, their total \Lbol\ would be inflated anywhere between 20\% and 60\% \citep[assuming that equal mass binaries lie 0.375 in $M_G$ above the single main sequence, see, e.g.,][]{Rampalli2022} compared to their hypothetical individual \Lbol\ values. All in all, for the scenario of binaries without inflated activity, the combined effect would be \LLX\ anywhere between 1.2 and 1.7$\times$ higher for binaries than for singles, or up to $\approx$0.2 orders of magnitude.

The top panel shows that binaries and singles have similar \LLX\ values---the largest differences are smaller than 0.5 orders of magnitude, and most differences are within 0.2 orders of magnitude, both above and below zero. We believe these minor differences can be explained mostly by the small sample sizes and by the inherent noise around measuring the X-ray emission of a main-sequence dwarf, by virtue of both short-term (e.g., flaring) and long-term (e.g., Solar-type activity cycles) variability. Such inherent noise in our result obscures the expected intrinsic over-luminosity of binaries in the hypothetical scenario described above.

The result in \LX\ space is almost identical, with two marginal exceptions. First, K5--K9 binaries in the Hyades are $\approx$0.7 orders of magnitude more luminous than their single counterparts. \citet{Pye1994} had found that the X-ray luminosity function of K binaries in the Hyades was almost one order of magnitude more luminous than that of their single counterparts. Our result, therefore, partly supports their conclusion. Second, M0--M3 binaries in Praesepe are $\approx$0.8 orders of magnitude more luminous than their single counterparts.

In both of these exceptions, the significant over-luminosity of the binary sample cannot be explained only by assuming that we are capturing two or more individual---and unresolved---X-ray emitters in our binary \LX\ measurements. Instead, such over-luminosity suggests that binaries have higher magnetic activity levels than single stars at those spectral type ranges. Their rotational information partly corroborates this: In our sample of K5--K9 Hyads, four out of the nine known binaries with known \prot\ are fast rotators ($\leq$5 d), whereas all seven singles with known \prot\ are slow rotators ($\geq$10 d). It is still puzzling that the highlighted exceptions are for only one specific spectral type range and for stars in only one of the two clusters. In any event, we caution the reader about the small sizes of our spectral type subsamples in Table~\ref{tbl_llxstats}, which may lead to uncertain statistical results.

A potential additional piece of evidence of inflated magnetic activity could be found in the fraction of X-ray variable/flaring stars in our full sample of X-ray detections. The most magnetically active stars are expected to have more common flaring episodes \citep[e.g.,][]{Kowalski2009}, and thus have a higher probability of getting caught flaring or displaying large variability during an X-ray observation. Of the 963 X-ray sources we have with a variability flag, 9\% are flagged as definitely variable: 36 are single members and 47 are binaries. As such, we find no strong preference for binaries to have signs of variability or flaring in X-rays over single members.

\begin{figure*}[t]
\centerline{\includegraphics{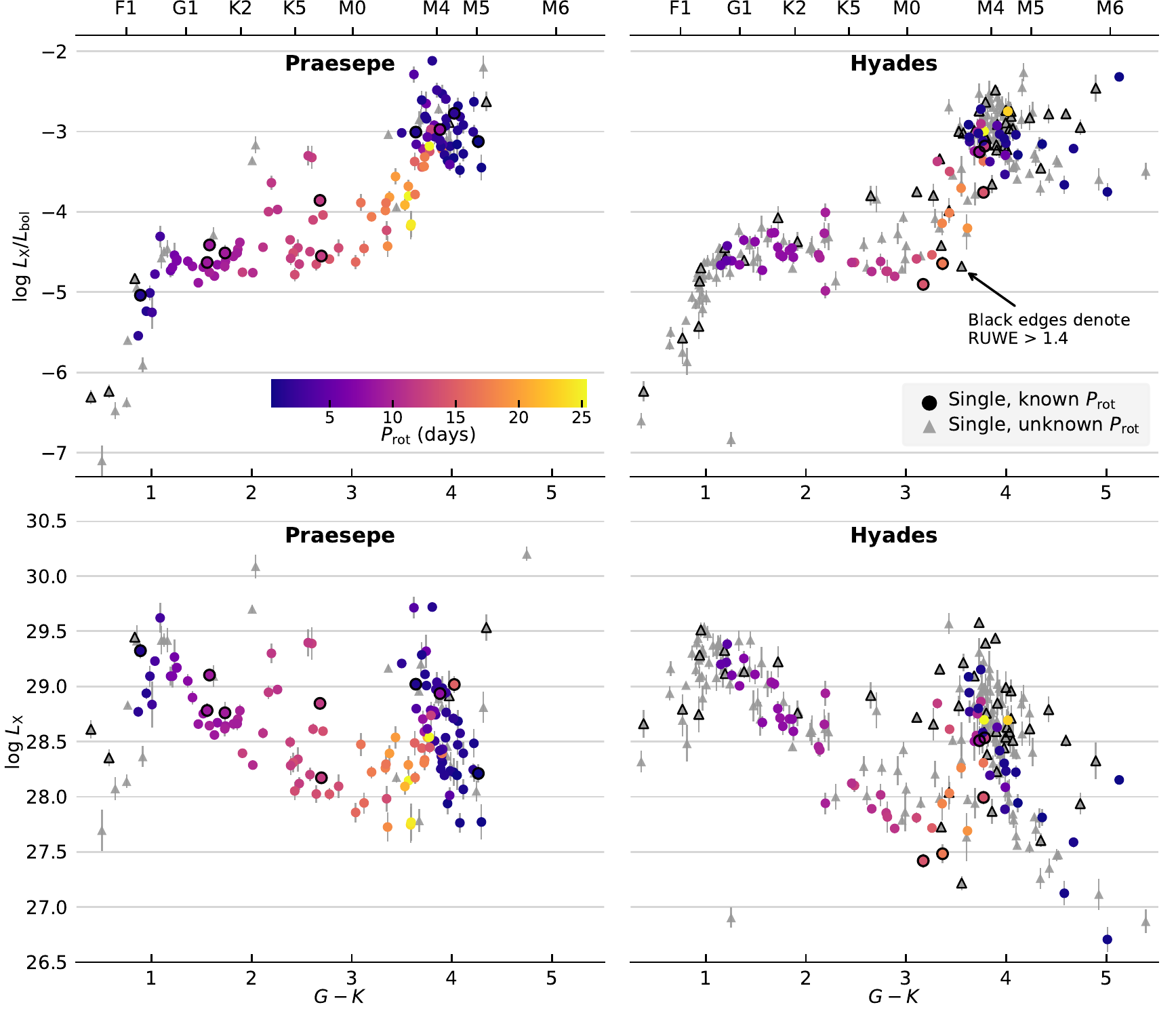}}
\caption{Same as Figure~\ref{fig_LLXb}, but plotting only stars with no binary flags. Circles indicate stars for which we have a \prot, and gray triangles, stars without \prot. The circles are color-coded according to the \prot\ value using the colorbar in the top right panel. Symbols with black edges are single stars with RUWE $>$ 1.4. Stars appear to plateau in \LLX\ between early G and early M spectral types, even as they display increasing \prot\ toward later types.}
\label{fig_LLXp}
\end{figure*}

Similar to Figure~\ref{fig_LLXb}, Figure~\ref{fig_LLXp} shows \LLX\ (upper panels) and \LX\ (lower panels) as a function of \gminusk\ color for stars in our sample, this time only including stars identified as single cluster members. Colored circles are stars with \prot\ measurements, identified by the color map in the upper left panel, and triangles are stars without \prot. Both clusters reveal a trend of increasing \prot\ with increasing \gminusk\ color for stars in the relatively flat \LLX\ region between early G and early M spectral types. For spectral types later than $\approx$M3, roughly coinciding with the transition between partially and fully convective stellar interiors \citep{Chabrier1997}, almost all stars display fast rotation (\prot\ $\ll$ 10~d) and high \LLX. Furthermore, both F and mid-to-late M dwarfs can have equivalent \prot\ values, and yet stars in the latter cohort display \LLX\ values that are two orders of magnitude higher than the F dwarfs in the same cluster.

Interestingly, and analogous to what we find in the color--\prot\ plane (Figure~\ref{fig_prots}), there is no significant offset in \LLX\ or \LX\ between single stars with RUWE $\leq$ 1.4 and $>$ 1.4. The latter---highlighted with black edges in Figure~\ref{fig_LLXp}---could be unresolved intermediate binaries (see Section~\ref{sec:bin}), which means that their X-ray detections could be a combination of X-ray emission from two (or more) separate sources. In the extreme case of two equally X-ray bright emitters, their \LLX\ and \LX\ values would appear over-luminous compared to their single counterparts by 0.3 in log space, everything else remaining equal. Such a relatively small difference is largely obscured by the intrinsic noise in our sample data. However, the fact that there is no widespread large X-ray over-luminosity in stars with RUWE $>$ 1.4 suggests that levels of magnetic activity in these potential binaries are not inflated compared to those in single stars.
%we find no clear evidence to conclude that stars with RUWE $>$ 1.4 are potentially systems with two or more X-ray emitting main-sequence dwarfs. More probably, the X-ray emission from these high-RUWE stars comes either predominantly or entirely from one of the members of the system.

\subsubsection{X-Ray Outlier in the Hyades}
The F8 Hyad HD 50554 (2MASS J06544283+2414441) appears under-luminous in X-rays by at least two orders of magnitude compared to its peers (log \LLX $=-6.83$, log [\LX/erg s$^{-1}$] $=26.9$; see right panels of Figure~\ref{fig_LLXb}). \citet{Sanz-Forcada2010} estimated an age of 12.2 Gyr for this star, and \citet{Kains2011}, 4.7 Gyr, based on a surrounding debris disk detected in the far infrared. Our faint \LX\ value for this star appears to corroborate an age much older than that of the Hyades. We included this star in our Hyades catalog because it is considered a trailing tail star in the \citet{Roser2019a} GDR2 Hyades catalog. However, these authors noted that almost 14\% of trailing tail stars in their catalog are statistically expected to be contaminants. Given the estimated ages in the literature and its significant under-luminosity in X-rays, we consider this star a likely contaminant in our Hyades membership catalog, and therefore exclude it from our calculations in Table~\ref{tbl_llxstats}.

\subsection{The Relationship Between X-Ray Emission and Rotation} \label{sec:rossbyresults}

A more direct way to analyze the link between coronal activity and rotation in low-mass stars is by characterizing the relationship between \LLX\ and \Ro\ (see section~\ref{rossby}). 
%Similar to other studies, we find in our sample of Praesepe and Hyades low-mass stars the following trends: a saturated regime, in which \LLX\ $\approx$10$^{-3}$ and independent from stellar rotation; and an unsaturated regime, in which \LLX\ depends on \Ro\ following a power law $\approx$ $-2$. % We also find that some single stars in Praesepe and Hyades with \Ro\ $\lapprox$ 0.01 appear to show decreased \LLX. 
We break down our sample of cluster stars into single and binary stars, both for each cluster and combined, to characterize the \Ro--\LLX\ relation. 

To obtain a very clean sample of single stars, we exclude those with RUWE $>$ 1.4, the latter considered to be likely unresolved intermediate binaries (see Section~\ref{sec:bin}). Instead, we include these high-RUWE stars in our sample of cluster binaries. As we noted in Section~\ref{sec:lxcolorbins}, the X-ray emission from high-RUWE stars may be dominated, partially or entirely, by only one of the members of the potential binary system. As we cannot resolve the system in X-rays or optically, we therefore consider these stars as potential contaminants in our analysis of the rotation-activity relation in single stars.

We also exclude cluster stars with $m \geq 1.3$ \Msun\ ($\approx$F5 type and earlier) to probe the \Ro--\LLX\ relation. Although main-sequence stars with $m$ up to $\approx$1.6 \Msun ($\approx$F0 type) can have an outer convective zone \citep{Boehm-Vitense1980}, their magnetic activity ceases to be correlated with rotation. It is, thus, not dominated by a solar-like dynamo, but by fossil magnetic fields instead \citep[e.g.,][]{Boehm-Vitense2002, Kochukhov2003, Walter1986, Wolff1987}.

We parameterize the \Ro--\LLX\ relationship as a flat region connected to a power law. For stars with \Ro\ $\leq \Ro_\mathrm{,sat}$, activity is constant---i.e., saturated---and equal to (\LLX)$_\mathrm{sat}$. Above $\Ro_\mathrm{,sat}$, activity declines as a power law with index $\beta$, and is, therefore, unsaturated. Functionally, this corresponds to
\begin{equation}\label{eq:rossby}
  \frac{L_{\mathrm{X}}}{L_{\mathrm{bol}}} = \left\{
  \begin{array}{l l}
    \left(\frac{L_{\mathrm{X}}}{L_{\mathrm{bol}}}\right)_{\mathrm{sat}} & \quad \textrm{if $R_\mathrm{o}\le R_{\mathrm{o,sat}}$}\\
    C R_\mathrm{o}^{\beta} & \quad \textrm{if $R_\mathrm{o}$ > $R_{\mathrm{o,sat}}$}
  \end{array} \right.
\end{equation}
where $C$ is a constant. This model has been widely used in the literature (e.g., \citeauthor{Randich2000b} \citeyear{Randich2000b}, \citeauthor{Wright2011} \citeyear{Wright2011}, \citeauthor{Douglas2014} \citeyear{Douglas2014}, \citeauthor{Nunez2015} \citeyear{Nunez2015}).

\begin{figure*}[t]
\centerline{\includegraphics{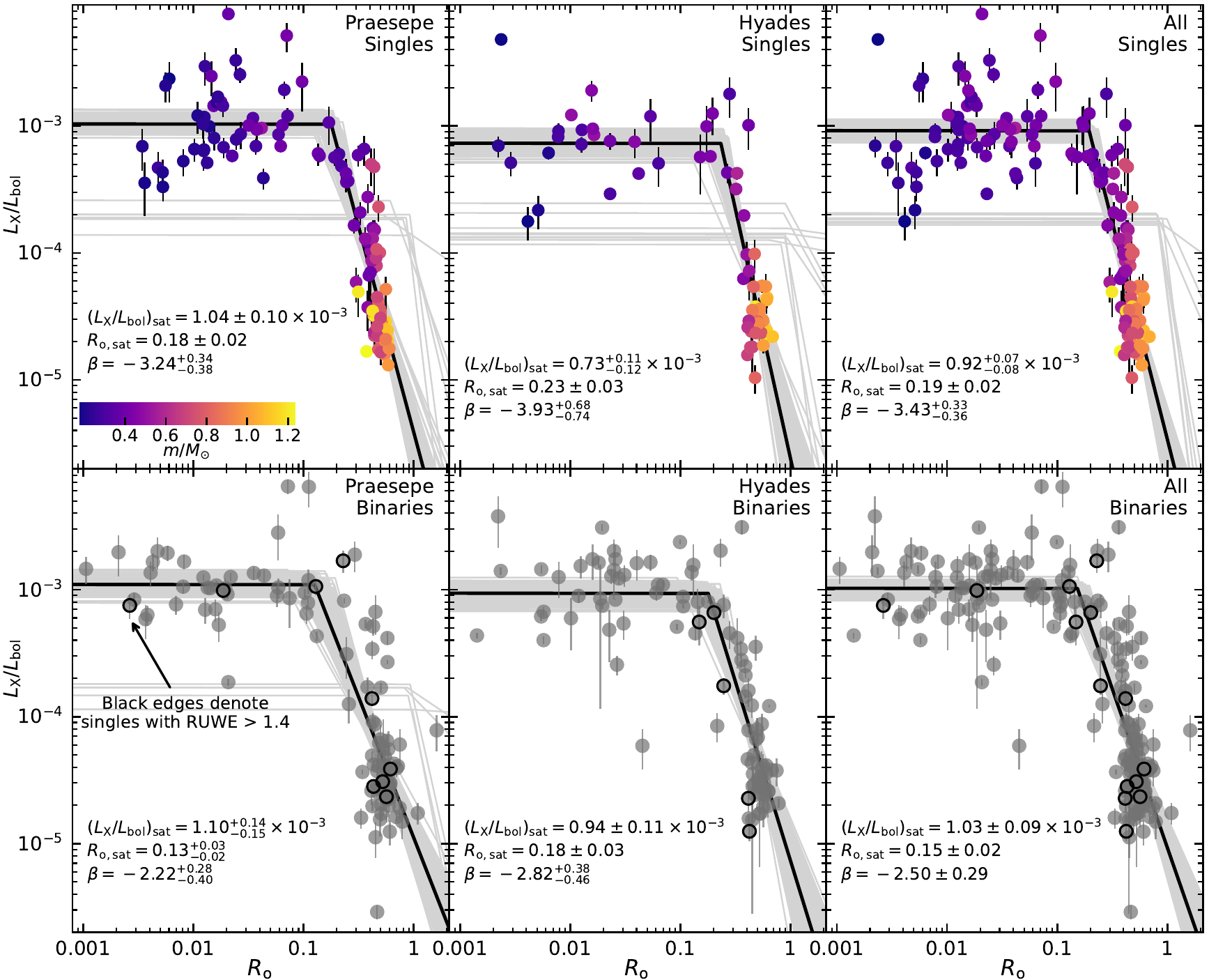}}
\caption{\LLX\ vs.~\Ro\ for stars in Praesepe (left column), Hyades (middle), and both clusters (right). The top row shows single stars, and the bottom row shows candidate/confirmed binaries. This latter set includes stars with no binary flags but with RUWE $>$ 1.4 (indicated with solid black circles). Single stars are color-coded by their $m$ following the colorbar in the top left panel. The solid black line in each panel is the maximum a posteriori fit from the MCMC algorithm, and the gray lines are 200 random samples from the posterior probability distributions. The fit result of the three parameters, (\LLX)$_\mathrm{sat}$, \Ro$_\mathrm{,sat}$, and $\beta$, are annotated in each panel. We show in the Appendix the marginalized posterior probability distributions from the MCMC analysis for each fit. The power-law index $\beta$ in the unsaturated regime for single stars is significantly steeper than that for binary stars, the latter being closer to the canonical $\beta \approx -2$ from other studies in the literature.}
\label{fig_rossbyfits}
\end{figure*}

We use the open-source Markov-chain Monte Carlo (MCMC) package \texttt{emcee} \citep{Foreman2013} to fit this three-parameter model to our data. Like in the \texttt{emcee} implementation by \citet{Magaudda2020}, we allow for a nuisance parameter $f$ to account for underestimated errors.\footnote{See \url{https://emcee.readthedocs.io/en/develop/user/line/}.} We assume flat priors over each parameter and use 300 walkers, each taking 5,000 steps in their MCMC chain, to infer maximum likelihood parameters. Our results are presented in Figure~\ref{fig_rossbyfits}. The posterior distributions for each parameter and 2D correlations between pairs of parameters from each fit
% for single and binary cluster members 
are included in a Figure Set in Appendix~\ref{app:main}; 200 random samples from these distributions are shown in Figure \ref{fig_rossbyfits}, along with the maximum a posteriori model. In all cases, the nuisance factor $f$ converges to $\approx$0.1, which suggests that our \LLX\ uncertainties are underestimated by no more than $\approx$10\%.

The (\LLX)$_\mathrm{sat}$, \Ro$_\mathrm{,sat}$, and $\beta$ parameters corresponding to the maximum a posteriori model are presented in Table~\ref{tbl_rossbies} for the six subsamples we show in Figure~\ref{fig_rossbyfits}, and are also annotated in each panel in the Figure. The stated values correspond to the 50$^{\rm th}$ percentiles of the results and the uncertainties are the 16$^{\rm th}$ and 84$^{\rm th}$ percentiles. We selected these percentiles to be consistent with 1$\sigma$ Gaussian uncertainties, even though our one-dimensional 1D posterior probability distributions are not Gaussian.
\input{tbl_Rossbies}

\subsubsection{X-Ray Emission and Rotation in Singles vs. Binaries}

\begin{figure*}[t]
\centerline{\includegraphics{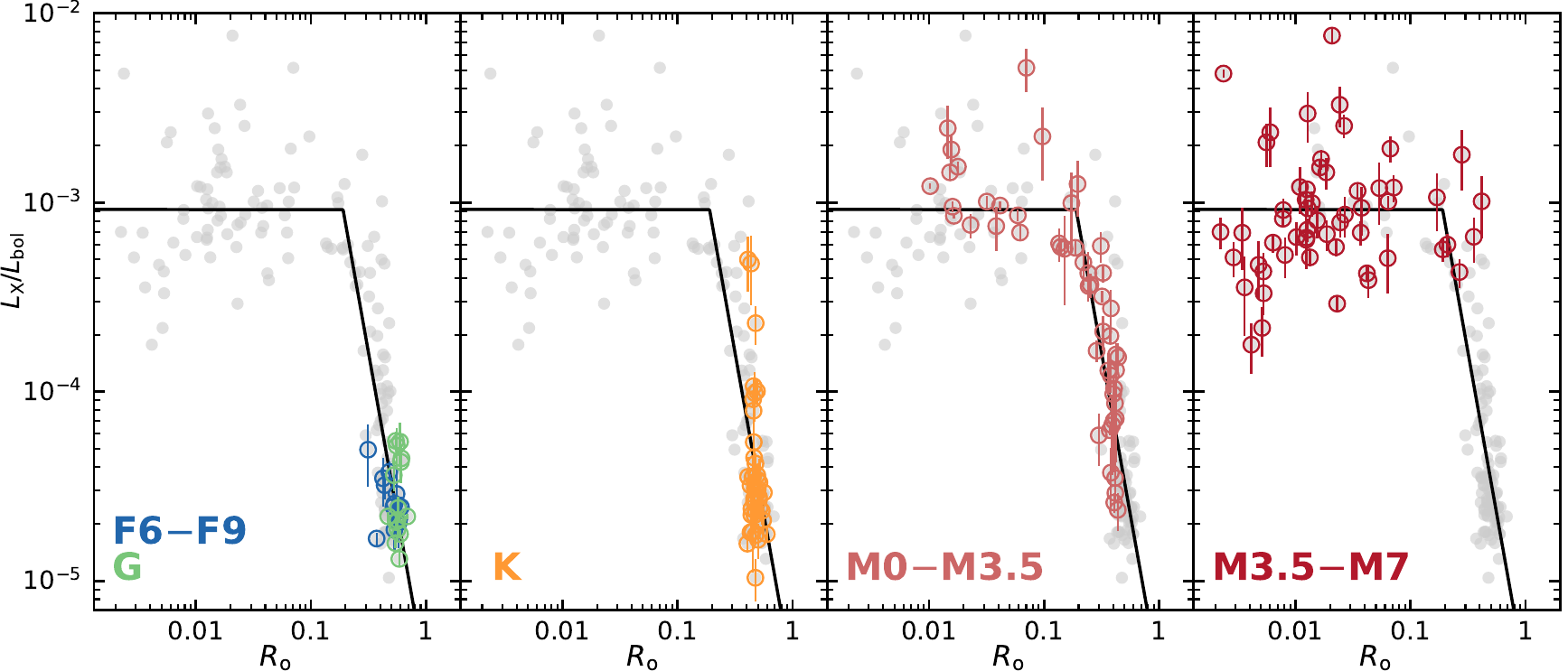}}
\caption{\LLX\ vs.~\Ro\ for single stars in Praesepe and Hyades, highlighting the different spectral types in each panel. The background gray symbols in all panels are the combined sample of single FGKM dwarfs in both Praesepe and Hyades, but excluding those with RUWE $>$ 1.4. The solid black line in all panels is the maximum a posteriori fit from the MCMC algorithm for the full sample of single Praesepe and Hyades stars (see top right panel of Figure~\ref{fig_rossbyfits}). At the age of Praesepe and Hyades, fully convective dwarfs (M3.5 and later) remain saturated in X-rays, whereas most early-to-mid M dwarfs and all late-F, G, and K dwarfs have spun down into the unsaturated regime.}
\label{fig_rossbyhighlights}
\end{figure*}

Stars in close binary systems may have their spin down history affected by the companion, leading to circularized orbits and synchronized \prot, the latter resulting in either faster or slower rotation than their single stellar counterparts at the same age \citep{Counselman1973, Meibom2005, Zahn2008, Fleming2019}. 
This could lead to higher or lower X-ray activity in binaries, compared to coeval single stars. For example, and as we noted in Section~\ref{sec:lxcolorbins}, \citet{Pye1994} found that K dwarf binaries in the Hyades are more than twice as luminous as their single counterparts.  
Also, if the binaries are close enough to each other, it is also possible that their magnetic fields may directly interact.

Based on our fits in Section~\ref{sec:rossbyresults}, the relationships between coronal activity and \Ro\ are mostly similar for both our single and binary subsamples. (\LLX)$_\mathrm{sat} \approx 10^{-3}$ in all subsamples, with singles vs. binaries agreeing within 1$\sigma$ in their saturation level. The \Ro$_\mathrm{,sat}$ thresholds fall within the approximate narrow range 0.13--0.23, also agreeing within 1$\sigma$. However, $\beta$ is steeper for single stars than for binaries: $\beta$ falls between $-3.24$ and $-3.93$ for single stars, and between $-2.22$ and $-2.82$ for binaries.

As we discussed in Section~\ref{sec:lxcolorbins}, if we assumed that both components in unresolved binaries are X-ray emitters, their combined X-ray emission and $M_G$ would translate into higher (\LLX)$_\mathrm{sat}$ for binaries than for singles. For unresolved binaries, their \Ro\ values are smaller than for equivalent single stars by anywhere between $\approx$10\% and 50\%, as our $\tau$ are derived from stellar masses, which in turn are derived from $M_G$ (see Section~\ref{sec:lbol}). The combined, almost uniform upward shift in \LLX\ and rightward shift in \Ro\ for binaries in the \Ro--\LLX\ plane would translate into a negligible change in $\beta$ compared to single stars. This shift would only affect \Ro$_\mathrm{,sat}$, increasing it by up to 50\%. 

In Table~\ref{tbl_rossbies} (and Figure~\ref{fig_rossbyfits}) we see that the binary subsamples have higher (\LLX)$_\mathrm{sat}$ than the single counterparts by no more than 1.3$\times$, and their values agree within 1$\sigma$ with one another. The large spread in activity in the saturated regime in all subsamples may be partially hiding what would otherwise be a more statistically significant over-luminosity of binaries over singles in this regime.

On the other hand, in the subsamples of Praesepe stars and of combined (i.e., ``All'') cluster stars, $\beta$ values are shallower for binaries than for singles, disagreeing by at least 1$\sigma$, and \Ro$_\mathrm{,sat}$ values are statistically equivalent for both binaries and singles. Only in the Hyades, the singles and binaries subsamples have $\beta$ within 1$\sigma$: $3.93^{+0.68}_{-0.74}$ vs. $-2.82^{+0.38}_{-0.46}$, respectively. \citet{Freund2020} found $\beta = -2.12\pm0.60$ in their Hyades study of \ffx\ vs. \Ro, which is statistically closer to our $\beta$ for binary Hyads than for single Hyads. We note that these authors did not exclude X-ray flares from their calculations of X-ray emission and did not systematically apply binary flags to their sample of stars. 

Both the persistent shallower $\beta$ values in our binaries subsamples and the indistinguishable \Ro$_\mathrm{,sat}$ values in both singles and binaries may be driven by the larger scatter---both in \LLX\ and \Ro---in their unsaturated regime compared to our subsamples of single stars, which leads to a poorer power-law fit. In our unresolved binaries, there is no way to appropriately assign \LX\ and \Lbol\ values to the individual components of the binary systems. Even if we assume a first order approach and split \LX\ evenly between the binary components, we are still left with the complication of assigning \Ro\ to only one of them, as the adopted \prot\ only captures one rotating signal. In short, the ambiguity in the X-ray emission level of stars flagged as binaries or with RUWE $>$ 1.4 makes the sample presented in the bottom row of Figure~\ref{fig_rossbyfits} inappropriate to study the \Ro--\LLX\ relation in stars with secular angular momentum evolution.

\subsubsection{Saturation and Supersaturation}\label{sec:sat_sup}

As we noted earlier, the level of saturation in \LLX\ lies at $\approx$10$^{-3}$ for all the subsamples we examined. There is a large spread, a little over one order of magnitude, in \LLX\ in the saturated regime, particularly in our sample of Praesepe stars. This finding is similar to the large sample in \citet{Wright2011}, in which a mix of cluster and field stars have a saturated \LLX\ value of $\approx$10$^{-3}$ and a spread of about 1.5 orders of magnitude.

We inspected the X-ray nature of stars that are either too under- or over-luminous in the saturated regime, to determine whether systematic issues were contributing to the spread in \LLX. We find no correlation between \LLX\ and X-ray observatory or source counts. That is, the under/over-luminosity is not a function of X-ray instrument or exposure times, which leads us to believe that the noise in \LLX\ in the saturated regime is intrinsic to the stars, due perhaps to unaccounted X-ray flaring or to activity cycles similar to that observed in the Sun.

To test for supersaturation, a phenomenon usually observed in stars with \Ro\ $\lapprox$ 0.02, we add a secondary power law of index $\beta_\mathrm{sup}$ connected to the flat region in Equation~\ref{eq:rossby} for \Ro\ $<$ \Ro$_\mathrm{,sup}$, where \Ro$_\mathrm{,sup}$ is the threshold \Ro\ value between supersaturation and saturation.
The results and figures of our MCMC routine on the modified five-parameter (the three parameters in Equation~\ref{eq:rossby} plus $\beta_\mathrm{sup}$ and \Ro$_\mathrm{,sup}$) model are included in Appendix~\ref{app:supersat}.

For our subsamples of single cluster stars, $\beta_\mathrm{sup}$ falls anywhere between 0.23 and 0.76 and \Ro$_\mathrm{,sup} \approx 0.01$. Such values of $\beta_\mathrm{sup}$ could indicate that stars with \Ro\ $<$ \Ro$_\mathrm{,sup}$ are supersaturated and, thus, experiencing a reduced amount of X-ray activity by virtue of their extremely fast rotation. However, we recognize that both $\beta_\mathrm{sup}$ and \Ro$_\mathrm{,sup}$ are not nearly as constrained as \Ro$_\mathrm{,sat}$ and $\beta$ in the unsaturated regime---both the random samples from the posterior probability distributions and the marginalized posterior probability distributions can attest to this. We therefore recognize these results as only tentative evidence for supersaturation.

Lastly, for our subsamples of binary cluster stars, $\beta_\mathrm{sup}$ is statistically indifferent from zero, and $\Ro_\mathrm{,sup}$ is virtually unconstrained. If it is true that binaries show no signs of supersaturation while single stars do, then this could be interpreted as evidence for inflated levels of coronal activity in binaries with extremely fast rotation.

% break up period: omega_br = sqrt(GM/ rR^3)
%Lastly, we find no convincing evidence for supersaturation, a phenomenon usually observed in stars with \Ro\ $\lapprox$ 0.02. We added a secondary power-law connected to the flat region in Equation~\ref{eq:rossby} for \Ro\ $<$ \Ro$_\mathrm{,supersat}$, where \Ro$_\mathrm{,supersat}$ is the threshold \Ro\ value between supersaturation and saturation. The results of our MCMC routine on this modified model return a slope in the secondary power-law statistically indifferent from zero, thus indicating no evidence for either increase or decrease in \LLX\ at \Ro\ $\lapprox$ 0.02. There is only a marginal indication (by visual inspection) of supersaturation in the subsamples of single stars in Praesepe and Hyades (top panels of Figure~\ref{fig_rossbyfits}). In these subsamples, some stars with \Ro\ $\lapprox$ 0.01 appear to have a level of \LLX\ up to one order of magnitude below (\LLX)$_\mathrm{sat}$. However, the number of stars in this \Ro\ regime is small and the data points too scattered in \LLX\ to make any conclusive remark.
%There is marginal visual evidence for supersaturation among single M3.5-M7 (0.27--0.09 \Msun) stars in Praesepe and the Hyades (top panels of Figure~\ref{fig_rossbyfits}). In these subsamples, some stars with \Ro\ $\lapprox$ 0.01 appear to have a level of \LLX\ up to one order of magnitude below (\LLX)$_\mathrm{sat}$. 

\subsubsection{Spectral Type vs. Rotation and Activity}

Figure~\ref{fig_rossbyhighlights} shows our sample of single Praesepe and Hyades stars combined (gray circles), highlighting with colored symbols in each panel different spectral types. As can be seen in the right-most panel, \emph{all} stars in the hypothetical supersaturated regime (\Ro\ $\lapprox$ 0.01) have spectral types later than M3.5, i.e., the fully convective stars. %For an M3.5 dwarf, \Ro\ = 0.01 corresponds to \prot\ $\approx$ 21 h. 
The smallest \Ro\ in our sample of single stars is 0.002 (\prot\ $\approx$ 7h), for the M5 and M7.5 dwarfs 2MASS J05003894+2422581 and J04351354+2008014, respectively. The latter, in particular, appears significantly brighter than its super fast rotating peers (\LLX\ = 4.8$\times10^{-3}$). This, in spite of the fact that we removed an X-ray flare from its detection to calculate its quiescent \LLX\ (see Section~\ref{sec:flares}). It is possible, therefore, that we observed this ultracool dwarf during an episode of large variability and flaring activity, which would explain its over-luminosity compared to the other M dwarfs in the supersaturated regime. % These results support previous findings suggesting that supersaturation in M dwarfs probably starts at \Ro\ values much lower than the conventional supersaturation threshold of \Ro\ $\approx$ 0.02.

We noted in Figure~\ref{fig_prots} the narrow slow-rotating sequence for single stars spanning the mid F to early M spectral types illustrating the one-to-one relationship between spectral type and rotation in Praesepe and the Hyades. Now, Figure~\ref{fig_rossbyhighlights} highlights the large spread in X-ray activity, as measured by \LLX, in these same stars. All late-F, G, and K dwarfs have practically the same \Ro, and yet they cover roughly one order of magnitude in \LLX. Early M dwarfs (M0--M3.5) cover one order of magnitude in \Ro\ space and two orders of magnitude in \LLX. Lastly, mid- and late M dwarfs (M3.5--M7), the fully convective cohort, span two orders of magnitude in \Ro\ space (with \prot\ values from $\approx$5 hr up to $\approx$30 d), while their spread in \LLX\ spans \emph{only} one and a half orders of magnitude (right-most panel in Figure~\ref{fig_rossbyhighlights}). 

As our sample of Praesepe and Hyades stars is approximately coeval, we can safely assume that within the same spectral type range, the cause of the spread in \LLX\ is not a function of age. In the current paradigm of angular momentum driving the level of magnetic activity in low-mass stars, our findings illustrate the strong dependence of \LLX\ on \Ro\ in unsaturated, partly convective dwarfs. On the other hand, in the cohort of fully convective dwarfs, a very large range in \Ro\ translates into only a modest range in \LLX. It would be easy to hold their fully convective nature responsible for this behavior. However, we may just be observing an age at which only the fully convective---and a few partially convective---stars remain saturated and, thus, restricted in their activity levels by whatever mechanisms drive the saturated regime.

\section{Conclusion}\label{sec:concl}

The Hyades and Praesepe form a crucial bridge between very young open clusters and field-age stars in the study of the evolution of rotation and activity in low-mass stars. 
We re-visit these two clusters to expand our understanding of the connection between rotation and coronal activity at their critical stellar age.

We start by combining several membership catalogs, including our own legacy catalogs and GDR2-based catalogs, and consolidating literature binary information on these stars. We produce a Praesepe catalog with 1739 members and a Hyades catalog with 1315 members. For these members, we collect GEDR3 $G$ and 2MASS $K$ and photometry and astrometry to derive stellar parameters, including stellar mass and \Lbol. Next, we consolidate \prot\ measurements from the literature for stars in both clusters.

% For a comprehensive probe on the coronal activity of stars in both clusters, 
To produce a comprehensive catalog of coronal activity, we search for X-ray detections of cluster members with the ROSAT, Chandra, Swift, and XMM observatories, both archival and new observations. 
We rely on the archival serendipitous catalogs of each observatory to obtain the X-ray flux of each detection. For ten XMM pointings, we perform our own data reduction to add 34 very faint X-ray sources that were not included in the XMM serendipitous catalog.
Additionally, we reduce seven new observations from Chandra and XMM that were taken after the cutoff dates of the serendipitous catalogs.

We derive \fx\ values for each X-ray detection from the instrumental count rates using our own ECFs. 
For 139 X-ray sources with sufficient counts, we perform spectral analysis instead to derive \fx\ along with coronal temperature and metal abundance. 
For the 32 sources in which we identify flares in their X-ray light curves, we first remove counts from the flare events before calculating their \fx. 
We find no difference in the coronal parameters of binary and nonbinary cluster members.
We also find a very weak correlation between spectral type and coronal temperature, and a slightly stronger positive correlation between \LLX\ and coronal temperature. 
More importantly, we find that coronal abundance displays a strong correlation with stellar spectral type: G and early-K dwarfs have typical abundance values of $\approx$0.4 Solar, while mid-K to mid-M dwarfs have values of 0.2 Solar or lower. At the age of Praesepe and Hyades, these two sets of stars coincide with the difference between slow and fast rotators, or high and low \Ro\ numbers, thus suggesting that coronal abundance is directly linked to \prot\ and, thus, to magnetic activity level.
% Stephanie's note to self: stopped reading here

We find that in both clusters \LLX\ increases from F through early G types, then flattens between early G and early M types, and then increases again beyond early M types. Binary stars show a larger spread than the single counterparts in this \LLX\ trend. In particular, late K dwarf binary Hyads are almost one order of magnitude more luminous in \LX\ than their single counterparts, partially supporting previous findings in the literature. Single stars with RUWE~$>$~1.4---a threshold we adopt to indicate unresolved binaries---have the similar \LLX\ levels as stars with RUWE~$\leq$~1.4, suggesting that X-ray emission in the former is not significantly inflated by binary interactions compared to single stars.

In the \Ro--\LLX\ plane of both clusters, we find a saturated regime, in which \LLX\ $\approx$10$^{-3}$ and independent from stellar rotation for stars with \Ro\ $\lapprox$ 0.1 or 0.2, and an unsaturated regime, in which \LLX\ depends on \Ro\ following a power law with slope $\beta$ between $-3.93$ and $-3.24$ for single stars, and $-2.82$ and $-2.22$ for binaries. As \LX, \Lbol, and \Ro\ measurements of unresolved binary systems are ambiguous and problematic, we reiterate that the \Ro--\LLX\ relation in stars with secular angular momentum evolution must be studied in samples that exclude unresolved binaries.

We find tentative evidence for supersaturation for single stars in both clusters with 0.002 $<$ \Ro\ $\lapprox$ 0.01. In this \Ro\ regime, stars follow a power-law with slope between 0.24 and 0.76. However, both the slope and the \Ro\ threshold between supersaturation and saturation are not as well constrained as the parameters in the saturated and unsaturated regimes.
%Although some single stars in Praesepe and Hyades with \Ro\ $\lapprox$ 0.01 have decreased \LLX\ compared to the rest of the saturated cohort, we find no statistical evidence of supersaturation in our samples of fast Praesepe and Hyades rotators, even though we probe X-ray activity down to \Ro\ = 0.002. 

Lastly, in our large coeval sample of single known rotators with unsaturated X-ray emission, a change of 0.5 orders of magnitude in \Ro\ translates into a change in up to two orders of magnitude in \LLX. This steep relationship highlights the sensitivity of coronal activity to the rotation of a star.
On the other hand, fully convective stars have small \Ro\ values spanning two orders of magnitude, and \LLX\ values, roughly one order of magnitude. The latter phenomenon may be driven by the fact that fully convective stars at the age of Praesepe and Hyades are fully within the saturated regime and, thus, restricted in their activity levels by whatever mechanisms drive saturation.

%\begin{acknowledgments}
\bigskip

A.N. acknowledges support provided by the NSF through grant 2138089. M.A.A.~acknowledges support provided by NASA through grant NNX17AD71G. J.J.D.~was supported by NASA contract NAS8-39073 to the Chandra X-ray Center and thanks the Director, P.~Slane, for continuing advice and support.

We estimated the spectral types used in our analysis and several figures using a \gminusk-spectral type relation based on the Stellar Color and Effective Temperature Sequence of E.~Mamajek, available at \url{http://www.pas.rochester.edu/~emamajek/EEM_dwarf_UBVIJHK_colors_Teff.txt} (v. 2021.03.02).

This work has made use of data from the European Space Agency (ESA) mission Gaia (\url{https://www.cosmos.esa.int/gaia}), processed by the Gaia
Data Processing and Analysis Consortium (DPAC, \url{https://www.cosmos.esa.int/web/gaia/dpac/consortium}). Funding for the DPAC has been provided by national institutions, in particular the institutions participating in the Gaia Multilateral Agreement.

This research has made use of the NASA/IPAC Infrared Science Archive, which is operated by the Jet Propulsion Laboratory, California Institute of Technology, under contract with the National Aeronautics and Space Administration. The Two Micron All-Sky Survey was a joint project of the University of Massachusetts and IPAC.

Funding for the Sloan Digital Sky Survey IV has been provided by the Alfred P. Sloan Foundation, the U.S. Department of Energy Office of Science, and the Participating Institutions. SDSS acknowledges support and resources from the Center for High-Performance Computing at the University of Utah. The SDSS website is www.sdss.org.

This research has made use of data obtained from the 4XMM XMM-Newton Serendipitous Source Catalog compiled by the 10 institutes of the XMM-Newton Survey Science Centre selected by ESA, and of data obtained from the Chandra Source Catalog, provided by the Chandra X-ray Center as part of the Chandra Data Archive.

This work made use of data supplied by the UK Swift Science Data Centre at the University of Leicester.

This research has made use of NASA's Astrophysics Data System Bibliographic Services, the SIMBAD database, operated at CDS, Strasbourg, France, the VizieR database of astronomical catalogs \citep{Ochsenbein2000}, and the ``Aladin Sky Atlas'', developed at CDS, Strasbourg Observatory, France.

%\end{acknowledgments}

\setlength{\baselineskip}{0.6\baselineskip}
\bibliography{main}
\bibliographystyle{aasjournal}
\setlength{\baselineskip}{1.667\baselineskip}

\appendix

\section{Marginalized Posterior Probability Distributions for the MCMC Analysis of Our \Ro--\LLX\ Model}\label{app:main}

\begin{figure}
\centerline{\includegraphics{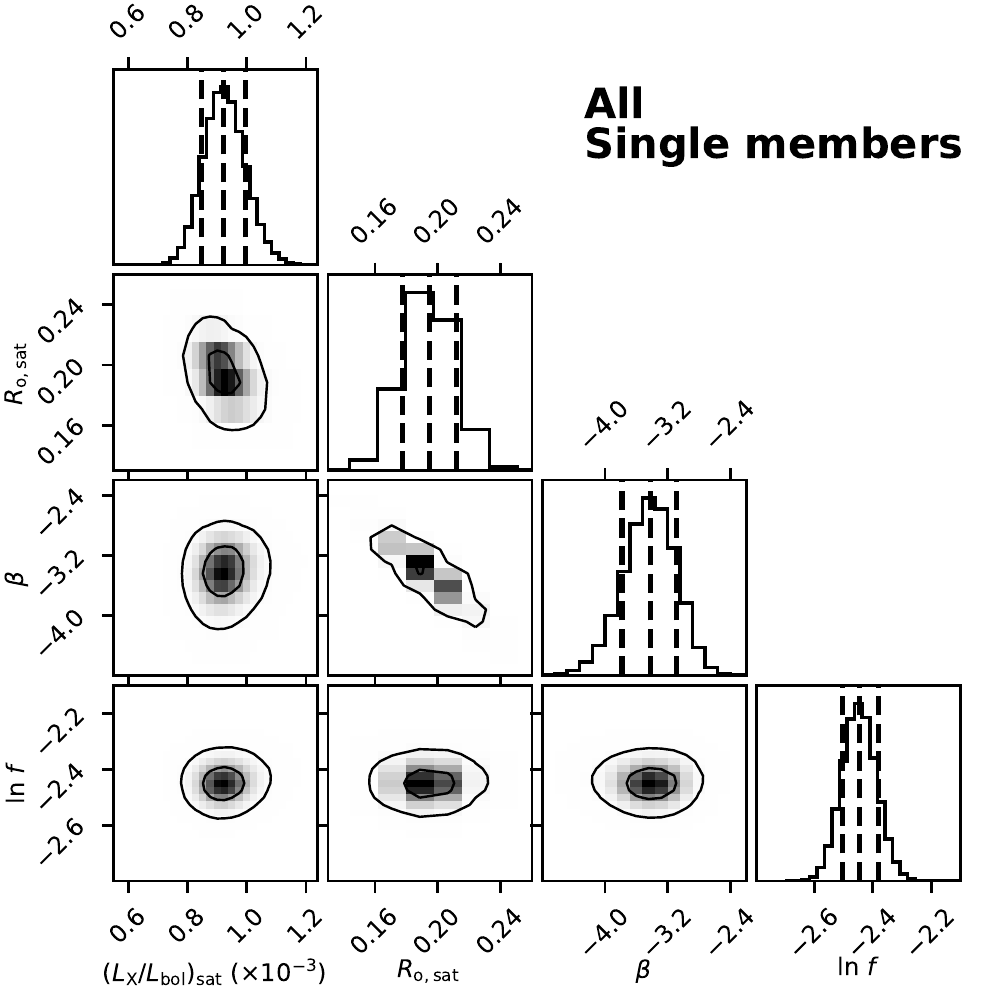}}
\caption{Marginalized posterior probability distributions from the MCMC analysis of our \Ro--\LLX\ model using \texttt{emcee} for single members in both Praesepe and Hyades. The parameter values of the a posteriori model are the peaks of the one-dimensional distributions; the vertical dashed lines approximate the median and 16$^{th}$, 50$^{th}$, and 84$^{th}$ percentiles. The two-dimensional distributions illustrate covariances between parameters; the contour lines approximate the 1$\sigma$ and 2$\sigma$ levels of the distributions. The complete figure set, which includes an image for each of the six subsamples in Table~\ref{tbl_rossbies}, is available in the online journal.}
\label{fig_posteriors}
\end{figure}

We present the marginalized posterior probability distributions from the MCMC analysis we performed on six different subsamples of Praesepe and Hyades stars: single members of each cluster, binary members for each cluster, single members of both clusters combined, and binary members of both clusters combined (see Section~\ref{sec:rossbyresults} and Table~\ref{tbl_rossbies}). The binary samples include candidate and confirmed binaries, including single stars with RUWE $>$ 1.4. Figure~\ref{fig_posteriors} shows an example of the marginalized posterior probability distributions for the subsample of single members from both clusters.

\figsetstart
\figsetnum{14}
\figsettitle{Marginalized posterior probability distributions from the MCMC analysis of Our \Ro--\LLX\ Model}

\figsetgrpstart
\figsetgrpnum{14.1}
\figsetgrptitle{Single members of Praesepe}
\figsetplot{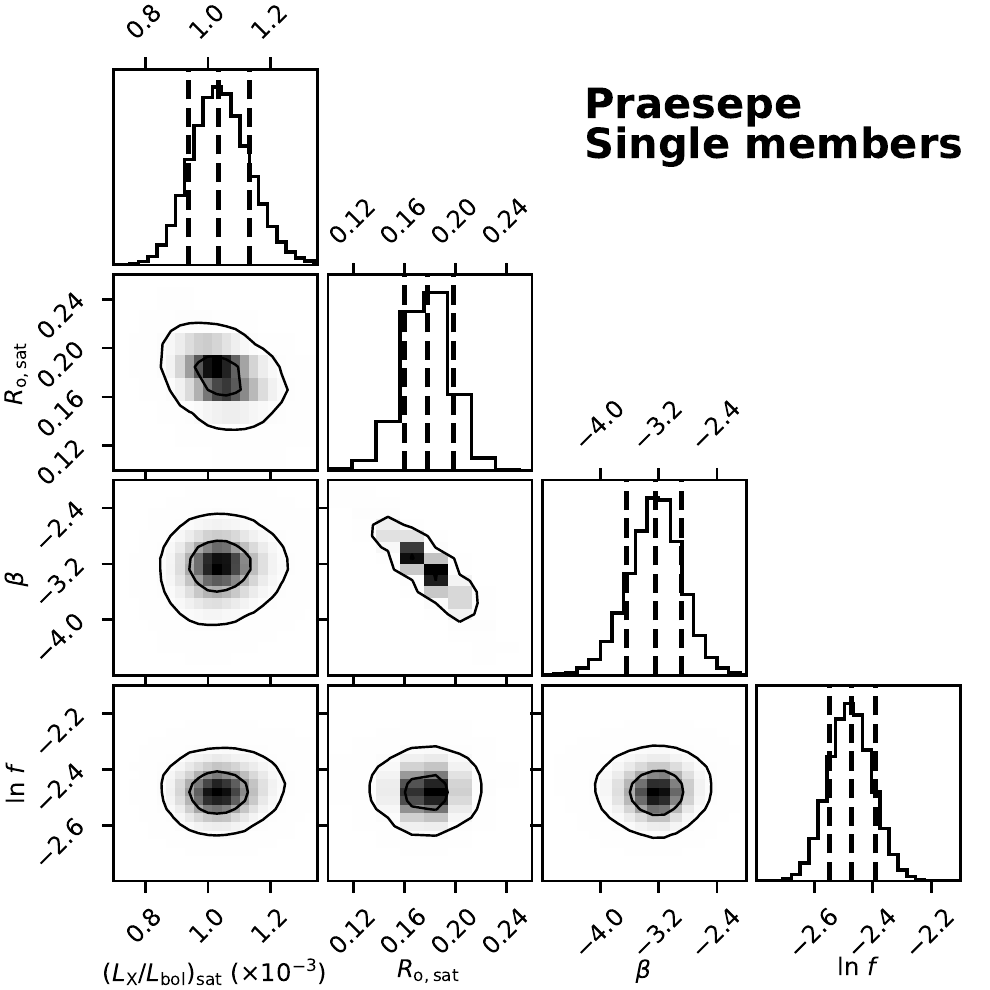}
\figsetgrpnote{Marginalized posterior probability distributions from the MCMC analysis of our \Ro--\LLX\ model using \texttt{emcee} for single members of Praesepe. The parameter values of the a posteriori model are the peaks of the one-dimensional distributions; the vertical dashed lines approximate the median and 16$^{th}$, 50$^{th}$, and 84$^{th}$ percentiles. The two-dimensional distributions illustrate covariances between parameters; the contour lines approximate the 1$\sigma$ and 2$\sigma$ levels of the distributions.}
\figsetgrpend

\figsetgrpstart
\figsetgrpnum{14.2}
\figsetgrptitle{Single members of Hyades}
\figsetplot{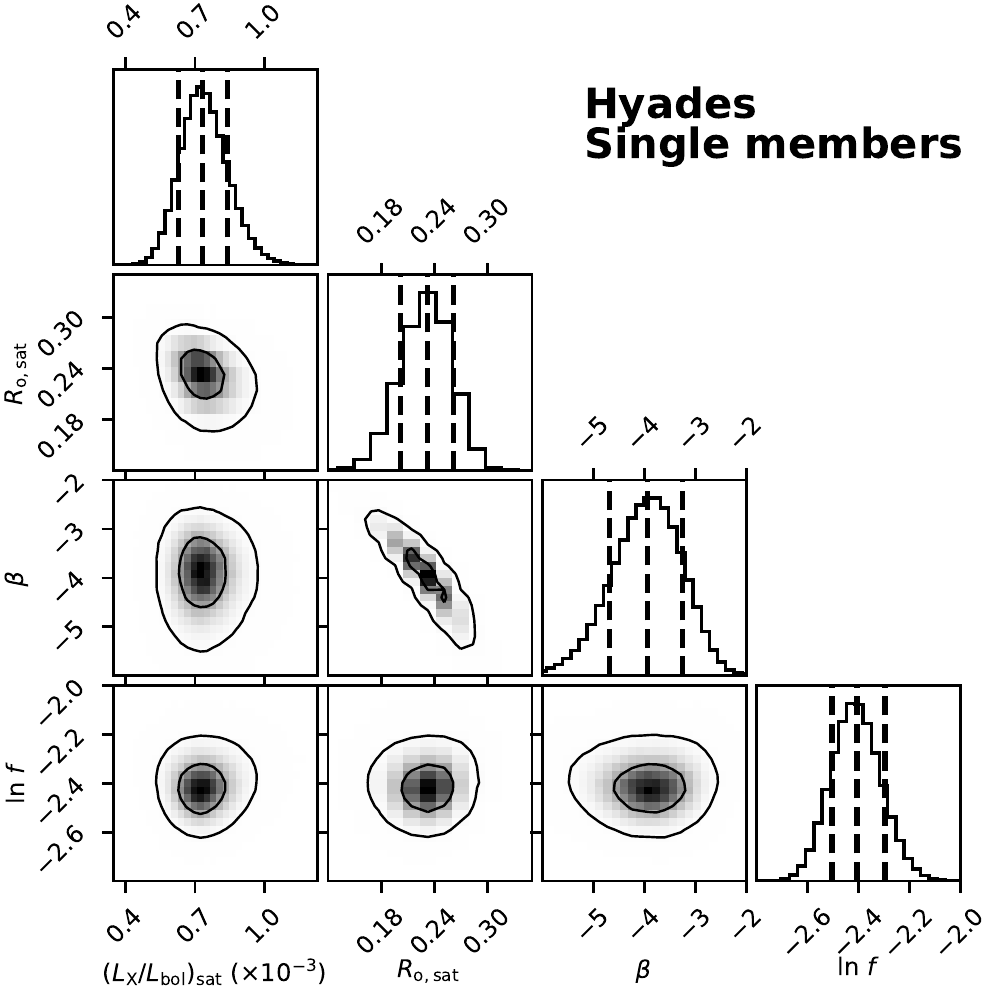}
\figsetgrpnote{Marginalized posterior probability distributions from the MCMC analysis of our \Ro--\LLX\ model using \texttt{emcee} for single members of Hyades. The parameter values of the a posteriori model are the peaks of the one-dimensional distributions; the vertical dashed lines approximate the median and 16$^{th}$, 50$^{th}$, and 84$^{th}$ percentiles. The two-dimensional distributions illustrate covariances between parameters; the contour lines approximate the 1$\sigma$ and 2$\sigma$ levels of the distributions.}
\figsetgrpend

\figsetgrpstart
\figsetgrpnum{14.3}
\figsetgrptitle{Single members of both clusters}
\figsetplot{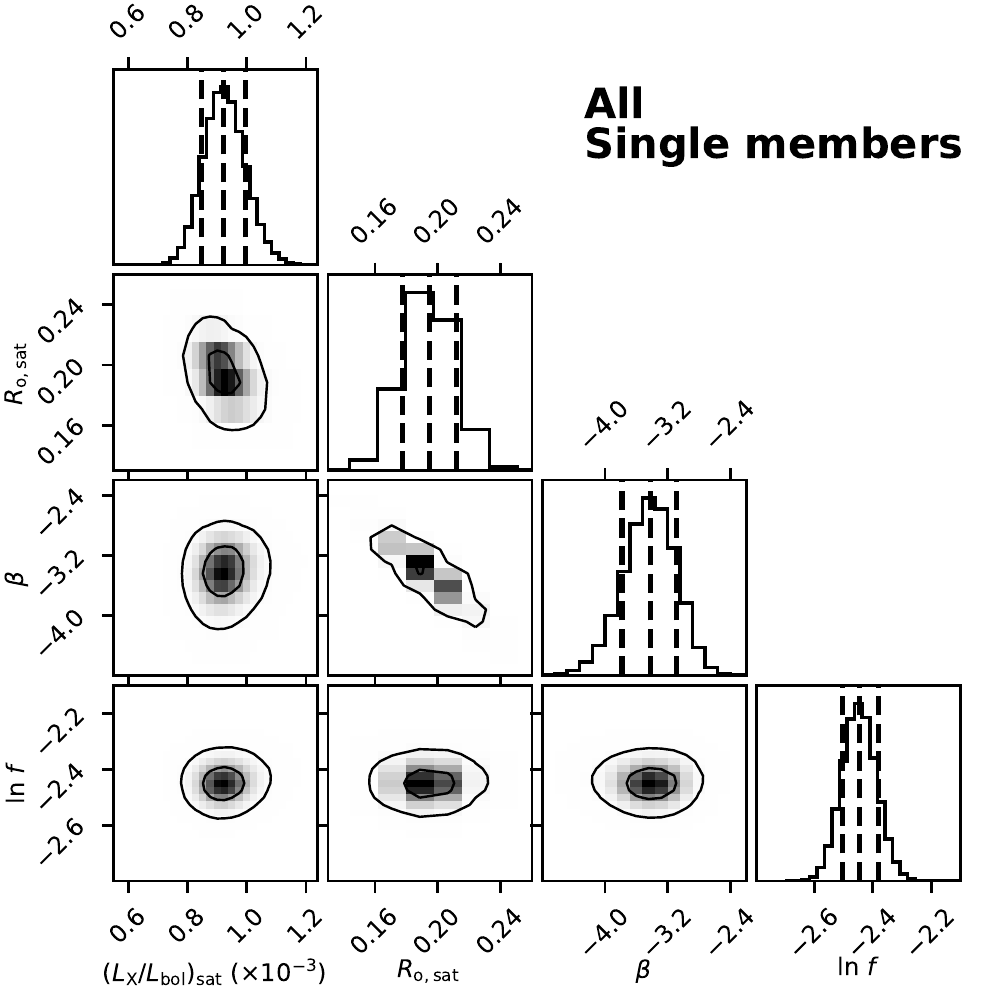}
\figsetgrpnote{Marginalized posterior probability distributions from the MCMC analysis of our \Ro--\LLX\ model using \texttt{emcee} for single members of Praesepe and Hyades together. The parameter values of the a posteriori model are the peaks of the one-dimensional distributions; the vertical dashed lines approximate the median and 16$^{th}$, 50$^{th}$, and 84$^{th}$ percentiles. The two-dimensional distributions illustrate covariances between parameters; the contour lines approximate the 1$\sigma$ and 2$\sigma$ levels of the distributions.}
\figsetgrpend

\figsetgrpstart
\figsetgrpnum{14.4}
\figsetgrptitle{Binary members of Praesepe}
\figsetplot{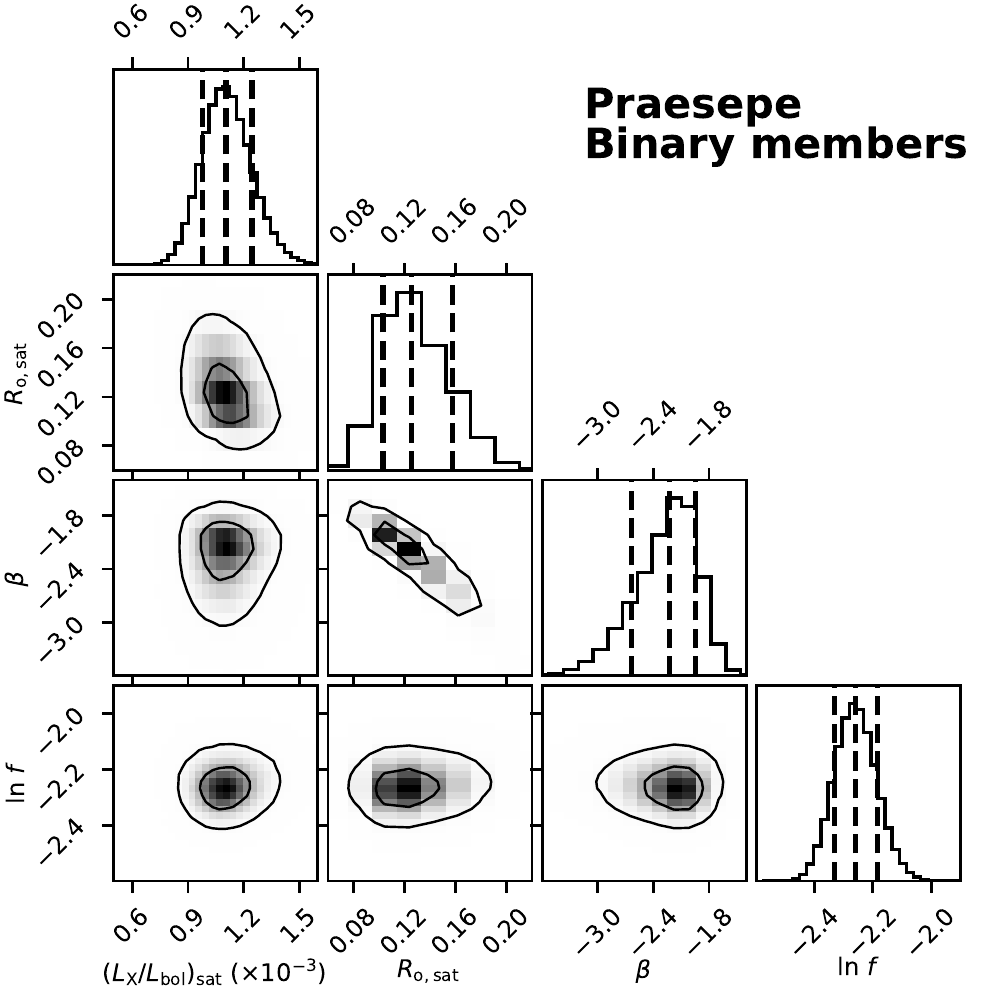}
\figsetgrpnote{Marginalized posterior probability distributions from the MCMC analysis of our \Ro--\LLX\ model using \texttt{emcee} for binary members of Praesepe. The parameter values of the a posteriori model are the peaks of the one-dimensional distributions; the vertical dashed lines approximate the median and 16$^{th}$, 50$^{th}$, and 84$^{th}$ percentiles. The two-dimensional distributions illustrate covariances between parameters; the contour lines approximate the 1$\sigma$ and 2$\sigma$ levels of the distributions.}
\figsetgrpend

\figsetgrpstart
\figsetgrpnum{14.5}
\figsetgrptitle{Binary members of Hyades}
\figsetplot{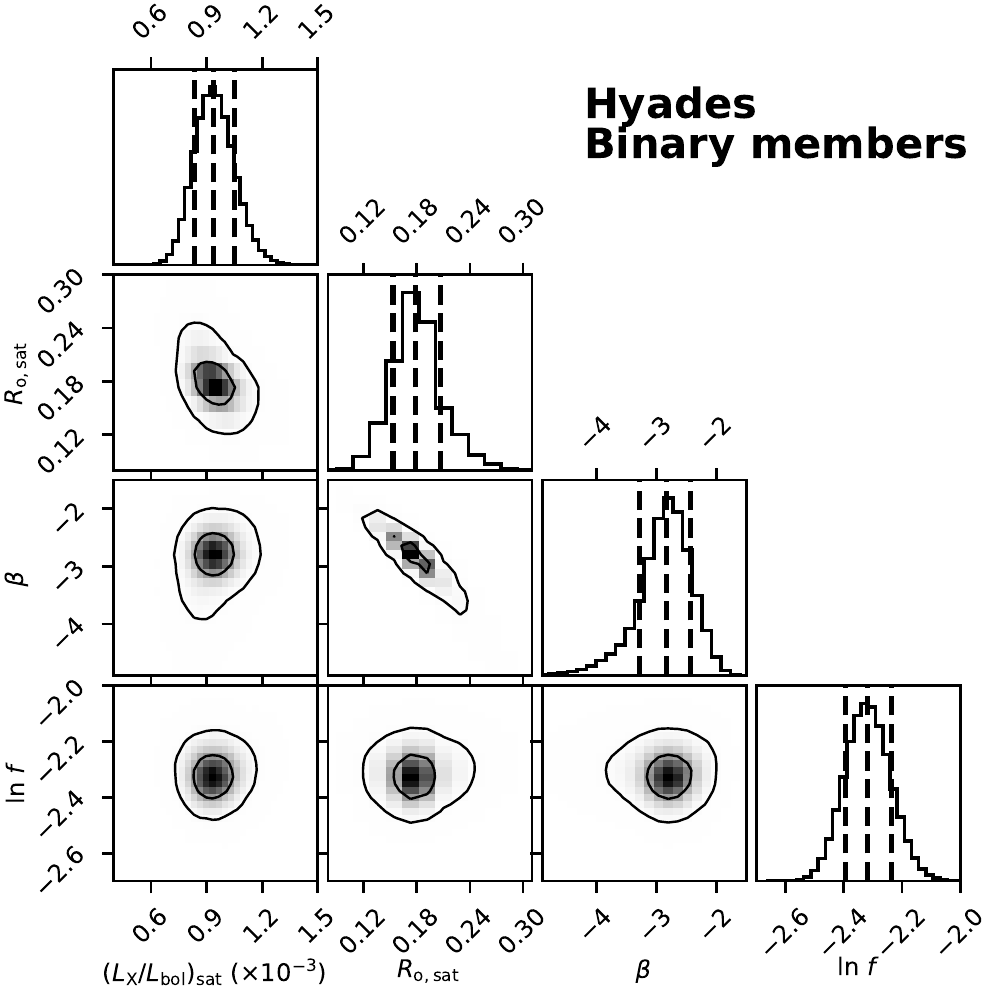}
\figsetgrpnote{Marginalized posterior probability distributions from the MCMC analysis of our \Ro--\LLX\ model using \texttt{emcee} for binary members of Hyades. The parameter values of the a posteriori model are the peaks of the one-dimensional distributions; the vertical dashed lines approximate the median and 16$^{th}$, 50$^{th}$, and 84$^{th}$ percentiles. The two-dimensional distributions illustrate covariances between parameters; the contour lines approximate the 1$\sigma$ and 2$\sigma$ levels of the distributions.}
\figsetgrpend

\figsetgrpstart
\figsetgrpnum{14.6}
\figsetgrptitle{Binary members of both clusters}
\figsetplot{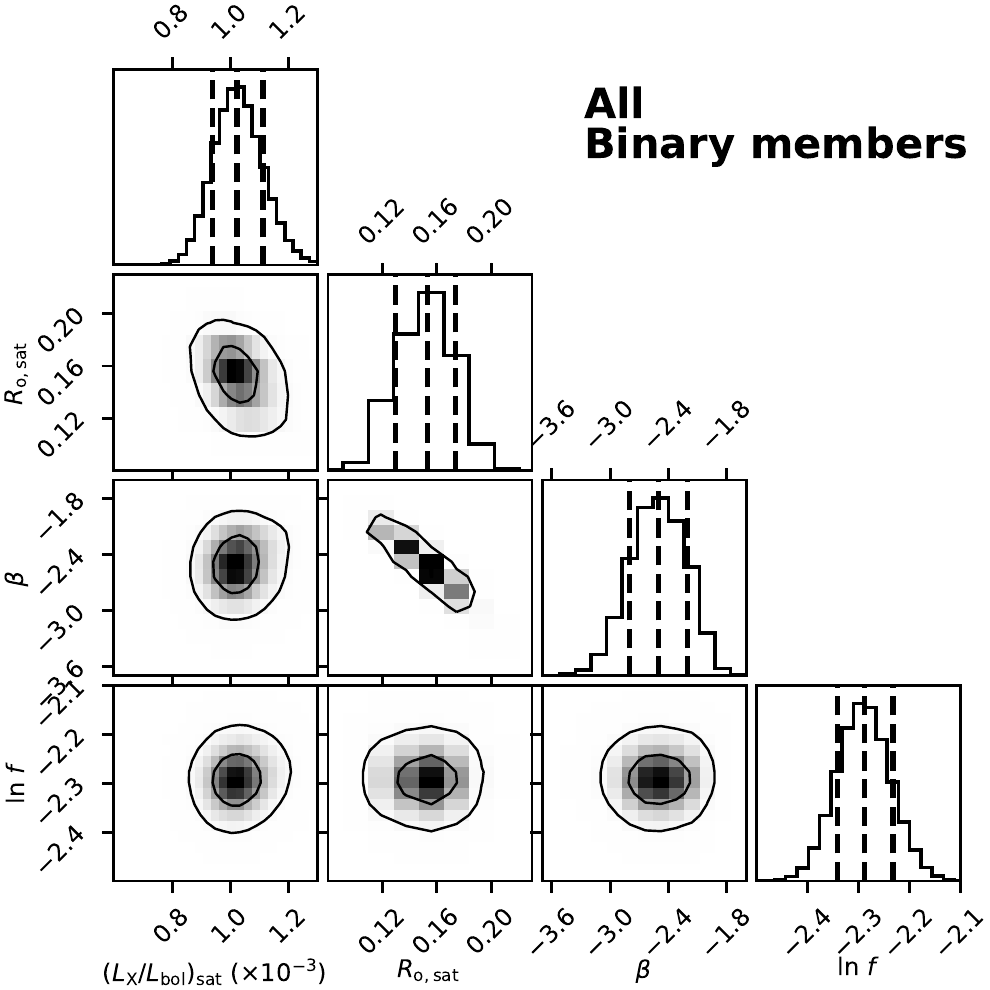}
\figsetgrpnote{Marginalized posterior probability distributions from the MCMC analysis of our \Ro--\LLX\ model using \texttt{emcee} for binary members of Praesepe and Hyades together. The parameter values of the a posteriori model are the peaks of the one-dimensional distributions; the vertical dashed lines approximate the median and 16$^{th}$, 50$^{th}$, and 84$^{th}$ percentiles. The two-dimensional distributions illustrate covariances between parameters; the contour lines approximate the 1$\sigma$ and 2$\sigma$ levels of the distributions.}
\figsetgrpend

\figsetend

% Sample figure for manuscript

\section{MCMC Results for Our Modified \Ro--\LLX\ Model with Supersaturation}\label{app:supersat}
\begin{figure*}[t]
\centerline{\includegraphics{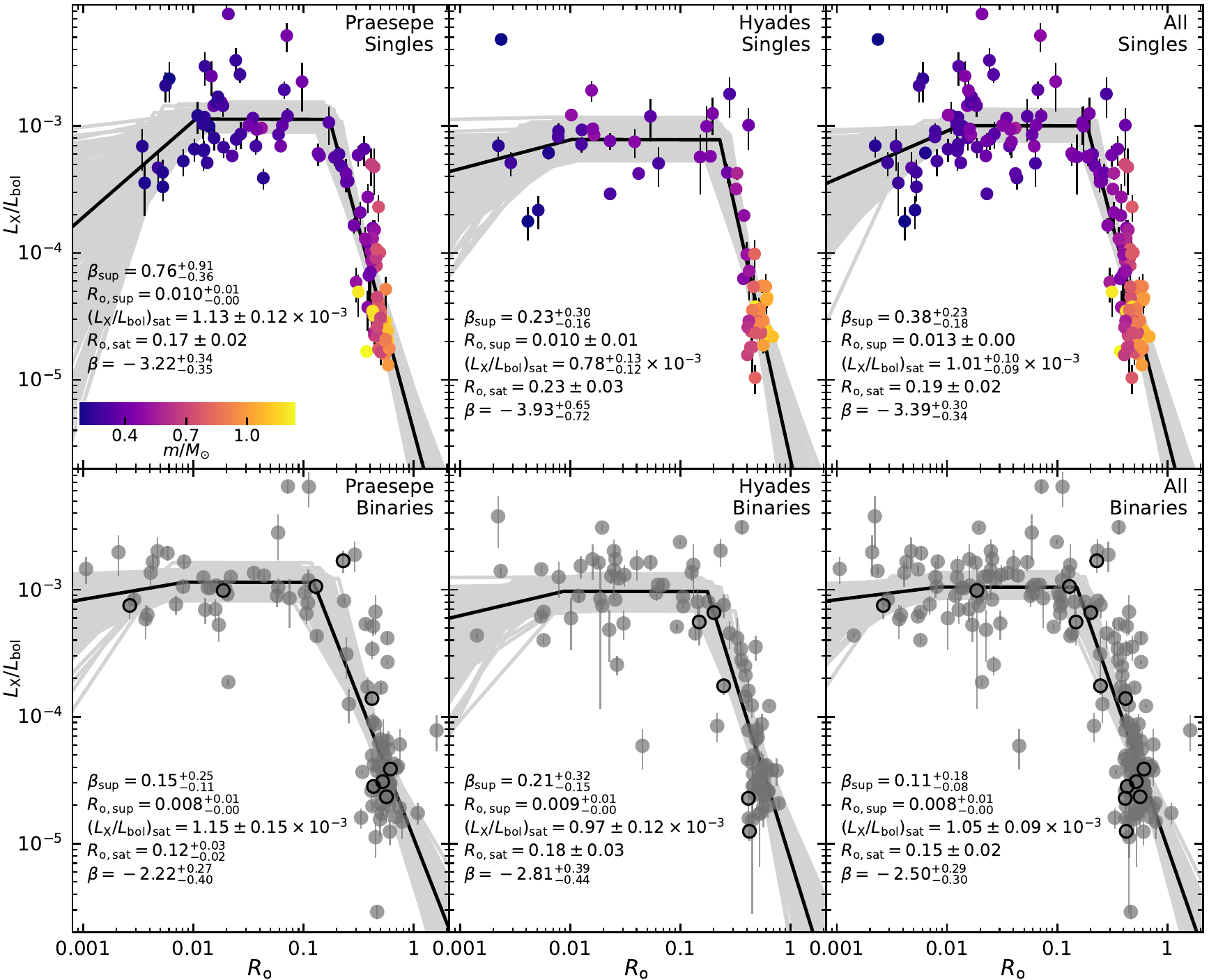}}
\caption{Same as Figure~\ref{fig_rossbyfits}, but for a modified \Ro--\LLX\ model that includes a secondary power-law with index $\beta_\mathrm{sup}$ for \Ro\ $<$ \Ro$_\mathrm{,sup}$, where \Ro$_\mathrm{,sup}$ is the threshold \Ro\ value between supersaturation and saturation. We find only tentative evidence for supersaturation in single stars with \Ro\ $\lapprox$ 0.01, for which the power law index $\beta_\mathrm{sup}$ is distinct enough from zero.}
\label{fig_rossbyfitssup}
\end{figure*}

\begin{figure}
\centerline{\includegraphics{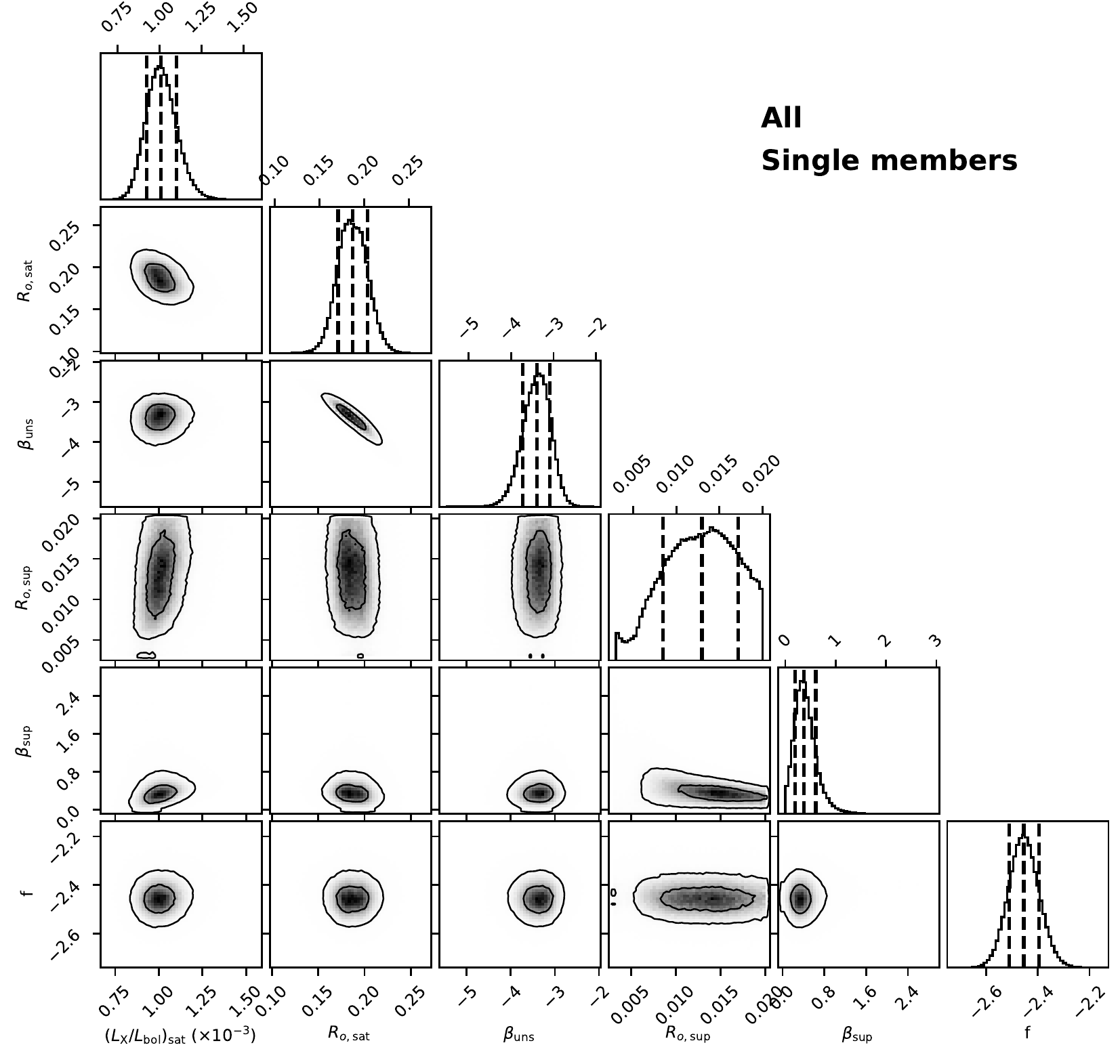}}
\caption{Same as Figure~\ref{fig_posteriors}, but from the MCMC analysis of our modified \Ro--\LLX\ model with supersaturation (see Section~\ref{sec:sat_sup}). The complete figure set, which includes an image for each of the six subsamples in Table~\ref{tbl_rossbies}, is available in the online journal.}
\label{fig_posteriors_sup}
\end{figure}

We present the results and marginalized posterior probability distributions from the MCMC analysis we performed on six different subsamples of Praesepe and Hyades stars: single members of each cluster, binary members for each cluster, single members of both clusters combined, and binary members of both clusters combined, using a modified \Ro--\LLX\ model that includes a secondary power-law at small \Ro\ to represent the supersaturated regime (see Section~\ref{sec:sat_sup}). Figure~\ref{fig_rossbyfitssup} shows the modified \Ro--\LLX\ model on the six subsamples, using the same format as in Figure~\ref{fig_rossbyfits}. Figure~\ref{fig_posteriors_sup} shows an example of the marginalized posterior probability distributions for the subsample of single members from both clusters.

\figsetstart
\figsetnum{16}
\figsettitle{Marginalized posterior probability distributions from the MCMC analysis of Our Modified Model with Supersaturation}

\figsetgrpstart
\figsetgrpnum{15.1}
\figsetgrptitle{Single members of Praesepe}
\figsetplot{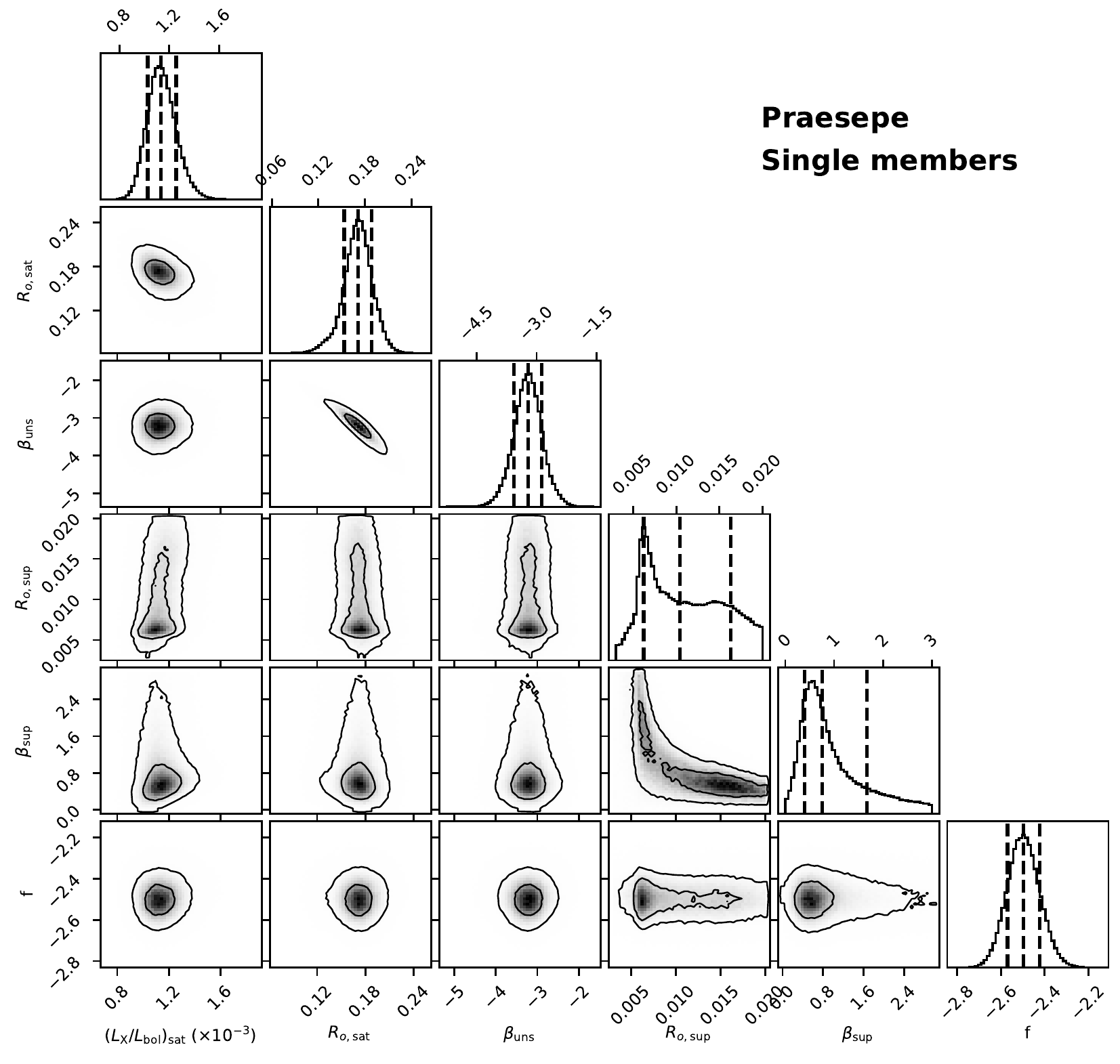}
\figsetgrpnote{Marginalized posterior probability distributions from the MCMC analysis of our modified \Ro--\LLX\ model with supersaturation using \texttt{emcee} for single members of Praesepe. The parameter values of the a posteriori model are the peaks of the one-dimensional distributions; the vertical dashed lines approximate the median and 16$^{th}$, 50$^{th}$, and 84$^{th}$ percentiles. The two-dimensional distributions illustrate covariances between parameters; the contour lines approximate the 1$\sigma$ and 2$\sigma$ levels of the distributions.}
\figsetgrpend

\figsetgrpstart
\figsetgrpnum{15.2}
\figsetgrptitle{Single members of Hyades}
\figsetplot{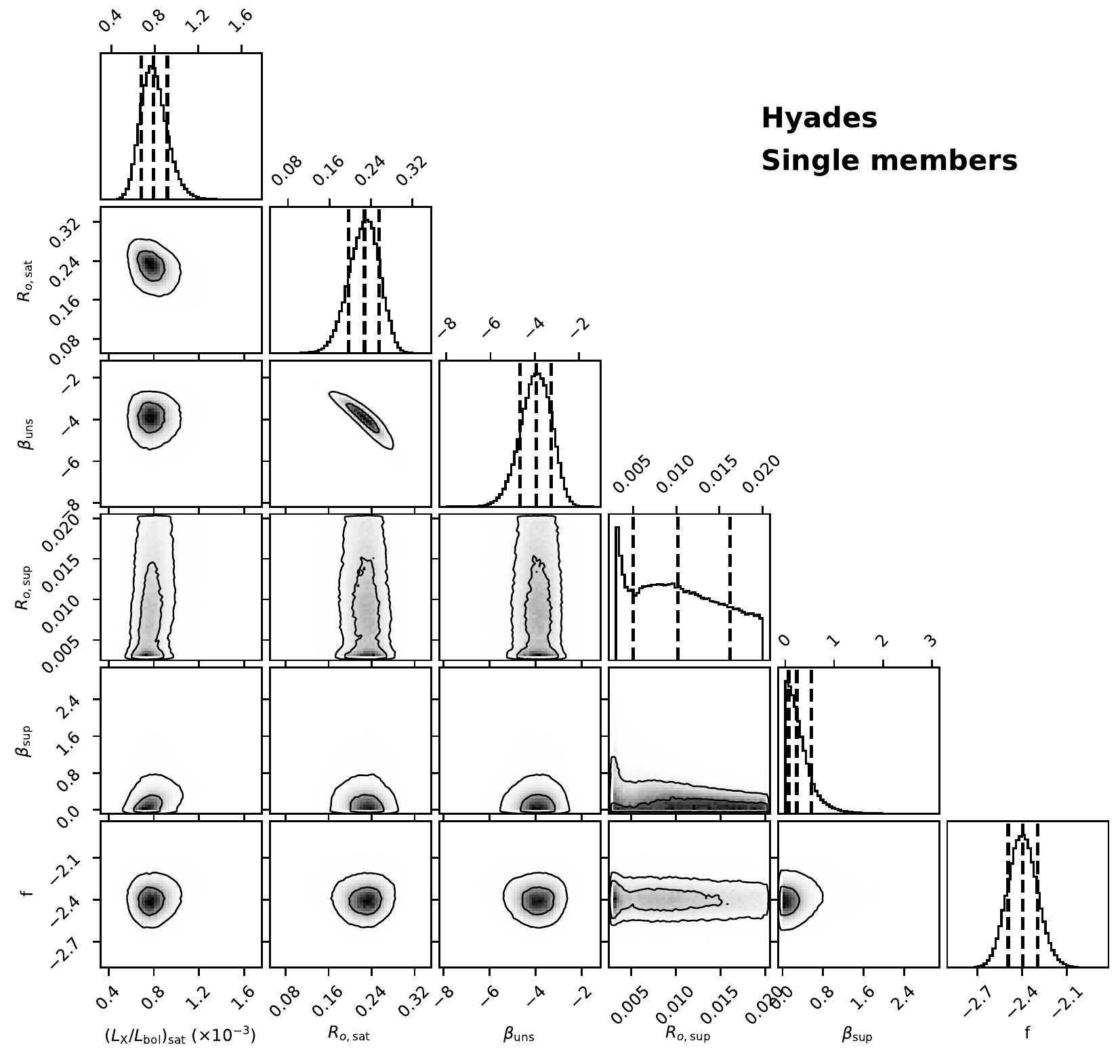}
\figsetgrpnote{Marginalized posterior probability distributions from the MCMC analysis of our modified \Ro--\LLX\ model with supersaturation using \texttt{emcee} for single members of Hyades. The parameter values of the a posteriori model are the peaks of the one-dimensional distributions; the vertical dashed lines approximate the median and 16$^{th}$, 50$^{th}$, and 84$^{th}$ percentiles. The two-dimensional distributions illustrate covariances between parameters; the contour lines approximate the 1$\sigma$ and 2$\sigma$ levels of the distributions.}
\figsetgrpend

\figsetgrpstart
\figsetgrpnum{15.3}
\figsetgrptitle{Single members of both clusters}
\figsetplot{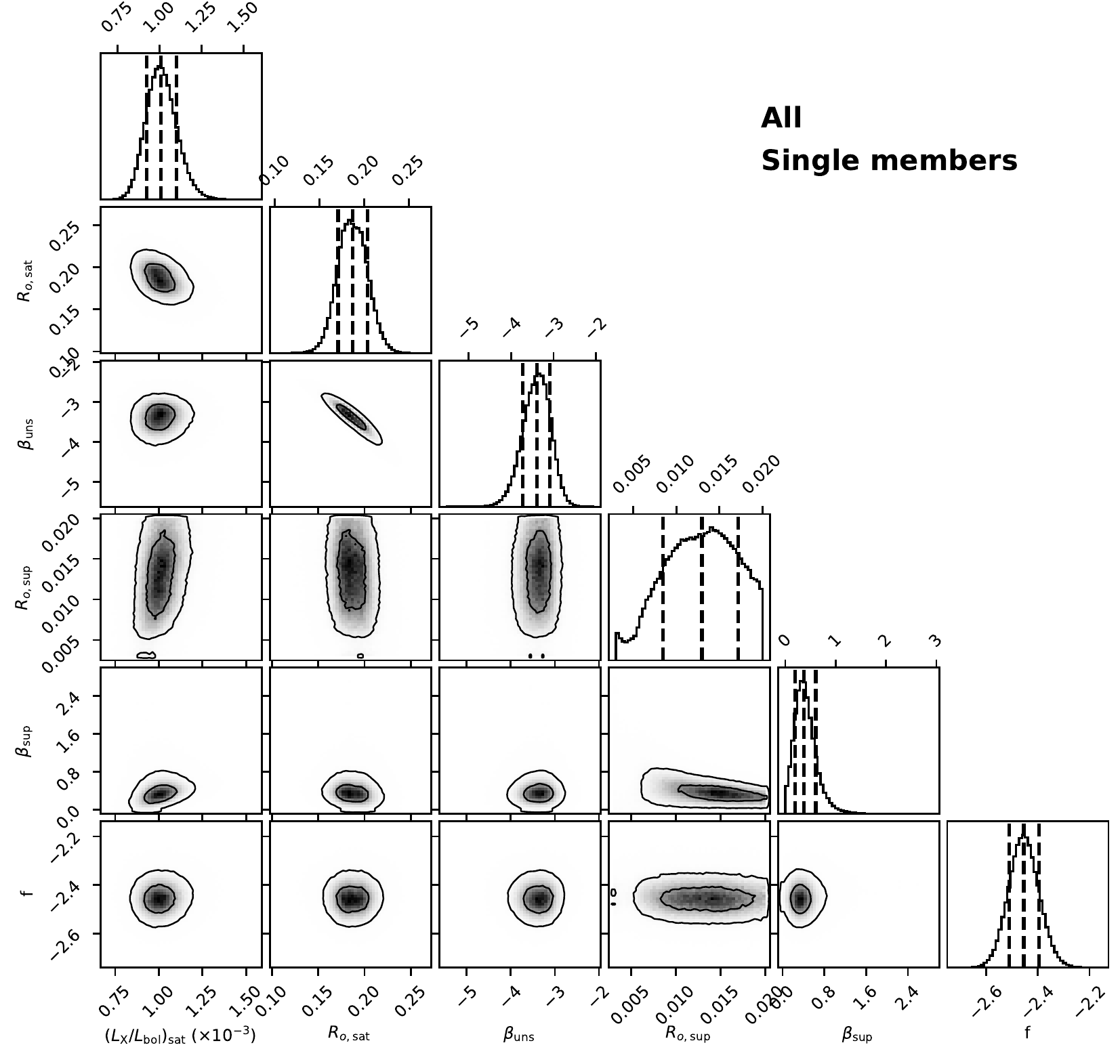}
\figsetgrpnote{Marginalized posterior probability distributions from the MCMC analysis of our modified \Ro--\LLX\ model with supersaturation using \texttt{emcee} for single members of Praesepe and Hyades together. The parameter values of the a posteriori model are the peaks of the one-dimensional distributions; the vertical dashed lines approximate the median and 16$^{th}$, 50$^{th}$, and 84$^{th}$ percentiles. The two-dimensional distributions illustrate covariances between parameters; the contour lines approximate the 1$\sigma$ and 2$\sigma$ levels of the distributions.}
\figsetgrpend

\figsetgrpstart
\figsetgrpnum{15.4}
\figsetgrptitle{Binary members of Praesepe}
\figsetplot{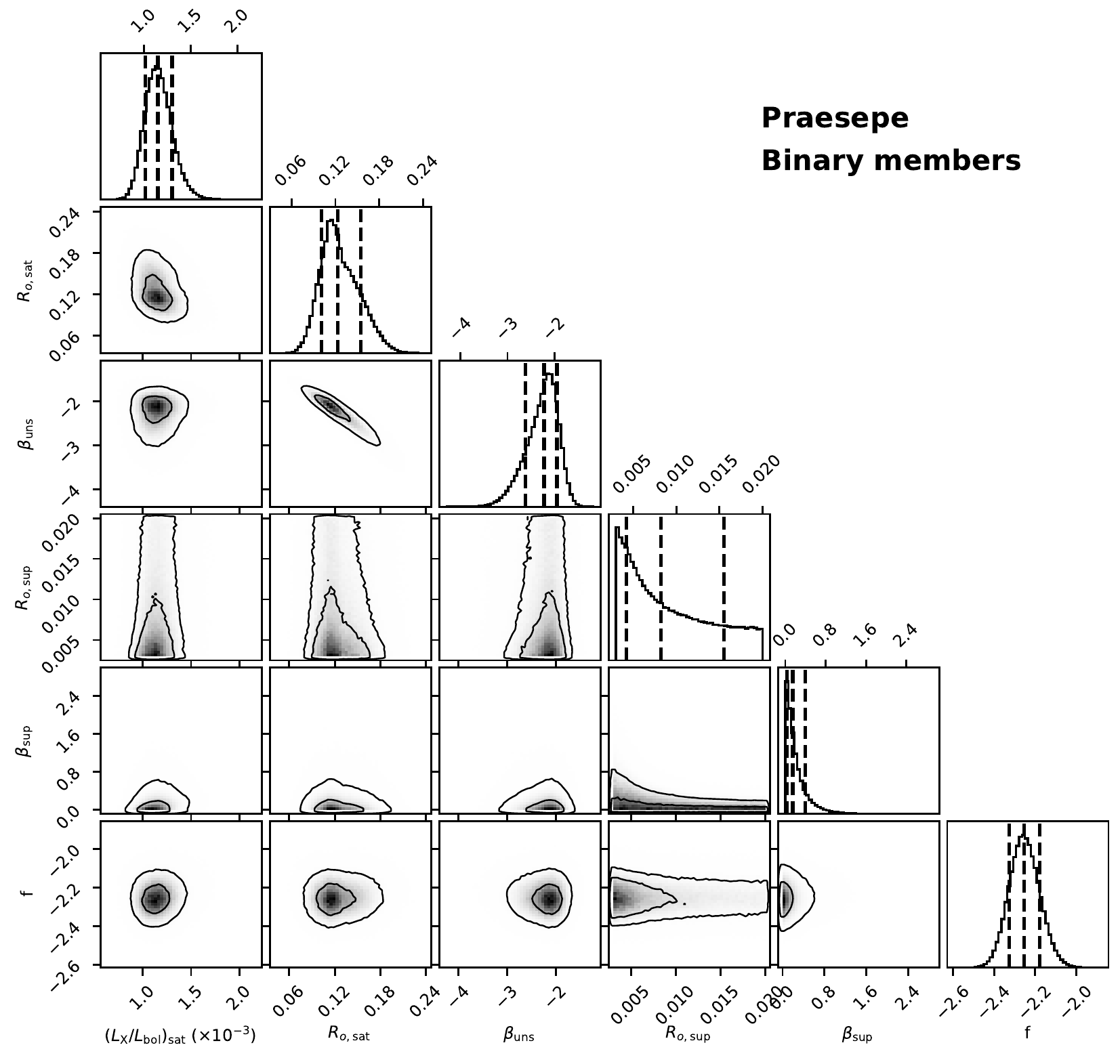}
\figsetgrpnote{Marginalized posterior probability distributions from the MCMC analysis of our modified \Ro--\LLX\ model with supersaturation using \texttt{emcee} for binary members of Praesepe. The parameter values of the a posteriori model are the peaks of the one-dimensional distributions; the vertical dashed lines approximate the median and 16$^{th}$, 50$^{th}$, and 84$^{th}$ percentiles. The two-dimensional distributions illustrate covariances between parameters; the contour lines approximate the 1$\sigma$ and 2$\sigma$ levels of the distributions.}
\figsetgrpend

\figsetgrpstart
\figsetgrpnum{15.5}
\figsetgrptitle{Binary members of Hyades}
\figsetplot{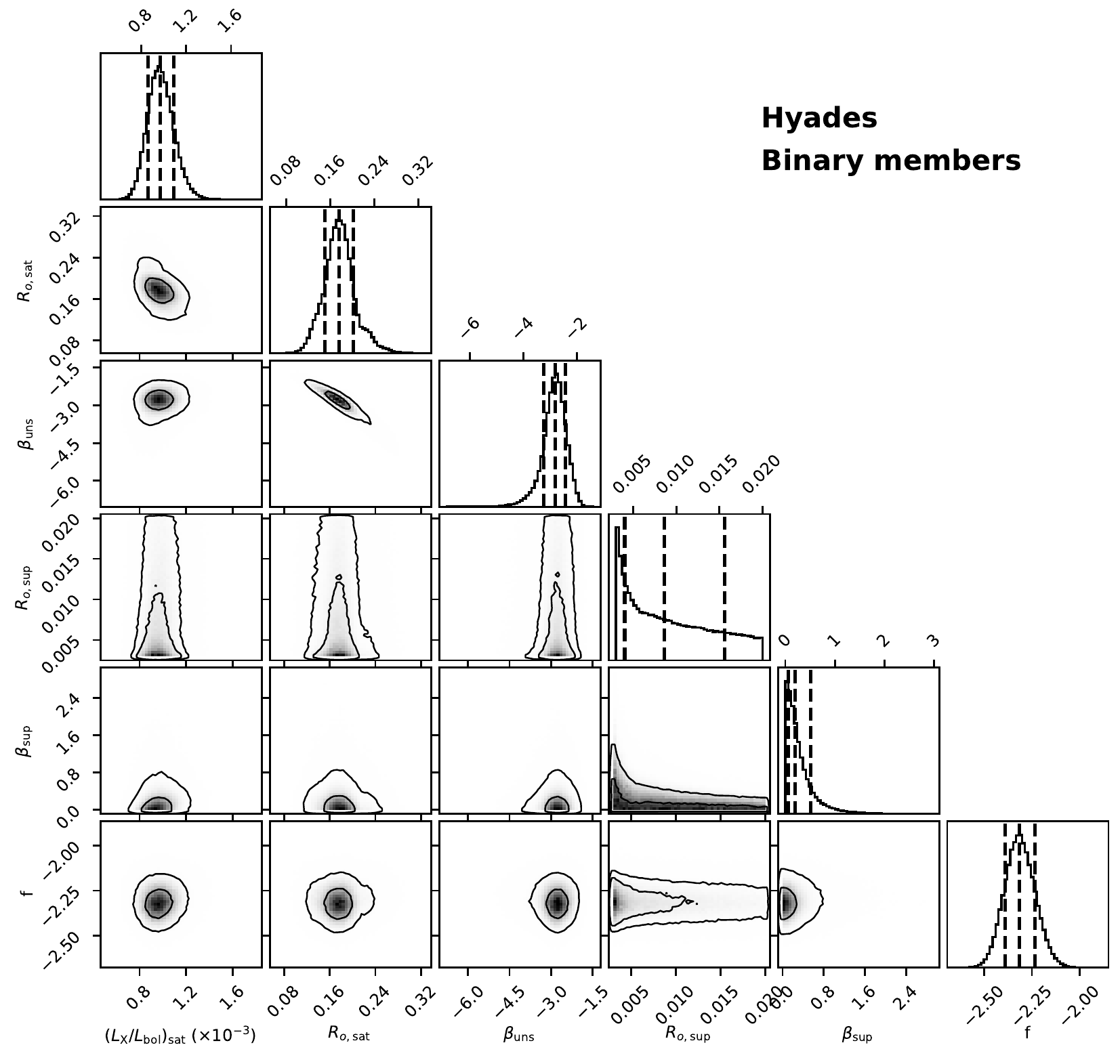}
\figsetgrpnote{Marginalized posterior probability distributions from the MCMC analysis of our modified \Ro--\LLX\ model with supersaturation using \texttt{emcee} for binary members of Hyades. The parameter values of the a posteriori model are the peaks of the one-dimensional distributions; the vertical dashed lines approximate the median and 16$^{th}$, 50$^{th}$, and 84$^{th}$ percentiles. The two-dimensional distributions illustrate covariances between parameters; the contour lines approximate the 1$\sigma$ and 2$\sigma$ levels of the distributions.}
\figsetgrpend

\figsetgrpstart
\figsetgrpnum{15.6}
\figsetgrptitle{Binary members of both clusters}
\figsetplot{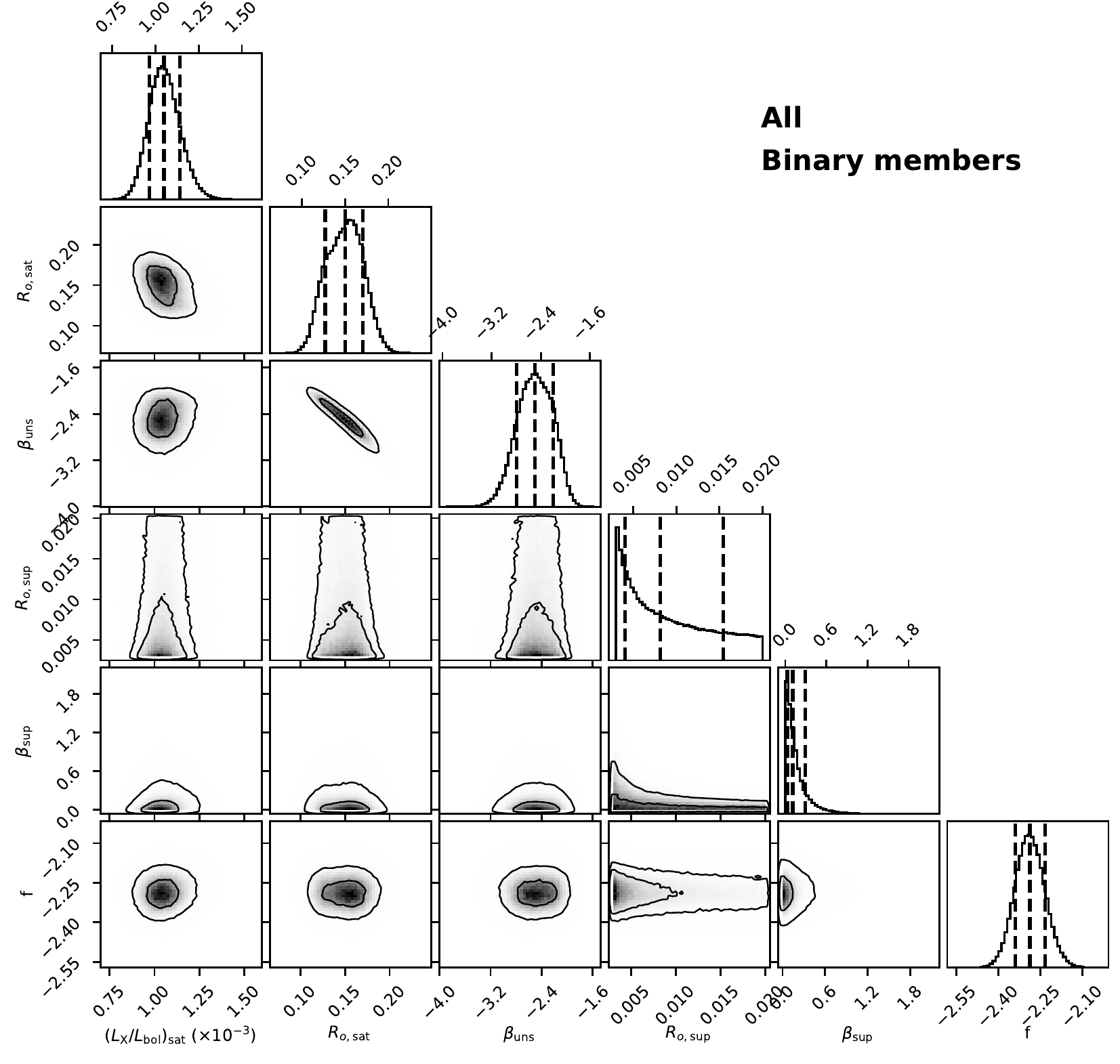}
\figsetgrpnote{Marginalized posterior probability distributions from the MCMC analysis of our modified \Ro--\LLX\ model with supersaturation using \texttt{emcee} for binary members of Praesepe and Hyades together. The parameter values of the a posteriori model are the peaks of the one-dimensional distributions; the vertical dashed lines approximate the median and 16$^{th}$, 50$^{th}$, and 84$^{th}$ percentiles. The two-dimensional distributions illustrate covariances between parameters; the contour lines approximate the 1$\sigma$ and 2$\sigma$ levels of the distributions.}
\figsetgrpend

\figsetend

% Sample figure for manuscript

\end{document}

%% file: tbl_mems.tex
\setlength{\tabcolsep}{2.5pt}
\begin{deluxetable}{lcc}
\tablewidth{0pt}
\tabletypesize{\small}
\tablecaption{Praesepe and Hyades Membership Catalogs Considered in This Work\label{tbl_mems}}
\tablehead{
\colhead{}     & \multicolumn{2}{c}{\# of Members}\\
\colhead{Provenance} & \colhead{Praesepe} & \colhead{Hyades} 
}
\startdata
Legacy Catalogs %\tablenotemark{a} 
& 1123 & 746\\ 
\citet{Cantat-Gaudin2018} & 719 & \nodata \\
\citet{GaiaCol2018} & 937 & 508 \\ 
\citet{Lodieu2019b} & 721 & \nodata \\
\citet{Lodieu2019a}\tablenotemark{a} & \nodata & 556 \\
\citet{Meingast2019} & \nodata & 238\\
\citet{Roser2019a} & \nodata & 976 \\
\citet{Roser2019b} & 1393 & \nodata \\
Ultra-cool Dwarfs\tablenotemark{b} & 12 & 21 \\
\hline
Consolidated \# of Members & 1739 & 1315
\enddata
\tablenotetext{a}{We did not include the set of stars farther than 18 pc from the cluster center, as these authors considered this set to have a significant level of contamination.}
\tablenotetext{b}{Late-M dwarfs and brown dwarfs found in the literature. See Section~\ref{sec:ultracool}.}
%\vspace{-0.2in}

\end{deluxetable}

%% file: tbl_catalogcols.tex
\begin{deluxetable}{@{}ll}
\tabletypesize{\footnotesize} 

\tablecaption{Overview of Columns in the Praesepe and Hyades Membership Catalog \label{tbl_catalogcols}}

\tablehead{
\colhead{Column} & \colhead{Description} %\\[-0.1 in]
}

\startdata
1      & Name \\
2      & 2MASS designation \\
3      & Gaia EDR3 designation \\
4, 5   & R.A., Decl. for epoch J2000\\
6      & Cluster to which the star belongs \\
7      & Member in legacy catalog? \\
8      & Member in \citet{Cantat-Gaudin2018}? \\
9      & Member in \citet{GaiaCol2018}? \\
10     & Member in \citet{Lodieu2019b}? \\
11     & Member in \citet{Lodieu2019a}? \\
12     & Member in \citet{Meingast2019}? \\
13     & Member in \citet{Roser2019b}? \\
14     & Member in \citet{Roser2019a}? \\
15     & Ultracool dwarf in literature? \\
16, 17 & Distance and 1$\sigma$ uncertainty \\
18     & Source of distance\tablenotemark{a} \\
19, 20 & $K$-band magnitude and 1$\sigma$ uncertainty \\
21     & Source of $K$ magnitude\tablenotemark{b} \\
22, 23 & Gaia EDR3 $G$ band magnitude and 1$\sigma$ uncertainty \\
24     & Source of $G$ magnitude\tablenotemark{c} \\
25     & Gaia EDR3 $G_\mathrm{BP} - G_\mathrm{RP}$ \\
26     & Binary flag: (0) no binary flag; (1) candidate binary; \\
       & (2) confirmed binary \\
27     & Gaia EDR3 RUWE \\
28     & Rotation period \prot \\
29, 30 & X-ray energy flux \fx\ (0.1--2.4 keV) and 1$\sigma$ uncertainty \\
31     & Stellar mass \\
32     & Convective turnover time $\tau$ \\
33, 34 & \Lbol\ and 1$\sigma$ uncertainty \\
\enddata

\tablenotetext{a}{BJ: \citet{BailerJones2021}; C: Cluster distance; D: \citet{Dittmann2014}; L: \citet{Lodieu2014}; P: \citet{PerezGarrido2018}; R11: \citet{Roser2011}; R16: \citet{Robert2016}; S: \citet{Schneider2017}; vL: \citet{vanLeeuwen2007}; Z: \citet{Zhang2010}.} 
\tablenotetext{b}{2M: 2MASS; B: \citet{Boudreault2010}; G: Derived from GEDR3 photometry; UL: UKIDSS LAS; UG: UKIDSS GCS.}
\tablenotetext{c}{2M: Derived from 2MASS photometry; G: GEDR3}
%\tablenotetext{d}{0: No binary flags; 1: candidate binary; 2: confirmed binary}
%29     & Source of \Prot \tablenotemark{e} \\
%\tablenotetext{e}{A: ASAS; A11: \citet{Agueros2011}; D: \citet{Delorme2011}; D16: \citet{Douglas2016}; D17: \citet{Douglas2017}; D19: \citet{Douglas2019}; H: \citet{Hartman2011}; K: \citet{Kovacs2014}; R: \citet{Radick1987,Radick1995}; S: \citet{Scholz2007}, \citet{Scholz2011}; T: TESS.} 

\vspace{-0.5in}

\end{deluxetable}

%% file: tbl_Xcols.tex
\begin{deluxetable}{@{}ll}
\tabletypesize{\footnotesize} 

\tablecaption{Overview of Columns in the X-ray Source Catalog \label{tbl_Xcols}}

\tablehead{
\colhead{Column} & \colhead{Description} %\\[-0.1 in]
}

\startdata
1      & External catalog source ID \\
2      & Provenance of X-ray information\tablenotemark{a}\\
3      & IAU Name \\
4      & Observation ID \\
5      & Instrument \\
6, 7   & R.A., Decl. for epoch J2000\\
8      & X-ray positional uncertainty \\
9      & Off-axis angle $\theta$\\
10     & Detection likelihood $L$\tablenotemark{b} \\
11     & Net counts in the broad band \\
12, 13 & Net count rate and 1$\sigma$ uncertainty in broad band \\
14, 15 & Net count rate and 1$\sigma$ uncertainty in soft band \\
16, 17 & Net count rate and 1$\sigma$ uncertainty in hard band \\
18$-$20  & Definition of broad, soft, and hard bands \\
21     & Hardness ratio: (hard band $-$ soft band) / \\
       & (hard band + soft band) \\
22     & Exposure time \\
23     & Variability flag: (0) no evidence for variability; \\
       & (1) possibly variable; (2) definitely variable \\
24, 25 & Unabsorbed energy flux and 1$\sigma$ uncertainty in the \\
       & 0.1--2.4 keV band \\
26     & Source of energy flux: (ECF) from applying ECF; \\
       & (SpecFit) from spectral fitting \\
27     & X-ray flare removed? \\
28     & Quality Flag\tablenotemark{c} \\
29     & Name of the optical counterpart \\
30     & Separation between X-ray source and optical counterpart \\
\enddata

\tablenotetext{a}{1RXH; 2RXP; 2RXS; CSC; CIAO: Reduction of Chandra observation with CIAO; R95: \citet{Randich1995}; S01: \citet{Stelzer2001}; S95: \citet{Stern1995}; SAS: Reduction of XMM observations with SAS; Swift.}
\tablenotetext{b}{For CIAO sources, it is the source significance; for all others, it is the maximum likelihood.}
\tablenotetext{c}{m: likely mismatch to optical counterpart; x: likely extended source.}
\end{deluxetable}

%% file: tbl_Xlog.tex
\begin{deluxetable*}{ccclcrccrc}
\centering 
\tabletypesize{\small} \tablewidth{0pt}

\tablecaption{Log of Chandra and XMM Observations of Praesepe and Hyades\label{tbl_Xlog}}

\tablehead{
\colhead{Cluster} &
\colhead{Instrument} &
\colhead{ObsIDs} &
\colhead{P.I.} &
\colhead{Start} & 
\colhead{Duration\tablenotemark{a}} & 
\multicolumn{2}{c}{Nominal Aimpoint} & 
\colhead{Roll} &
\colhead{Filter/}\\[-0.05 in]
\cline{7-8}\\[-0.2 in]
\colhead{} &
\colhead{} &
\colhead{} & 
\colhead{} & 
\colhead{Date} & 
\colhead{(s)} & 
\colhead{$\alpha_{\rm J2000}$} & 
\colhead{$\delta_{\rm J2000}$} & 
\multicolumn{1}{c}{ (\arcdeg)} &
\colhead{Data Mode}
}

\startdata
Praesepe & EPIC & 0101440401 & Pallavicini & 2000-11 & 48310 & 08:39:58.0 & +19:32:29.0 & 106.2 & Thick/Med\tablenotemark{b}\\
Praesepe & EPIC & 0721620101 & Ag\"ueros & 2013-10 & 71800 & 08:39:00.0 & +19:57:00.0 & 103.0 & Thin1\\
Praesepe & ACIS-I & 17254--17257 & Drake, J. & 2015-05 & 191560 & 08:39:50.0 & +19:31:41.0 & 255.6--256.2 & VFaint\\
Praesepe & EPIC & 0863710201 & N\'u\~nez & 2021-04 & 32000 & 08:39:32.0 & +20:39:20.3 & 283.6 & Medium\\
Hyades & HRC-1 & 2554 & Ayres & 2001-12 & 17900 & 04:35:55.2 & +16:30:33.0 & 288.3 & \nodata\\
Hyades  & EPIC & 0762760601 & Ag\"ueros & 2015-09 & 17000 & 04:33:27.0 & +13:02:43.5 & 82.3 & Thin1
\enddata

\tablenotetext{a}{Exposure time before any filtering is applied.}
\tablenotetext{b}{The first filter is for the MOS cameras and the second for the pn camera.}
\tablecomments{This table is available in its entirety in the electronic edition of the \apj. Six rows are shown here for guidance regarding its form and content.}

\vspace{-0.2in}

\end{deluxetable*}

%% file: tbl_spec.tex
\begin{deluxetable*}{@{}lrrrrrrr@{}}
%\tabletypesize{\scriptsize}

\tablecaption{Spectral Fits for the Highest X-Ray Count Chandra, Swift, and XMM Sources\label{tbl_spec}}

\tablehead{
\colhead{Source ID} & \colhead{IAU Name} & \colhead{$kT$} & \colhead{Metal} & \colhead{$N_\mathrm{H}$} & \colhead{Flux} & \colhead{$\chi^2_{\nu}$} & \colhead{d.o.f.} \\[-0.06 in]
 & & & \colhead{Abund.} & \colhead{($10^{21}$)} & \colhead{($10^{-14}$)} & & \\[-0.08 in]
 & & \colhead{(keV)} & & \colhead{(cm$^{-2}$)} & \colhead{(\ergcms}) & & \\[-0.2 in]
}
\startdata
107851101010001 & 4XMM J041802.0+181522 & 0.68$\pm$0.02 & 0.54$\pm$0.08 & \nodata & 41.36$\pm$3.05 & 1.99 & 227\\
107619211010001 & 4XMM J083656.2+185747 & 0.76$\pm$0.03 & 0.09$\pm$0.02 & \nodata & 5.83$\pm$0.45 & 1.06 & 121\\
107216201010012 & 4XMM J083915.7+200413 & 0.61$\pm$0.05 & 0.05$\pm$0.02 & \nodata & 3.58$\pm$0.56 & 0.81 & 53\\
\nodata & XMMU J083952.7+203046 & 0.72$\pm$0.09 & 0.05$\pm$0.03 & 1.08$\pm$0.36 & 4.68$\pm$1.44 & 0.71 & 49\\
\nodata & CXOU J042138.5+201809 & 0.98$\pm$0.07 & 0.27$\pm$0.13 & \nodata & 4.61$\pm$1.25 & 1.51 & 24\\
\enddata

\tablecomments{Results of all acceptable spectral fits for our highest-count X-ray sources using a one-temperature APEC model. All fits included the XSPEC model {\sc tbabs} to account for ISM extinction, and we give the best fit atomic hydrogen column density $N_{\mathrm{H}}$ in the cases where freeing this parameter resulted in a better fit. We also show the derived unabsorbed flux in the 0.1--2.4 keV band. Finally, we give for each fit its reduced chi-square ($\chi^2_{\nu}$) and degrees of freedom (d.o.f.). This table is available in its entirety in the electronic edition of the \apj. Five rows are shown here for guidance regarding its form and content.}
\end{deluxetable*}

%% file: tbl_ECFs.tex
\begin{deluxetable*}{cccllcrc}
\centering 
\tabletypesize{} \tablewidth{0pt}

\tablecaption{ECFs Used For the Low-Count X-Ray Sources \label{tbl_ECFs}}

\tablehead{
\colhead{Observatory} &
\colhead{Instrument} &
\colhead{Energy} &
\colhead{Other} &
\multicolumn{2}{c}{Praesepe} &
\multicolumn{2}{c}{Hyades} \\[-0.19cm]
\cmidrule(lr){5-6}\cmidrule(lr){7-8} \\[-0.6cm]
\colhead{} &
\colhead{} &
\colhead{Band\tablenotemark{a}} &
\colhead{Describers\tablenotemark{b}} & 
\colhead{Abundance 0.2} & 
\colhead{Abundance 0.4} & 
\colhead{Abundance 0.2} & 
\colhead{Abundance 0.4}
}

\startdata
Chandra & ACIS-I & 0.5--7.0 & Cycle 4 & \nodata & \nodata & 7.314E$-$12 & 6.462E$-$12 \\
 & & & Cycle 8 & 9.956E$-$12 & 8.842E$-$12 & 9.672E$-$12 & 8.584E$-$12 \\
 & & & Cycle 9 & 9.972E$-$12 & 8.856E$-$12 & \nodata & \nodata \\
 & & & Cycle 15 & \nodata & \nodata & 1.278E$-$11 & 1.140E$-$11 \\
 & & & Cycle 16 & 1.529E$-$11 & 1.370E$-$11 & \nodata & \nodata \\
 & ACIS-S & 0.5--7.0 & Cycle 3 & \nodata & \nodata & 4.240E$-$12 & 3.798E$-$12 \\
 & & & Cycle 8 & \nodata & \nodata & 5.354E$-$12 & 4.740E$-$12 \\
 & & & Cycle 9 & 5.554E$-$12 & 4.918E$-$12 & \nodata & \nodata \\
 & & & Cycle 13 & \nodata & \nodata & 6.048E$-$12 & 5.342E$-$12 \\
 & & & Cycle 14 & 6.084E$-$12 & 5.554E$-$12 & 6.082E$-$12 & 5.372E$-$12 \\
 & & & Cycle 20 & \nodata & \nodata & 2.020E$-$11 & 1.843E$-$11 \\
 & & & Cycle 21 & \nodata & \nodata & 1.808E$-$11 & 1.647E$-$11 \\
 & & & Cycle 22 & \nodata & \nodata & 2.120E$-$11 & 1.947E$-$11 \\
 & HRC & 0.1--10.0 & Cycle 3 & \nodata & \nodata & 8.218E$-$12 & 8.126E$-$12 \\
 & & & Cycle 4 & \nodata & \nodata & 8.140E$-$12 & 7.980E$-$12 \\
\cline{1-8}
ROSAT & PSPC & 0.1--2.0 & 2RXS/2RXP & 1.141E$-$11 & 1.101E$-$11 & 6.950E$-$12 & 7.314E$-$12 \\
 &  & 0.4--2.0 & R95 & 1.630E$-$11 & 1.431E$-$11 & \nodata & \nodata \\
 &  & 0.1--1.8 & S95 & \nodata & \nodata & 6.928E$-$12 & 7.294E$-$12 \\
 &  & 0.1--2.0 & S01 & \nodata & \nodata & 6.904E$-$12 & 7.266E$-$12 \\
 & HRI & 0.2--2.4 & 1RXH & 3.578E$-$11 & 3.258E$-$11 & 3.074E$-$11 & 2.880E$-$11 \\
\cline{1-8}
Swift & XRT & 0.3--10.0 & PC & 3.202E$-$11 & 2.976E$-$11 & 2.998E$-$11 & 2.804E$-$11 \\
\cline{1-8}
XMM & pn & 0.2--12.0 & Thin & 1.559E$-$12 & 1.511E$-$12 & 1.322E$-$12 & 1.315E$-$12 \\
 & & & Medium & 1.632E$-$12 & 1.560E$-$12 & 1.425E$-$12 & 1.390E$-$12 \\
 & & & Thick & 2.248E$-$12 & 2.094E$-$12 & 2.055E$-$12 & 1.935E$-$12 \\
 & MOS & 0.2--12.0 & Thin & 7.186E$-$12 & 6.800E$-$12 & 6.452E$-$12 & 6.204E$-$12 \\
 & & & Medium & 7.454E$-$12 & 6.996E$-$12 & 6.800E$-$12 & 6.462E$-$12 \\
 & & & Thick & 9.322E$-$12 & 8.634E$-$12 & 8.686E$-$12 & 8.102E$-$12
\enddata

\tablenotetext{a}{The original energy band of the instrumental count rate values.}
\tablenotetext{b}{Cycle number for Chandra; catalog for ROSAT: R95 is \citet{Randich1995}, S95 is \citet{Stern1995}, and S01 is \citet{Stelzer2001}; operation mode for Swift; and filter type for XMM.}
\tablecomments{All ECFs have units erg cm$^{-2}$ cts$^{-1}$ and produce unabsorbed \fx\ values in the 0.1--2.4 keV band. We calculated two sets of ECFs using a 1T APEC model: one set adopted a metal abundance of 0.2 (columns 5 and 7), and another set adopted a metal abundance of 0.4 (columns 6 and 8). See Section~\ref{lx} for more details.}

\end{deluxetable*}

%% file: tbl_LLXstats.tex
\begin{deluxetable}{@{}llrcc@{}}

\tablecaption{\LLX\ and \LX\ Statistics\label{tbl_llxstats}}

\tablehead{
\colhead{Sample} & \colhead{Sp. Type} & \colhead{$N_\star$} & \colhead{log \LLX} & \colhead{log \LX} \\[-0.2 in]
}
\startdata
\multicolumn{5}{c}{\it Praesepe} \\
Singles  & F0--F4   & 8  & $-5.39^{+0.44}_{-0.48}$ & 28.88$^{+0.52}_{-0.47}$ \\
         & F5--F9   & 12 & $-4.60^{+0.12}_{-0.23}$ & 29.20$^{+0.22}_{-0.11}$ \\
         & G0--G4   & \nodata & \nodata & \nodata \\
         & G5--G9   &  8 & $-4.67^{+0.23}_{-0.12}$ & 28.71$^{+0.34}_{-0.07}$ \\
         & K0--K4   & 19 & $-4.51^{+0.58}_{-0.18}$ & 28.66$^{+0.35}_{-0.35}$ \\
         & K5--K9   & 12 & $-4.45^{+0.72}_{-0.15}$ & 28.27$^{+0.71}_{-0.19}$ \\         
         & M0--M3.0 & 37 & $-3.43^{+0.59}_{-0.65}$ & 28.44$^{+0.68}_{-0.47}$ \\         
         & M3.5--M5 & 50 & $-3.00^{+0.33}_{-0.24}$ & 28.48$^{+0.46}_{-0.27}$ \\         
Binaries & F0--F4   & 11 & $-5.51^{+0.64}_{-0.40}$ & 29.17$^{+0.16}_{-0.50}$ \\
         & F5--F9   & 18 & $-4.80^{+0.34}_{-0.57}$ & 29.22$^{+0.31}_{-0.34}$ \\
         & G0--G4   & 12 & $-4.48^{+0.22}_{-0.22}$ & 29.19$^{+0.60}_{-0.11}$ \\
         & G5--G9   & 12 & $-4.55^{+0.38}_{-0.14}$ & 29.05$^{+0.31}_{-0.14}$ \\
         & K0--K4   & 22 & $-4.45^{+0.26}_{-0.31}$ & 28.68$^{+0.43}_{-0.28}$ \\
         & K5--K9   & 11 & $-4.39^{+0.88}_{-0.20}$ & 28.35$^{+0.86}_{-0.17}$ \\
         & M0--M3.0 & 26 & $-3.23^{+0.39}_{-0.54}$ & 29.21$^{+0.21}_{-0.56}$ \\
         & M3.5--M5 & 33 & $-2.97^{+0.25}_{-0.18}$ & 28.57$^{+0.51}_{-0.33}$ \\
\multicolumn{5}{c}{\it Hyades} \\
Singles  & F0--F4   & 17 & $-5.06^{+0.24}_{-0.43}$ & 29.15$^{+0.21}_{-0.38}$ \\
         & F5--F9   & 20 & $-4.61^{+0.11}_{-0.17}$ & 29.27$^{+0.12}_{-0.17}$ \\
         & G0--G4   &  7 & $-4.41^{+0.07}_{-0.19}$ & 29.16$^{+0.26}_{-0.03}$ \\
         & G5--G9   & 11 & $-4.37^{+0.14}_{-0.28}$ & 29.02$^{+0.13}_{-0.26}$ \\
         & K0--K4   & 20 & $-4.45^{+0.12}_{-0.12}$ & 28.61$^{+0.15}_{-0.16}$ \\
         & K5--K9   & 17 & $-4.62^{+0.44}_{-0.12}$ & 28.07$^{+0.27}_{-0.23}$ \\
         & M0--M3.0 & 50 & $-3.30^{+0.37}_{-0.92}$ & 28.60$^{+0.43}_{-0.68}$ \\        
         & M3.5--M5 & 95 & $-3.02^{+0.29}_{-0.35}$ & 28.52$^{+0.34}_{-0.71}$ \\    
Binaries & F0--F4   &  8 & $-5.51^{+0.10}_{-0.24}$ & 28.92$^{+0.08}_{-0.12}$ \\
         & F5--F9   & 15 & $-4.53^{+0.13}_{-0.26}$ & 29.41$^{+0.17}_{-0.08}$ \\
         & G0--G4   & 10 & $-4.60^{+0.16}_{-0.15}$ & 29.16$^{+0.22}_{-0.12}$ \\
         & G5--G9   & 18 & $-4.46^{+0.13}_{-0.25}$ & 29.07$^{+0.41}_{-0.31}$ \\
         & K0--K4   & 32 & $-4.42^{+0.36}_{-0.15}$ & 28.74$^{+0.44}_{-0.19}$ \\
         & K5--K9   & 18 & $-4.20^{+1.06}_{-0.41}$ & 28.75$^{+0.91}_{-0.75}$ \\
         & M0--M3.0 & 33 & $-3.14^{+0.25}_{-0.52}$ & 29.00$^{+0.36}_{-0.37}$ \\
         & M3.5--M5 & 53 & $-2.86^{+0.26}_{-0.40}$ & 28.59$^{+0.45}_{-0.56}$ \\[0.04 in]
\enddata

\tablecomments{The quoted values are median, 16$^{\rm th}$, and 84$^{\rm th}$ percentiles.}

\end{deluxetable}

%% file: tbl_Rossbies.tex
\begin{deluxetable}{@{}rrccc@{}}
%\tabletypesize{\scriptsize}

\tablecaption{Rossby--\LLX\ Relation Fitting Results\label{tbl_rossbies}}

\tablehead{
\colhead{Sample} & \colhead{$N_\star$} & \colhead{(\LLX)$_\mathrm{sat}$} & \colhead{$R_\mathrm{o,sat}$} & \colhead{$\beta$} \\[-0.06 in]
 & & \colhead{($10^{-3}$)} & \\[-0.2 in]
}
\startdata
\multicolumn{5}{c}{\it Singles} \\
Praesepe & 114 & 1.04$\pm$0.10 & 0.18$\pm$0.02 & $-3.24^{+0.34}_{-0.38}$ \\
Hyades   &  63 & 0.73$^{+0.11}_{-0.12}$ & 0.23$\pm$0.03 & $-3.93^{+0.68}_{-0.74}$ \\
All      & 177 & 0.92$^{+0.07}_{-0.08}$ & 0.19$\pm$0.02 & $-3.43^{+0.33}_{-0.36}$ \\[0.02 in]
\multicolumn{5}{c}{\it Binaries \& with RUWE $>$ 1.4} \\
Praesepe & 107 & 1.10$^{+0.14}_{-0.15}$ & 0.13$^{+0.03}_{-0.02}$ & $-2.22^{+0.28}_{-0.40}$ \\
Hyades   &  98 & 0.94$\pm$0.11 & 0.18$\pm$0.03 & $-2.82^{+0.38}_{-0.46}$ \\
All      & 205 & 1.03$\pm$0.09 & 0.15$\pm$0.02 & $-2.50\pm$0.29 \\
\enddata

%\tablecomments{}
\end{deluxetable}